\documentclass[fleqn,usenatbib]{mnras}
\usepackage{newtxtext,newtxmath}
\usepackage[T1]{fontenc}
\usepackage{ae,aecompl}
\usepackage{graphicx}	
\usepackage{amsmath}	
\usepackage{amssymb}	

\title[Tidally induced warps of spirals in IllustrisTNG]{Tidally induced warps of spiral galaxies in IllustrisTNG}

\author[M. Semczuk et al.]{
Marcin Semczuk,$^{1,2}$
Ewa L. {\L}okas,$^{2}$
Elena D'Onghia,$^{3,4}$
E. Athanassoula,$^{5}$
\newauthor Victor P. Debattista,$^{6}$
Lars Hernquist$^{7}$
\vspace*{0.2cm}\\
$^{1}$Department of Physics and Astronomy, University of Leicester, Leicester LE1 7RH, UK\\
$^{2}$Nicolaus Copernicus Astronomical Center, Polish Academy of Sciences, Bartycka 18, 00-716 Warsaw, Poland\\
$^{3}$Department of Astronomy, University of Wisconsin, 475 North Charter Street, Madison, WI 53706, USA \\
$^{4}$Center for Computational Astrophysics, Flatiron Institute,162 Fifth Avenue, New York, NY 10010, USA \\
$^{5}$Aix Marseille Univ, CNRS, CNES, LAM, Marseille, France \\
$^{6}$Jeremiah Horrocks Institute, University of Central Lancashire, Preston PR1 2HE, UK\\
$^{7}$Center for Astrophysics | Harvard \& Smithsonian, 60 Garden Street, Cambridge, MA 02138, USA\\
}

\date{Accepted XXX. Received YYY; in original form ZZZ}

\pubyear{2020}

\begin{document}
\label{firstpage}
\pagerange{\pageref{firstpage}--\pageref{lastpage}}
\maketitle

\begin{abstract}
Warps are common features in both stellar and gaseous disks of nearby spiral galaxies with the latter usually
easier to detect. Several theories have been proposed in the literature to explain their formation and prevalence, including tidal interactions with external galaxies. Observational correlations also suggest the importance of
tides for warp formation. Here, we use the TNG100 run from the magnetohydrodynamical cosmological simulation suite IllustrisTNG to
investigate the connection between interactions and the formation of gas warps. We find that in the sample of
well-resolved gas-rich spiral galaxies ($10^{10}\lesssim\mathrm{M_{*}/M_{\odot}}\lesssim10^{11}$ at $z=0$) from the simulation TNG100-1, about 16\% possess the characteristic S-shaped
warp. Around one third of these objects have their vertical morphology induced by interactions with
other galaxies. Half of these interactions end with the perturber absorbed by the host. Warps formed in
interactions are more asymmetrical than the remaining sample, however after the interaction the asymmetry
decreases with time. We find that warps induced by interactions survive on average for $<1$ Gyr. The angle between
the orbital angular momentum of the perturber and the angular momentum of the host's disk that most likely leads to 
warp formation is around 45 degrees. While our main goal is to investigate tidally induced warps, we find that during
interactions in addition to tides, new gas that is accreted from infalling satellites
also can contribute to warp formation.
\end{abstract}


\section{Introduction}

Galactic warps are vertical distortions of stellar and gaseous disks of spiral galaxies. Observationally, warps are
found in a large fraction of nearby edge-on, or close to edge-on, galaxies (\citealt{bosma}; \citealt{garciaruiz}).
They are more prominent in HI observations since usually they appear at radii where stellar disks are already fading
(\citealt{briggs}). The most common type of warp has one side of the disk bent upwards and the other downwards.
This kind of warp is often referred to as "integral-sign" or S-shaped. All three massive spirals of the Local Group have
S-shaped warps. The gas warp of the Milky Way (MW) was the first discovered warp back in the 1950s (\citealt{Kerr};
\citealt{Burke}), while the warps of M33 and M31 were detected two decades later (\citealt{rogstad}; \citealt{newton}).
Recent results from the stellar surveys of the MW shed more light and attention on the subtle stellar warp of our
Galaxy (e.g. \citealt{skowron}; \citealt{MWwarp0}; \citealt{merce}; \citealt{walter}).

Several theories have been proposed in the literature trying to explain the formation and frequency of galactic warps. They
relate the formation of warps to e.g. the misalignment between the angular momentum of the halo and the disc
(\citealt{victor0}), the cosmic accretion of the material onto the galactic disk (e.g. \citealt{juntai}), the
misalignment between the inner disk and the hot gas halo that leads to the asymmetric accretion (\citealt{rok}), the
interaction between the gaseous disk and extragalactic magnetic fields (\citealt{battaner0}; \citealt{battaner1}),
or the gravitational coupling between the spiral arms and the warp (\citealt{masset}). In this paper we
investigate a scenario that was also proposed earlier, i.e. that warps originate from tidal interactions with
external galaxies, as occur naturally within the framework of cosmological simulations.

It was demonstrated in various numerical simulations that a fly-by perturber can tidally induce vertical distortions
in disks of host galaxies. The condition for such a scenario to occur is that the orbit of the perturber is inclined
with respect to the plane of the host's disc. The perturber can be either a less massive satellite
on a tight orbit (\citealt{elena0}) or a more massive galaxy on a wide orbit (\citealt{vesperini}; \citealt{kim}).
Cosmological simulations of MW-like galaxies show that interactions with satellites can not only create vertical distortions due to tidal forces (\citealt{gomez0}) but also that the impact of the satellite can be amplified by the dark
halo wake (\citealt{gomez1}), which can generate additional corrugation patterns in the stellar disk. \cite{m33}
showed that M33 could have had its stellar and gaseous warps induced by a recent interaction with M31, where
the gas warp was shaped not only by the tidal field, but also by the mild effect of ram pressure from the hot gas halo of
M31. \cite{kim} used idealized simulations to study the process of stellar warp formation via tidal interactions.
Interestingly, these authors argue that the parameter that strongly affects the strength and longevity of warps is
the incident angle between the angular momenta of the perturber on its orbit and the stars in the host's disk. They found
that the strongest and most persistent warps are formed when this angle takes values of $45^{\circ}$ and
$135^{\circ}$, which optimizes the integration time of the interaction with the vertical component of the tidal force.

A few observational studies also indicate a connection between warps and tidal interactions. \cite{koll} argued that the
very strong stellar warp of Mkn 306 originated from the interaction with the close-by Mkn 305. The proximity of the
perturber and the lack of gas argue against other explanations (e.g. gas accretion or interaction of gas
with magnetic fields) of this warp's formation. Using a sample of 540 optical images of galaxies, \cite{combes} found
that the frequency of warps is higher in galaxies having a companion or interacting than in isolated
objects. \cite{schwarz} discovered that stellar warps are on average 2.4 times bigger in interacting galaxies than
in isolated ones. \cite{annpark} and \cite{garciaruiz} showed that stellar and gas warps are
more asymmetric when they are associated with interactions or higher density environments, respectively. This
correlation can possibly be used as a diagnostic for the formation scenario of a given warp.

In this paper we use a simulated universe of galaxies from IllustrisTNG (\citealt{tng0}; \citealt{nelson};
\citealt{springel}; \citealt{naiman}; \citealt{marinacci}; \citealt{nelson2019}; \citealt{pillepich2019}) to check whether 
these simulations reproduce interesting warp features of disk morphology and to investigate how important tidal
interactions are in their formation. Cosmological simulations like IllustrisTNG that take into account baryonic
physics, while having several shortcomings and difficulties of their own, are often successful in reproducing
the observed Universe and galaxies within, therefore they can be useful in understanding what happens in the observed
Universe. In section 2 we briefly describe the simulations we used and the sample of warped galaxies we
selected from all galaxies in the simulation. In section 3 we discuss the properties of a subtype of warped galaxies,
i.e. those with S-shaped warps and we present our results on how important tidal interactions are in shaping them.
Section 4 contains the discussion and conclusions of this manuscript.

\section{The sample of simulated spiral galaxies}
\subsection{The IllustrisTNG simulation}

\begin{figure*}
\centering

\includegraphics[width=5.8cm]{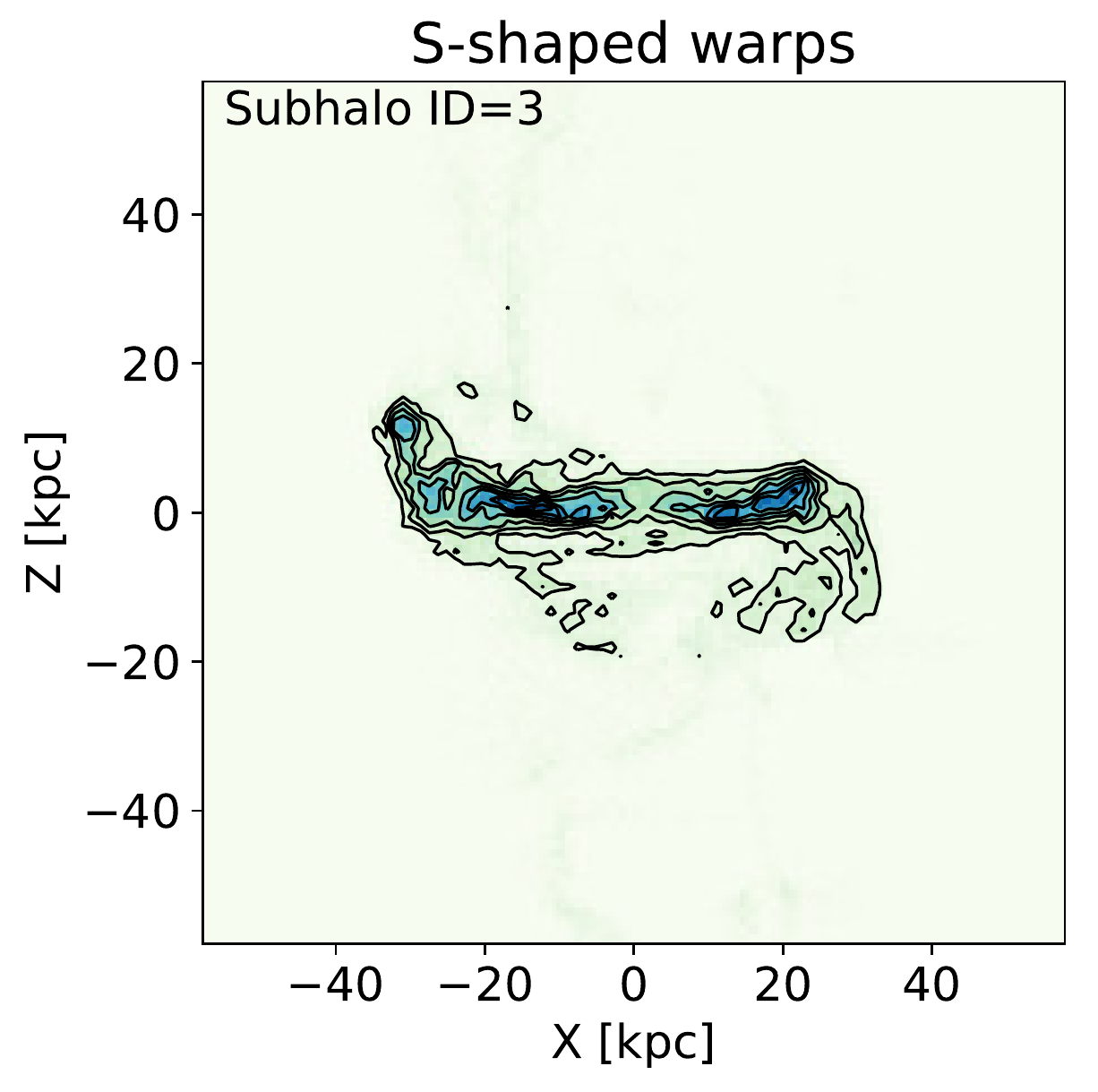}
\includegraphics[width=5.8cm]{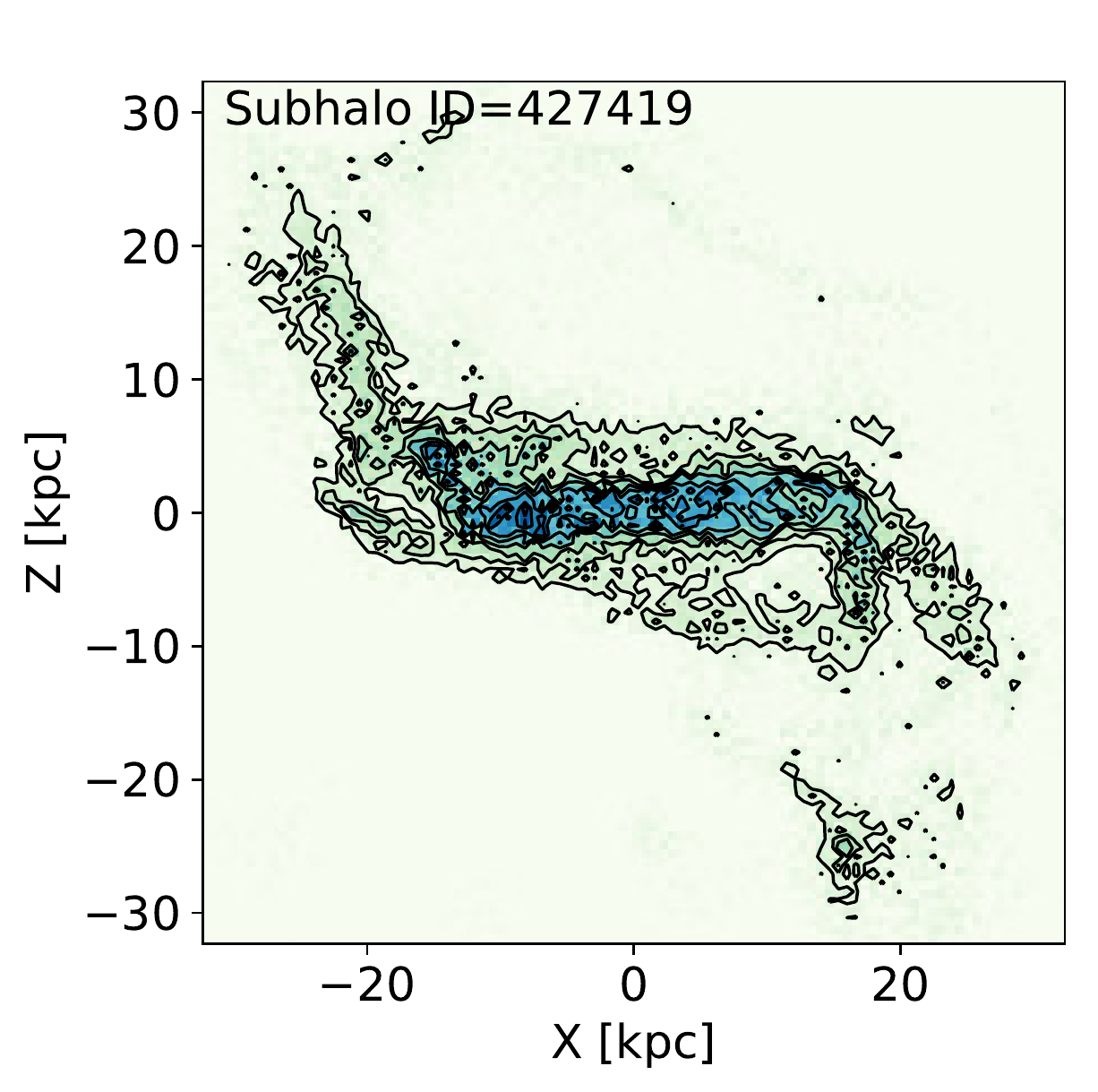}
\includegraphics[width=5.8cm]{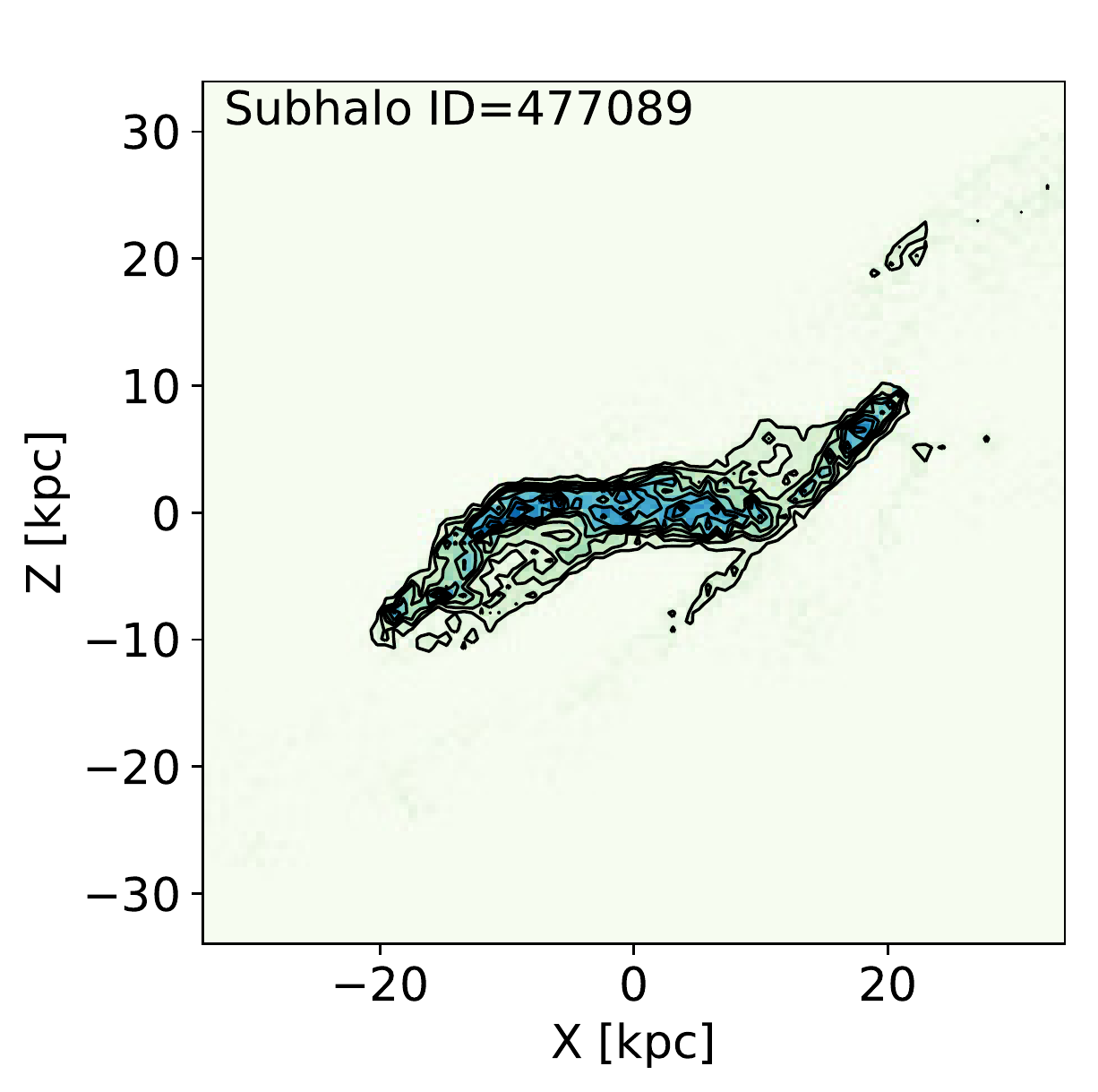}

\includegraphics[width=5.8cm]{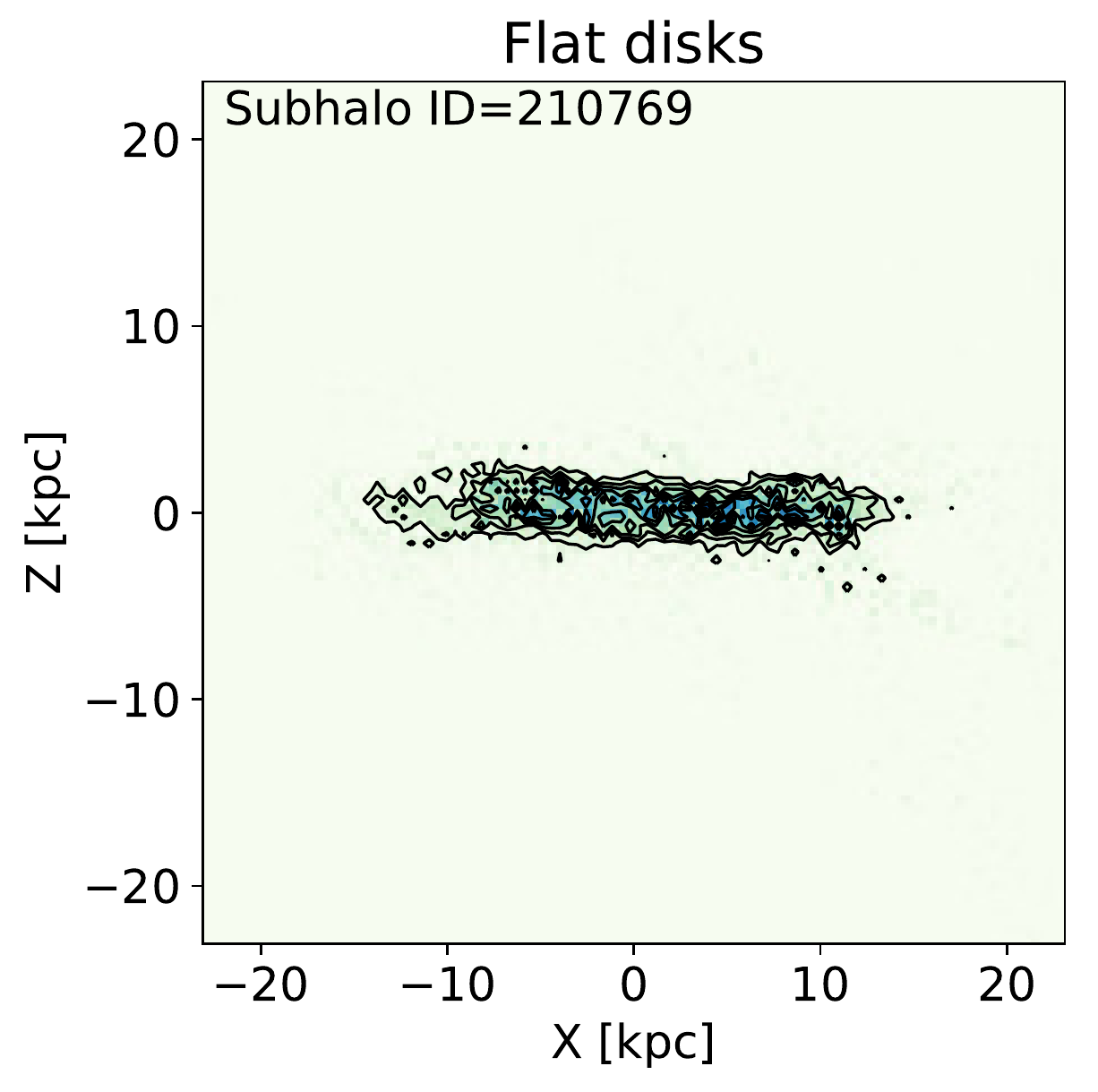}
\includegraphics[width=5.8cm]{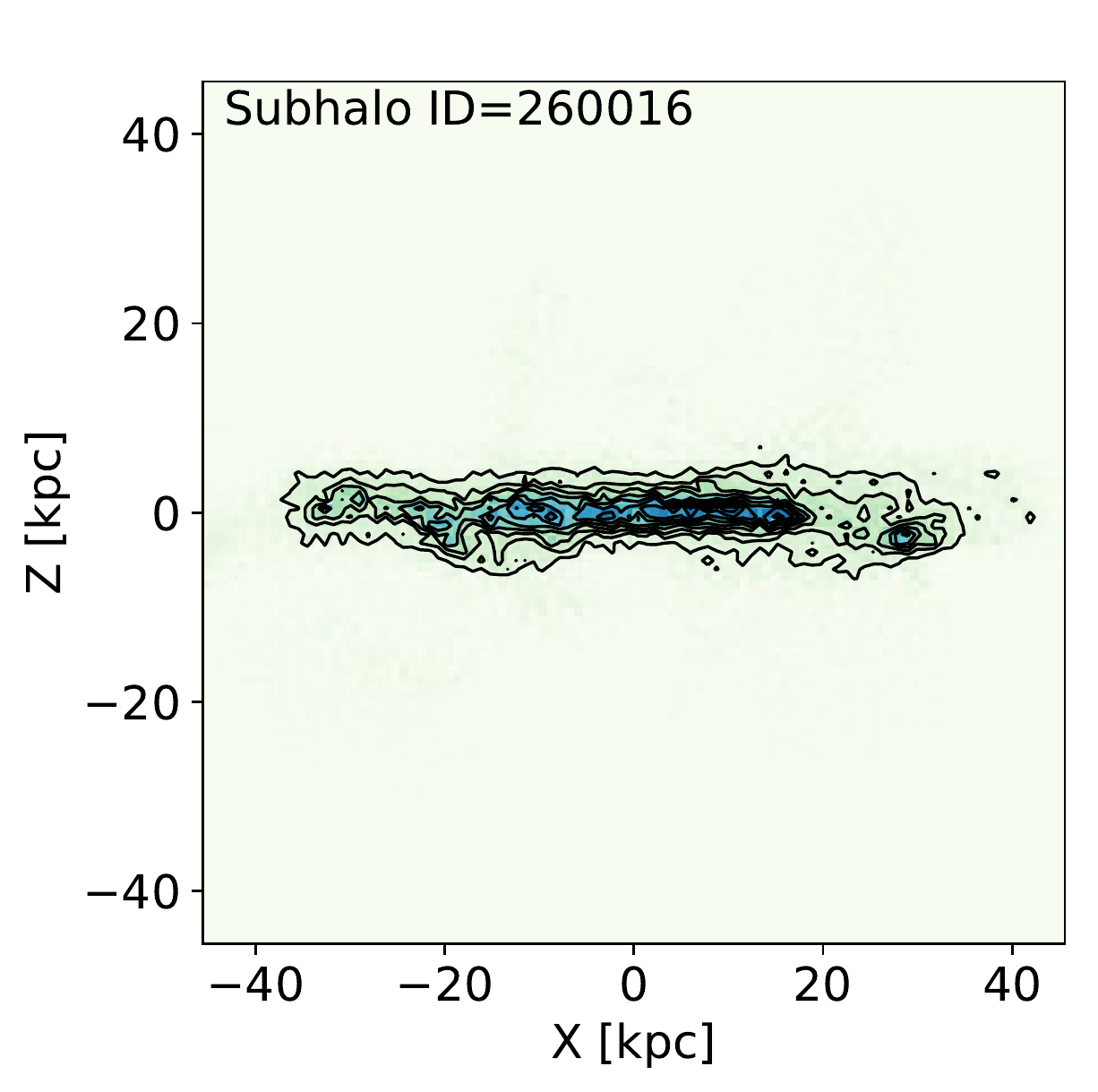}
\includegraphics[width=5.8cm]{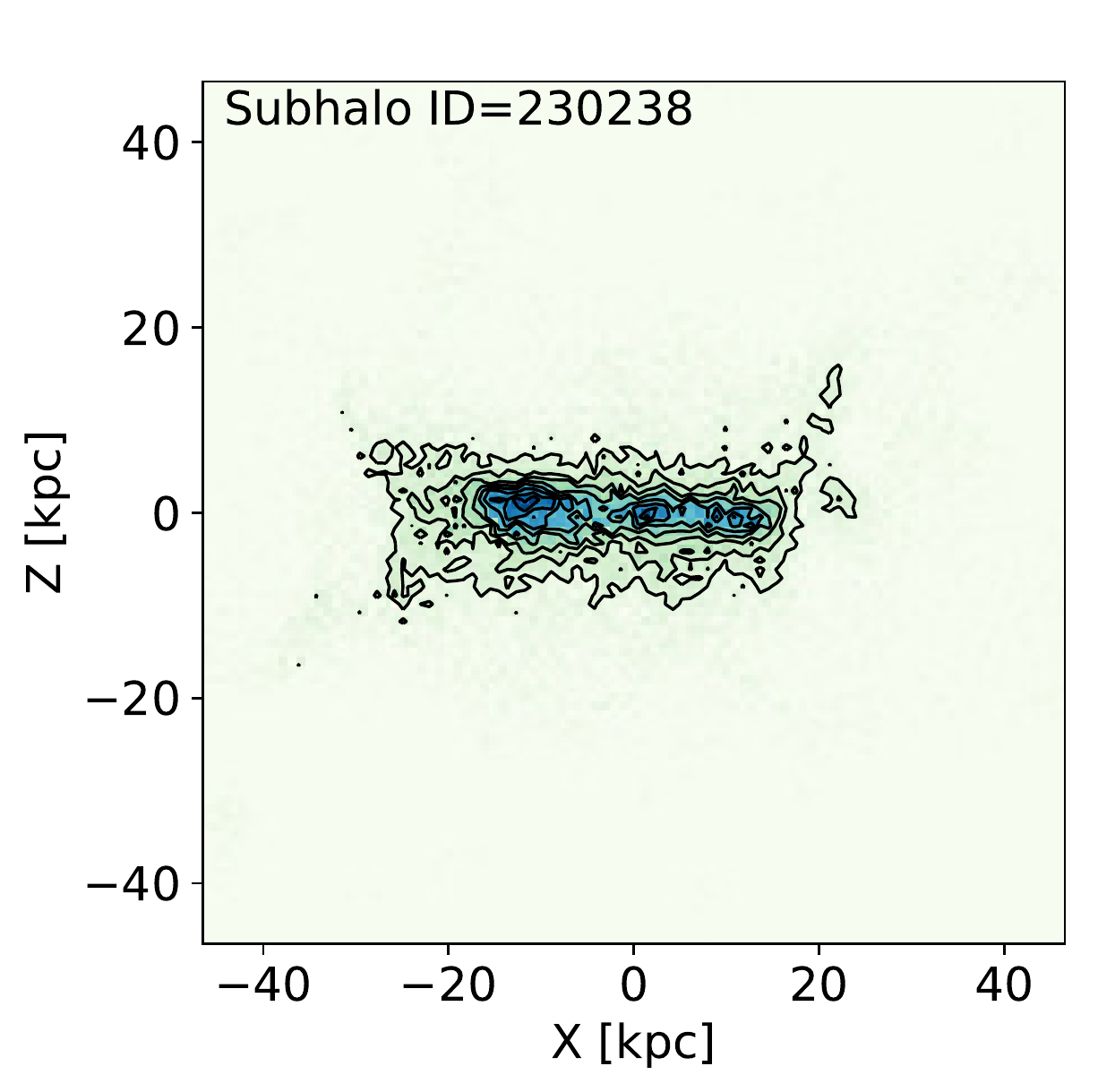}

\includegraphics[width=5.8cm]{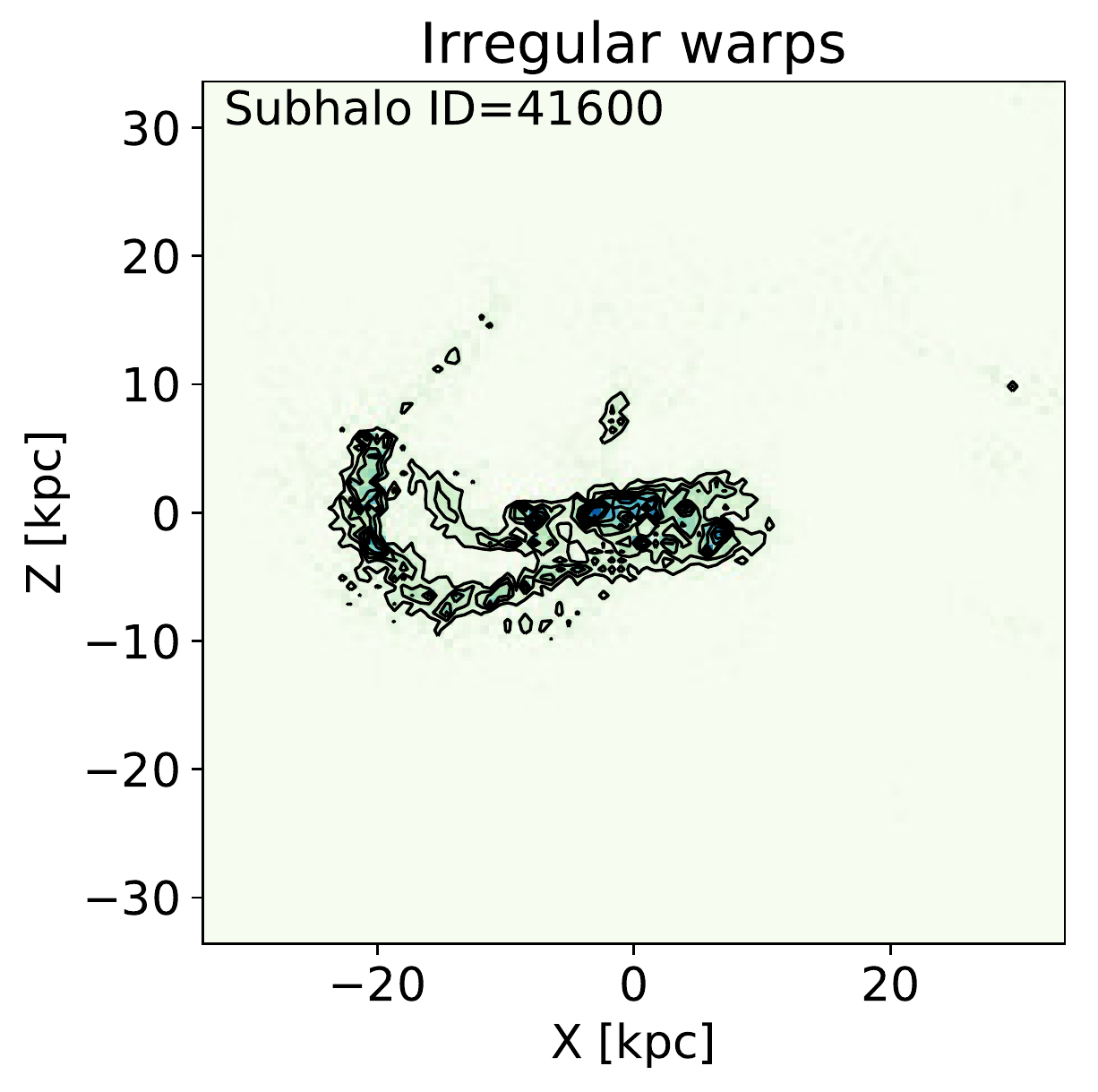}
\includegraphics[width=5.8cm]{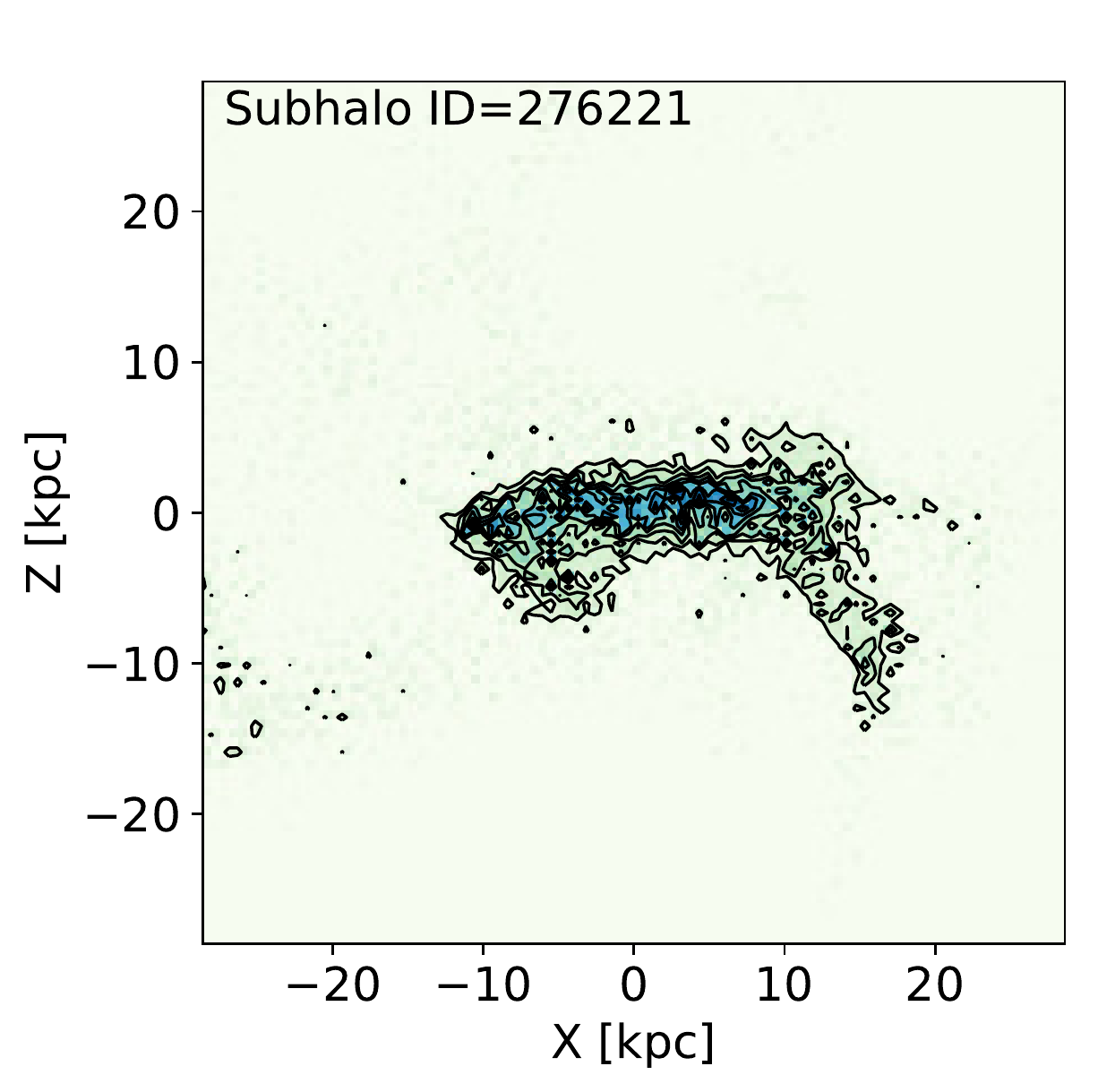}
\includegraphics[width=5.8cm]{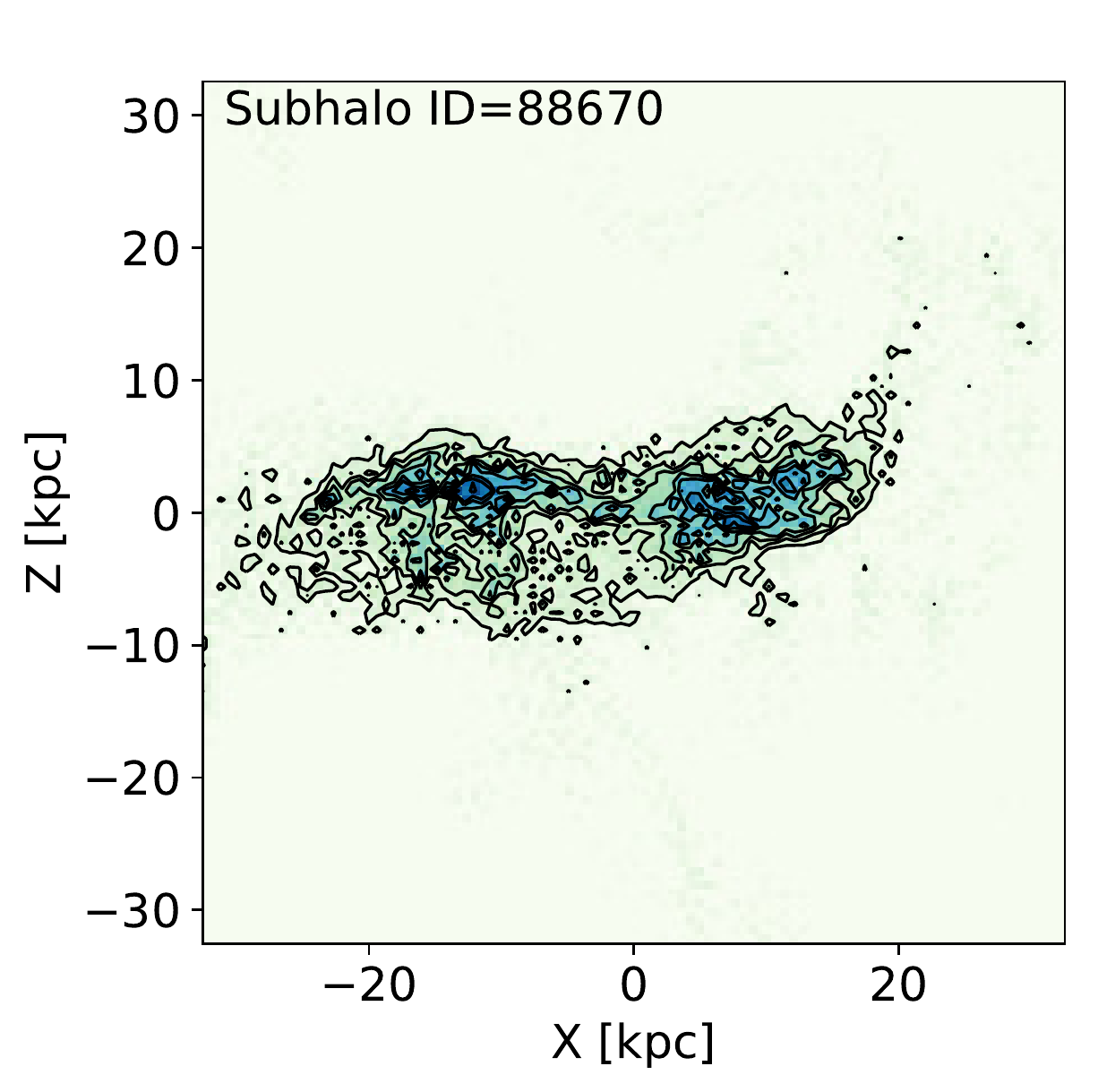}

\includegraphics[width=5.8cm]{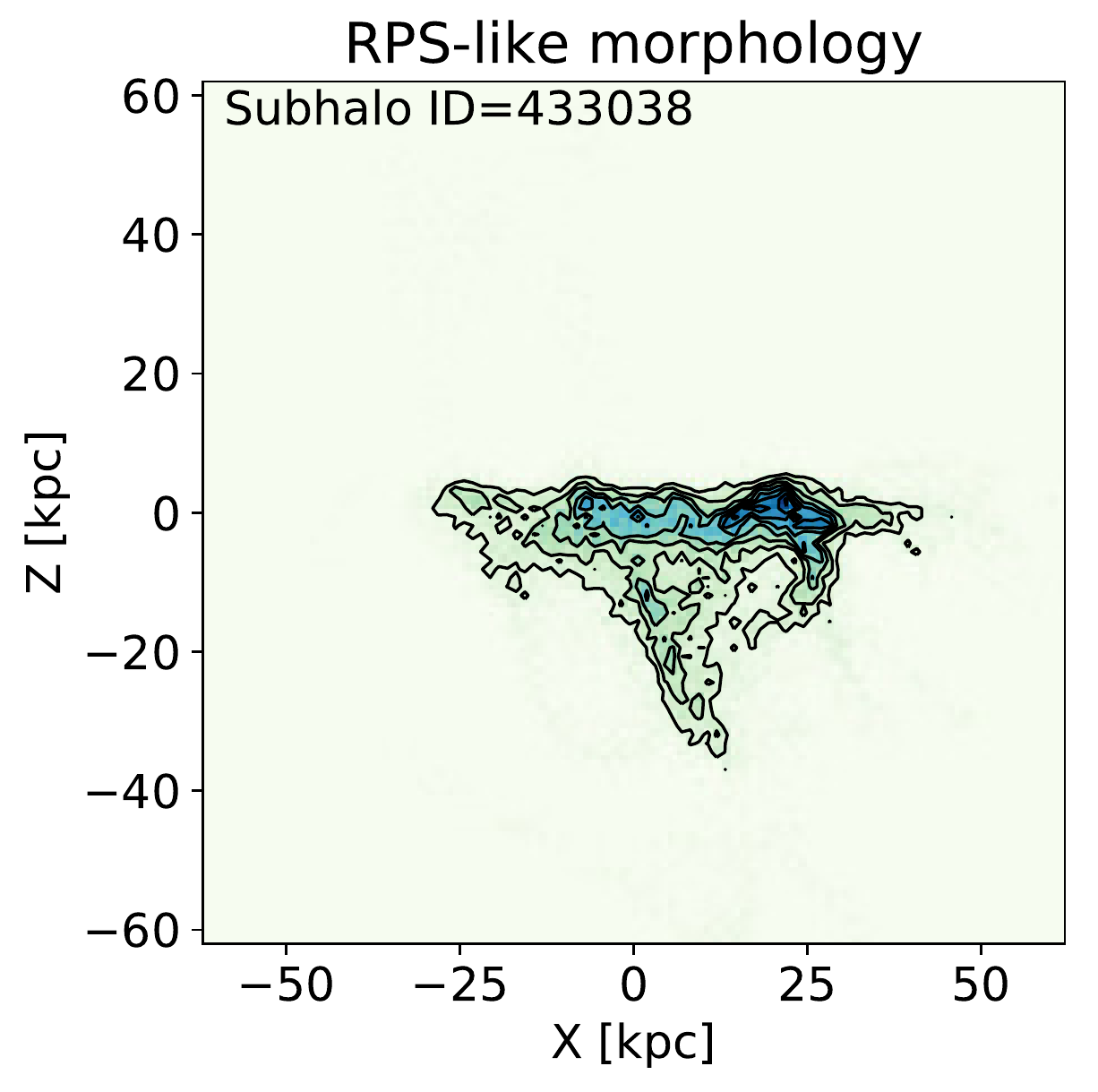}
\includegraphics[width=5.8cm]{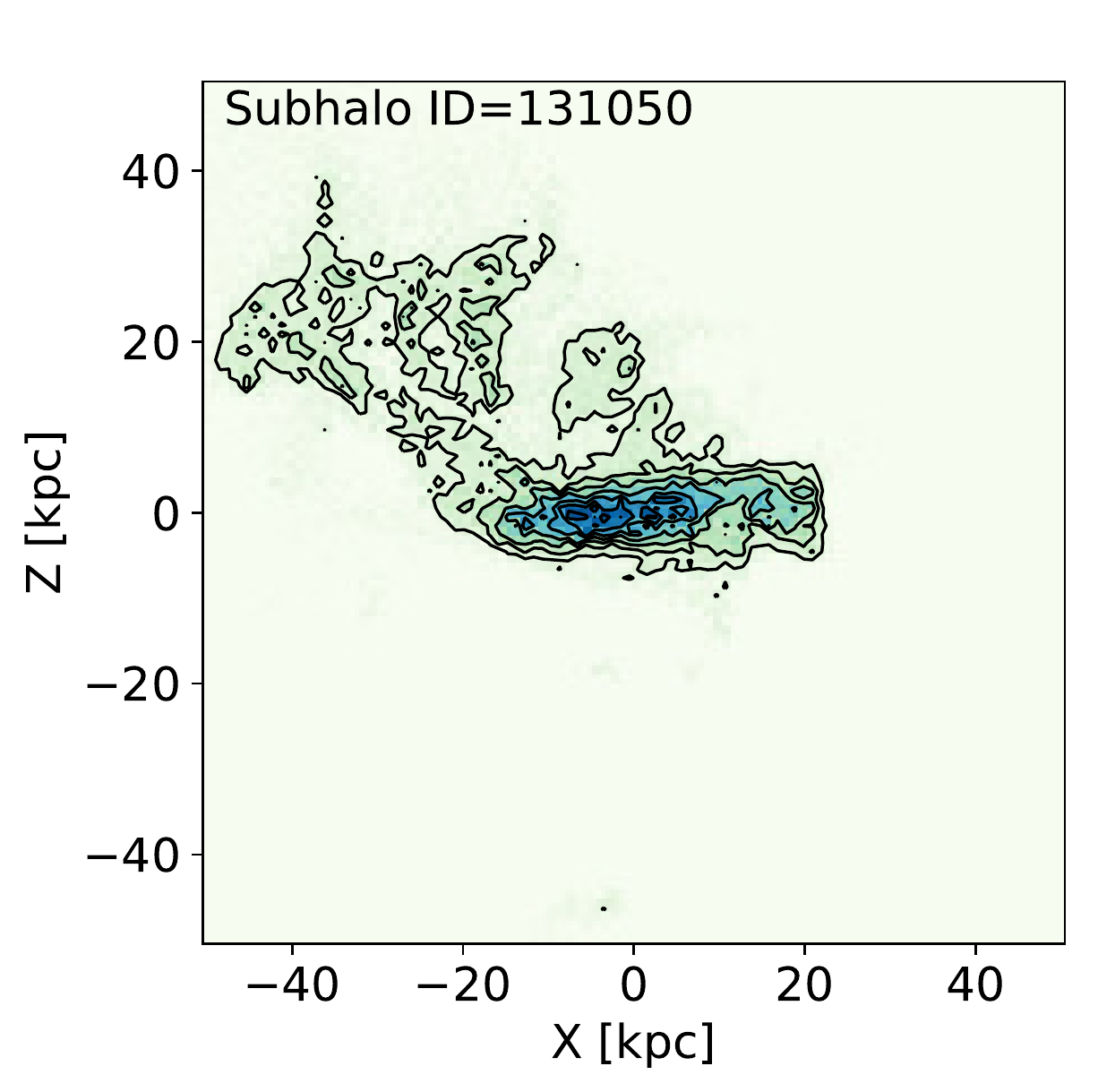}
\includegraphics[width=5.8cm]{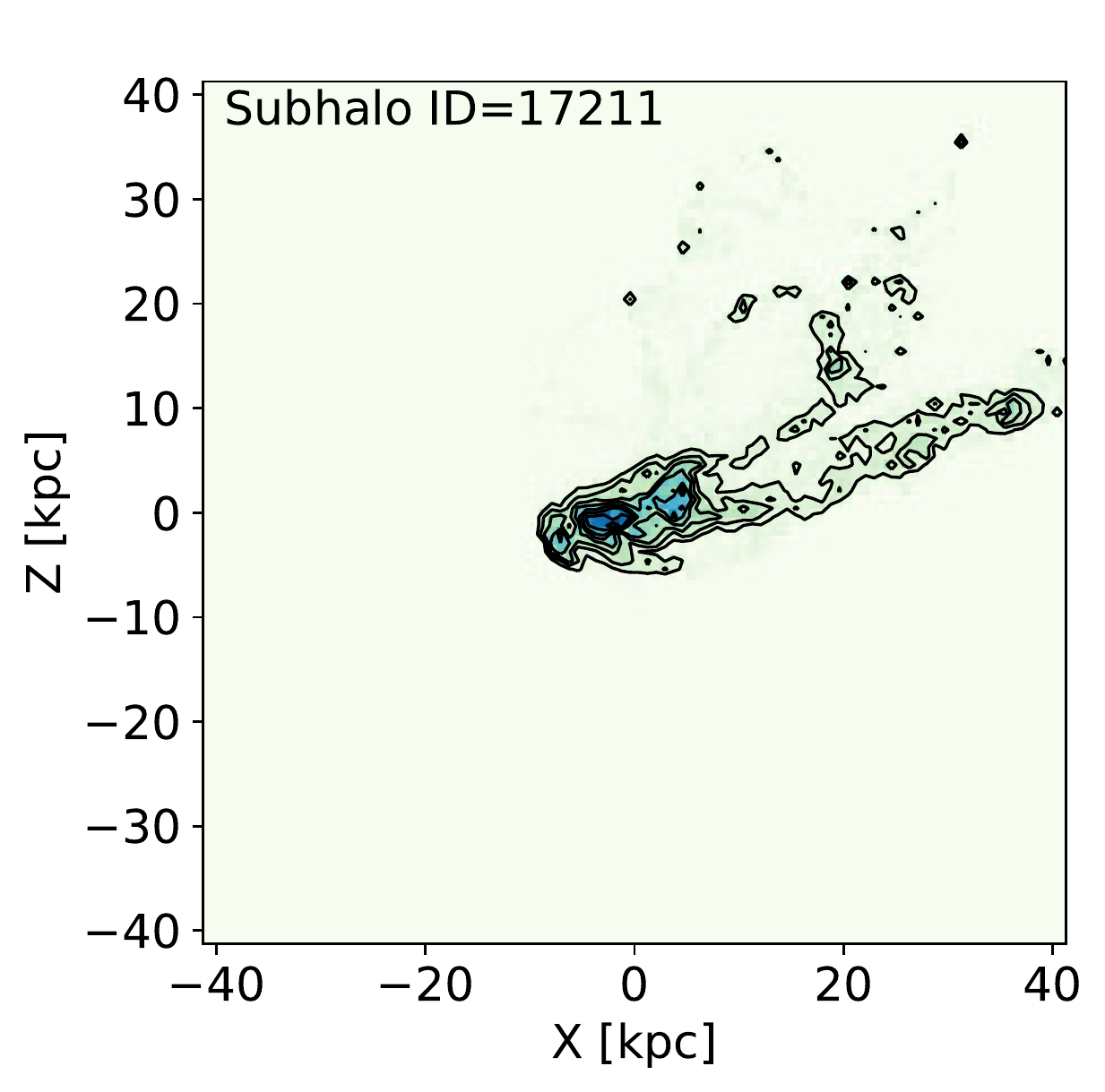}

\caption{Neutral hydrogen edge-on distributions for four different morphological types of discs considered in this
paper. Each of the four rows presents three examples of a given morphology: namely (from top to bottom) S-shaped warps, flat disks, irregular warps and the
galaxies with RPS-like morphologies. The color scale was set to be linear in gas density to highlight better the
morphological features.}
\label{morph}
\end{figure*}

To investigate warping of galactic disks we used the TNG100 simulation of the IllustrisTNG project (\citealt{tng0}; \citealt{nelson}; \citealt{springel}; \citealt{naiman}; \citealt{marinacci}), which follows the formation and evolution of galaxies within the $\Lambda$CDM cosmological paradigm.
IllustrisTNG is a continuation of the original Illustris project (\citealt{vog}; \citealt{voga}; \citealt{genel}) and uses the hydrodynamical
moving-mesh code AREPO (\citealt{arepo}). The code was modified for the purpose of the IllustrisTNG simulations to
take into account various physical phenomena relevant to the galaxy formation. These phenomena included radiative gas
cooling modulated by a time-dependent UV background, star formation via a sub-grid model as described in \cite{sh03},
galactic winds coming from metal enriching supernovae (\citealt{SNe}), two distinct channels of black hole feedback (\citealt{weinberger}), and growing black holes fed by
accretion. Cosmological parameters used in the simulation were adopted from the Planck
mission results (\citealt{planck}). In addition to the models of galactic and extragalactic physics, magnetic fields (\citealt{pakmor})
were also coupled to the hydrodynamics. The TNG100 data of the IllustrisTNG project is publicly available (\citealt{public}).

IllustrisTNG provides an improvement in the context of galactic morphology in comparison to the original Illustris.
\cite{nelson} showed that mock optical colors of simulated galaxies qualitatively resemble well the SDSS colors of galaxies.
\cite{rodriguez} compared synthetic images of IllustrisTNG galaxies with Pan-STARSS observations and found a good agreement in terms of non-parametric morphological diagnostics like e.g. Gini-$M_{20}$ diagram.
\cite{Nicolas} and \cite{rosas} used respectively Illustris and IllustrisTNG simulations to investigate bar formation in disk galaxies and both found the increase in the bar fraction with stellar mass that is in agreement with observations.
\cite{yun} used the box of 100 Mpc from the IllustrisTNG suite to investigate the so-called jellyfish galaxies that
have their morphologies shaped by the ram pressure stripping (RPS) in the hot gas halos. They found a plethora of
stripped galaxies and carefully investigated their properties and dependences on the relative orbits and positions
within clusters, which showed that IllustrisTNG has enough resolution to track the evolution of gas morphologies of
galaxies. \cite{pop} and \cite{popa} showed that the shells commonly seen in massive elliptical galaxies
naturally arise in the simulations from major mergers.  Based on TNG100 \cite{zhu} argued that the low-surface brightness
disk of Malin-1 may have originated from an interaction with a binary companion.  
Recently, \cite{kelly} analyzed pairs of interacting galaxies in IllustrisTNG, and demonstrated that
signatures of galaxy-galaxy interactions are also present in the simulation.

\subsection{Sample selection}

In observations, gas warps are found to be more pronounced and easier to detect than stellar warps. This is probably
because gas disks are less bound than stellar ones and they interact with the environment via more channels than only
through gravity. For these reasons we decided to focus our study on gas warps. To select galaxies with warped gas
disks from IllustrisTNG we classified by eye edge-on morphologies of gaseous galactic disks. This classification is
described in the next subsection. We chose the simulation TNG100-1 because the box of 100 Mpc length was the one with
the highest resolution available at the time this project was started and high resolution is crucial to investigating
galactic morphologies. In order to reduce the number of galaxies for visual classification we narrowed our sample to galaxies 
that at redshift $z=0$ have disky shape, contain enough gas material, and have sufficiently high
resolution to reproduce morphological features. These requirements were met by applying the following selection criteria:
\begin{itemize}
\item the number of stellar particles $>10^4$,
\item the number of gas cells $>10^4$,
\item the fractional mass of stars behaving kinematically as a disk (i.e. with circularity $\epsilon>0.7$),
$f_{\epsilon}>0.2$ (following \citealt{Nicolas}),
\item the flatness of stellar distribution, defined as $\gamma=M_1/\sqrt{M_2 M_3}$, where $M_1<M_2,M_3$ are eigenvalues
of the stellar mass tensor, $\gamma<0.7$ (\citealt{Nicolas}),
\item the gas mass fraction $M_{\mathrm{gas}}/M_{\mathrm{stars}}>0.01$ within twice the half-mass stellar radius
(following \citealt{yun}),
\item no peculiarities in the face-on gas distribution (e.g. big holes), judged by visual inspection.
\end{itemize}
Applying all the above criteria yielded a sample of 1593 disky gas-rich galaxies with enough resolution to analyze their
morphology. We refer to these galaxies as our default sample. The stellar mass range of this sample is $10<\log(\mathrm{M_{*}}/\mathrm{M}_{\odot})<11.89$ with the median being $10^{10.33}\;\mathrm{M_{\odot}}$.

\subsection{Morphological classification}

\begin{table}
\begin{center}
\caption{Results of the visual morphological classification}
\begin{tabular}{lll}
\hline
Type          & Number of galaxies & Percentage\\ \hline
S-shaped warps & 249 & 15.6 \\
Flat disks & 855 & 53.7 \\
Irregular warps & 454 & 28.5 \\
RPS-like morphologies & 35 & 2.2 \\ \hline
\end{tabular}
\label{tab}
\end{center}
\end{table}

We generated images of the edge-on distributions of neutral hydrogen in 6 equally spaced orientations for the galaxies
of the default sample to visually classify their morphology. The neutral hydrogen fractions were given in snapshots of the simulations and were calculated by including photoionizing rate from the ultraviolet background of \citealt{UVB} that accounts for self-shielding of ionizing photons (\citealt{rahmati}). These fractions neglect molecular hydrogen, which is mostly relevant for the centers of galaxies, but not the outer parts of disks.
 We have done the visual classification twice to reduce
human bias. Similarly to observational studies (e.g. \citealt{kuijken}) we divide edge-on gas morphologies into 4
different types:
\begin{enumerate}
\item S-shaped warps (also referred to in the literature as `integral sign' warps),
\item Flat disks,
\item RPS-like morphologies (having clear features of a jellyfish galaxy e.g. tails),
\item Irregular warps (not having any of the above three characteristics).
\end{enumerate}
Examples of galaxies of each type are presented in Fig.~\ref{morph}. We note that all images were generated in
square fields of side length equal to 12 times the stellar half-mass radius. Such a size was chosen to focus more on the
characteristic morphology of the disks rather than on external structures. This was motivated by the plan to
focus our analysis on S-shaped warps, indicative of tidal interactions. However, the small size of the fields
could lead to omission of some extended features like tails or streams and therefore we could miss some of the
RPS-like objects. We refer to the work of \cite{yun} who already extensively analyzed jellyfish galaxies of TNG100
simulations and in this paper we narrow our focus to S-shaped warps.

Table~\ref{tab} presents the results of the morphological classification in terms of relative abundances of the
different types. Roughly half of all gas disks turned out to be vertically disturbed with the other half approximately
flat. This is in agreement with the seminal observational finding called First Bosma's Law (\citealt{bosma};
\citealt{garciaruiz}) stating that at least half of the HI gas disks are warped. Among the whole sample, the RPS-like
galaxies seem to be underrepresented. In comparison \cite{yun} found 196 jellyfish galaxies at $z=0$, which is 5.6 times more than in this study. This discrepancy arises due to many differences between the classification processes done here and by \cite{yun}, from which the most significant seem to be the cut in numbers of particles. Their 3000 stellar particles extend the lower bound of the stellar mass to $10^{9.5}\;\mathrm{M}_{\odot}$, which gives much more objects that have the highest 'jellyfish fraction' ($\sim 0.5$).

We checked that the fact that galaxies belong to the specific morphological warp type does not correlate with global
parameters of these galaxies like e.g. total mass, half-mass radius or gas fraction.

\section{Characteristics of S-shaped warps}

\subsection{Warp parametrization}

\begin{figure*}
\centering
\includegraphics[width=18cm]{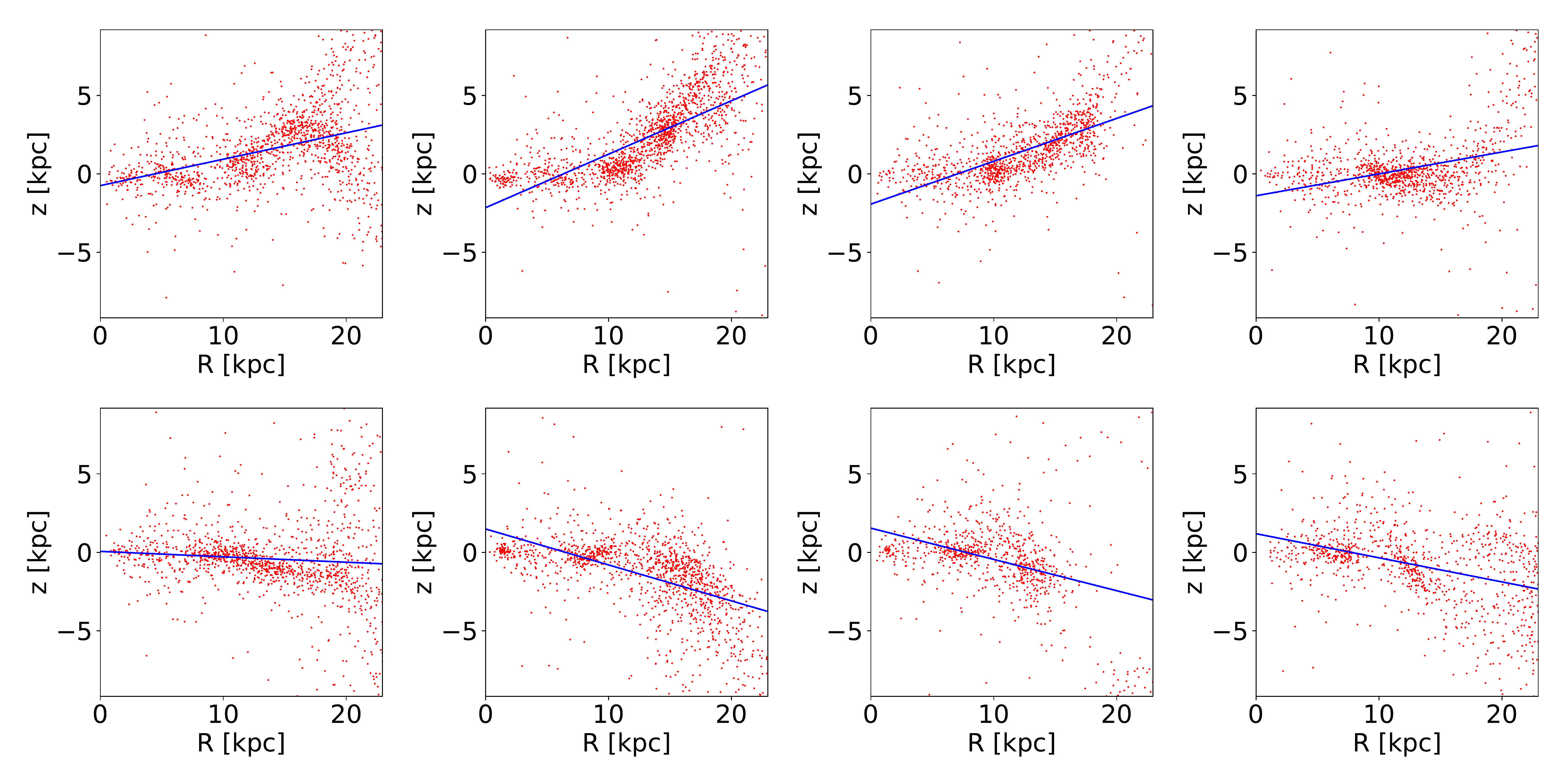}
\caption{Eight slices of gas cell distributions for an example galaxy from the S-shaped (Subhalo ID=494487 at $z=0$) sample in cylindrical coordinates with fitted lines described by equation~(\ref{we2}). Each plot represents gas cells from one-eighth of the disk that is equally sliced in azimuth. 
This figure is an example of the procedure described in
subsection 3.1.}
\label{example}
\end{figure*}

\begin{figure}
\centering
\includegraphics[width=9cm]{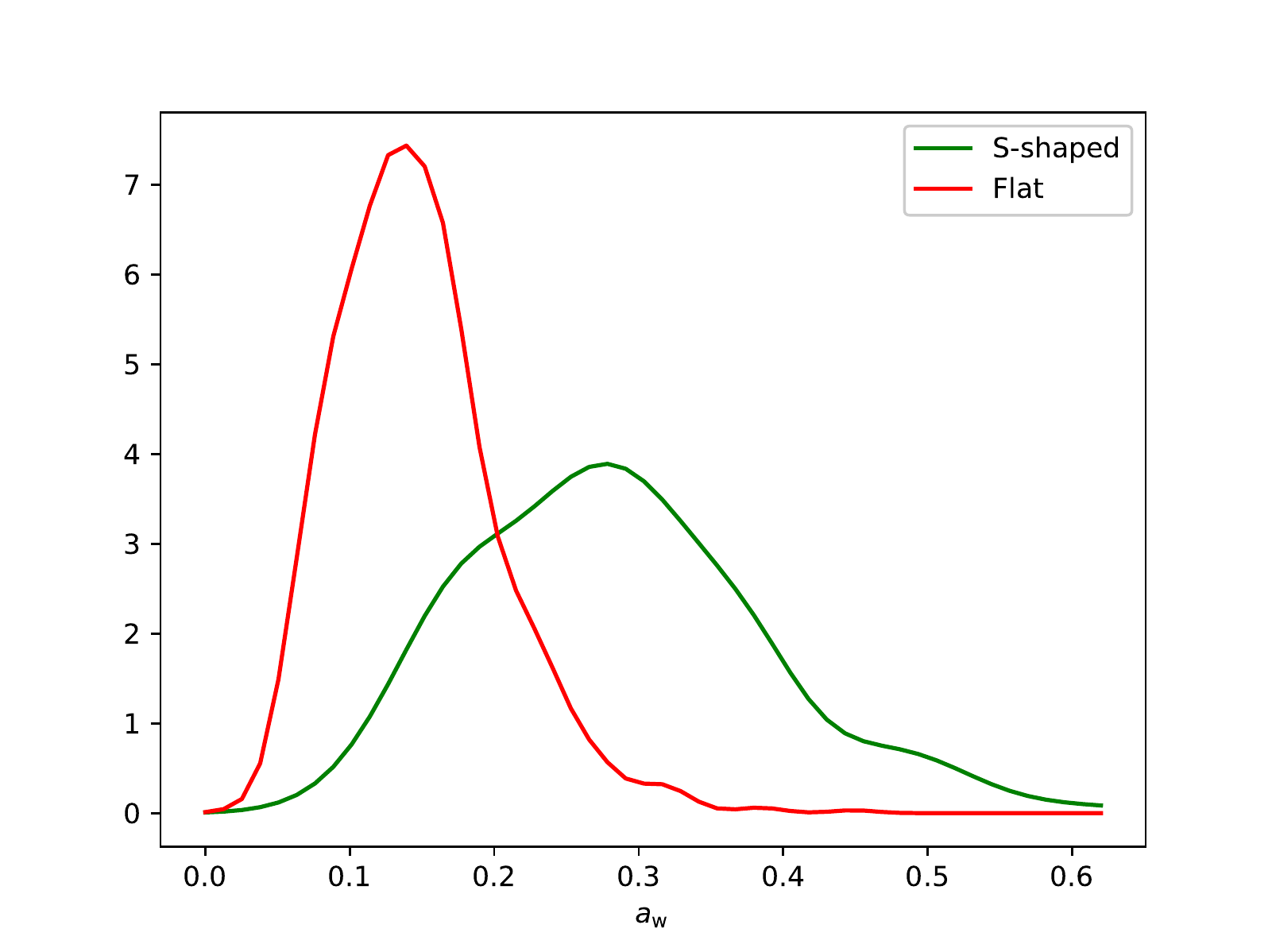}
\caption{Normalized histograms of $a_\mathrm{w}$ for visually classified S-shaped warped galaxies and flat disk galaxies calculated at
$z=0$ with optimal cuts $R_{\mathrm{cut}}=5\; R_{1/2*}$ and $z_{\mathrm{cut}}=2\; R_{1/2*}$.}
\label{awh}
\end{figure}

To find out what processes are most often responsible for warp formation we need to know how warps are evolving. In
order to characterize warp evolution we decided to use the so-called warp equation (as in e.g.
\citealt{MWwarp0}):
\begin{equation}
	z(R)=a_{\mathrm{w}}(R-R_{\mathrm{w}})^b \sin (\phi-\phi_{\mathrm{w}}),
\label{we}
\end{equation}
where $z$ and $R$ are cylindrical coordinates, and $a_{\mathrm{w}},\; R_{\mathrm{w}},\; \phi_{\mathrm{w}}$ and $b$ are
warp parameters characterizing, respectively, its inclination, onset radius, line of nodes (LON), and shape. For simplicity, we will
assume the linear approximation for warps, which means that $b\equiv 1$. The parameter that is of most interest for us
is $a_{\mathrm{W}}$ as it describes how strongly the disk it vertically warped. To find $a_{\mathrm{W}}$ for a given
gas disk at a given snapshot, we first level the galaxy by aligning its stellar angular momentum with the $z$ axis of
the coordinate system and later we divide the gas cells into eight equally spaced slices in cylindrical azimuth. The
result of this division is that some slices contain parts of the disk that are flatter and some of them have the tips
of the warp. To each slice, we fit the reduced equation~(\ref{we}), i.e.
\begin{equation}
	z(R)=a_{\mathrm{w}}(R-R_{\mathrm{w}}).
\label{we2}
\end{equation}
The factor $\sin (\phi-\phi_{\mathrm{w}})$ disappeared as slicing the disk into parts is equivalent to
deriving the azimuthal dependence. An example of the fit is presented in Fig.~\ref{example}. To optimally measure the
warp's vertical distortion we average only the two maximal values of $|a_\mathrm{w}|$ to include the parts with the
tips of the warp on both sides.

The method of quantifying warps described above is dependent on and sensitive to two parameters:
$R_{\mathrm{cut}}$ and $z_{\mathrm{cut}}$, the cuts in cylindrical coordinates from which the gas cells are
assigned to the given galaxy. Taking these values too big may result in including some cells that are part of
e.g. an external gas filament and can result in a warp fitted badly. Taking these cuts too small may end up in
retaining only the inner flat part of the disk and obtaining $|a_\mathrm{w}|$ too small, perhaps not even detecting the
warp. In order to find the optimal values of $R_{\mathrm{cut}}$ and $z_{\mathrm{cut}}$ we used visually classified
samples of flat disks and S-shaped warps and for $z=0$ fitted equation~(\ref{we2}) on a grid of values of these two
parameters. The goal of this procedure was to find the best values that would separate these two morphological types.
We found that cuts equal to $R_{\mathrm{cut}}=5 \; R_{1/2*}$ and $z_{\mathrm{cut}}=2\; R_{1/2*}$, where $R_{1/2*}$ is
the stellar half-mass radius, separated best the distributions of $a_\mathrm{w}$ for these two different types (see
Fig.~\ref{awh}, where color lines represent visually classified samples and their correspondence to the defined parameter).

Histograms for flat disks and S-shaped warps in Fig.~\ref{awh} have a significant overlap, which means that for some
values the parameter $a_\mathrm{w}$ is not uniquely classifying the galaxy as one or the other type. In order to make
sure that we analyze the clearest S-shaped warps we reduced our sample by selecting only those galaxies for which
$a_\mathrm{w}>0.21$, where the threshold value was chosen as the one at the intersection of the two normalized
histograms. This reduction limited the number of S-shaped warped galaxies to 187. We will refer to this
sample of galaxies as our reduced S-shaped sample.

\subsection{Evolution of tidal warps}

\begin{figure*}
\centering
\includegraphics[width=5.8cm]{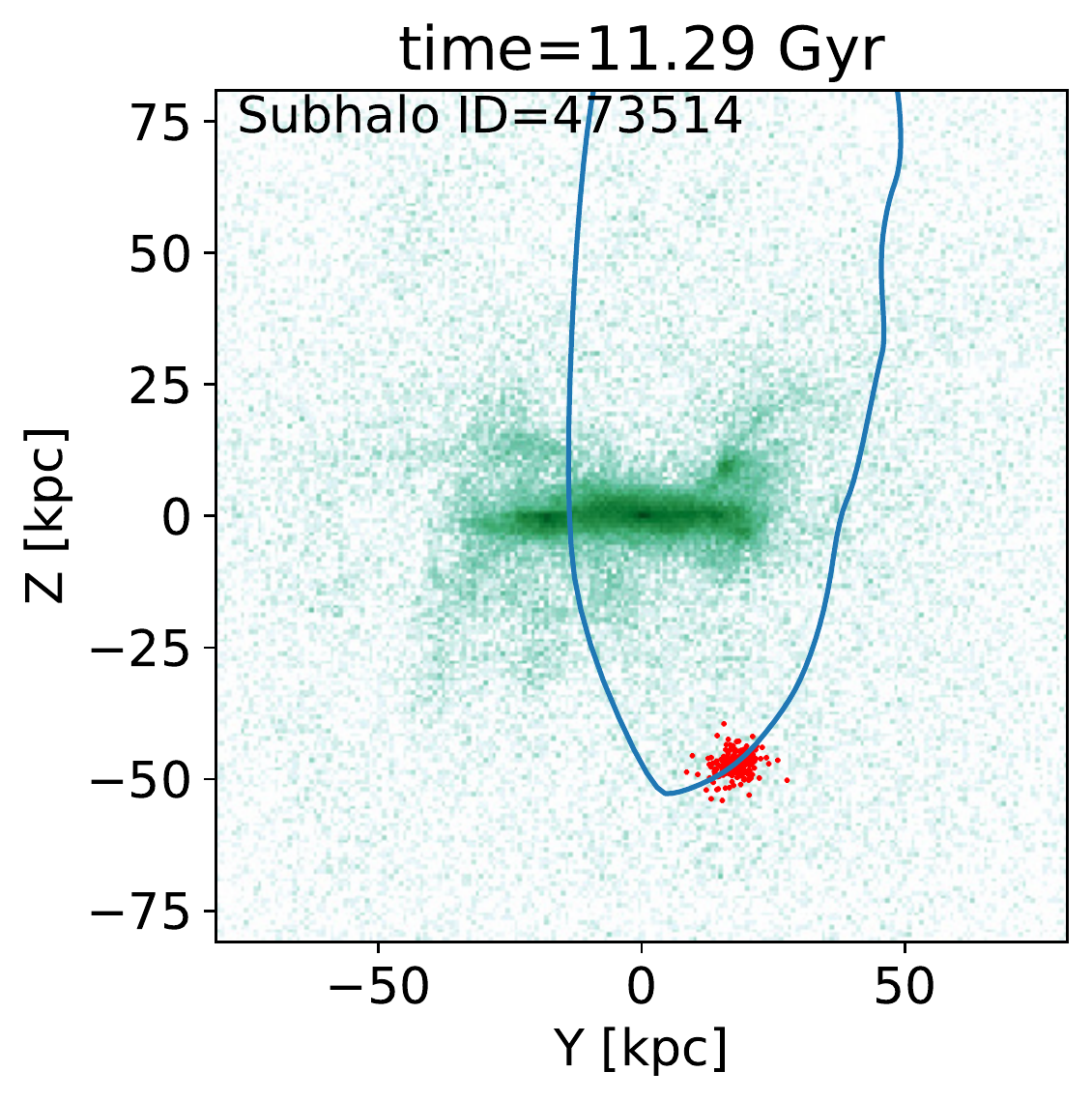}
\includegraphics[width=5.8cm]{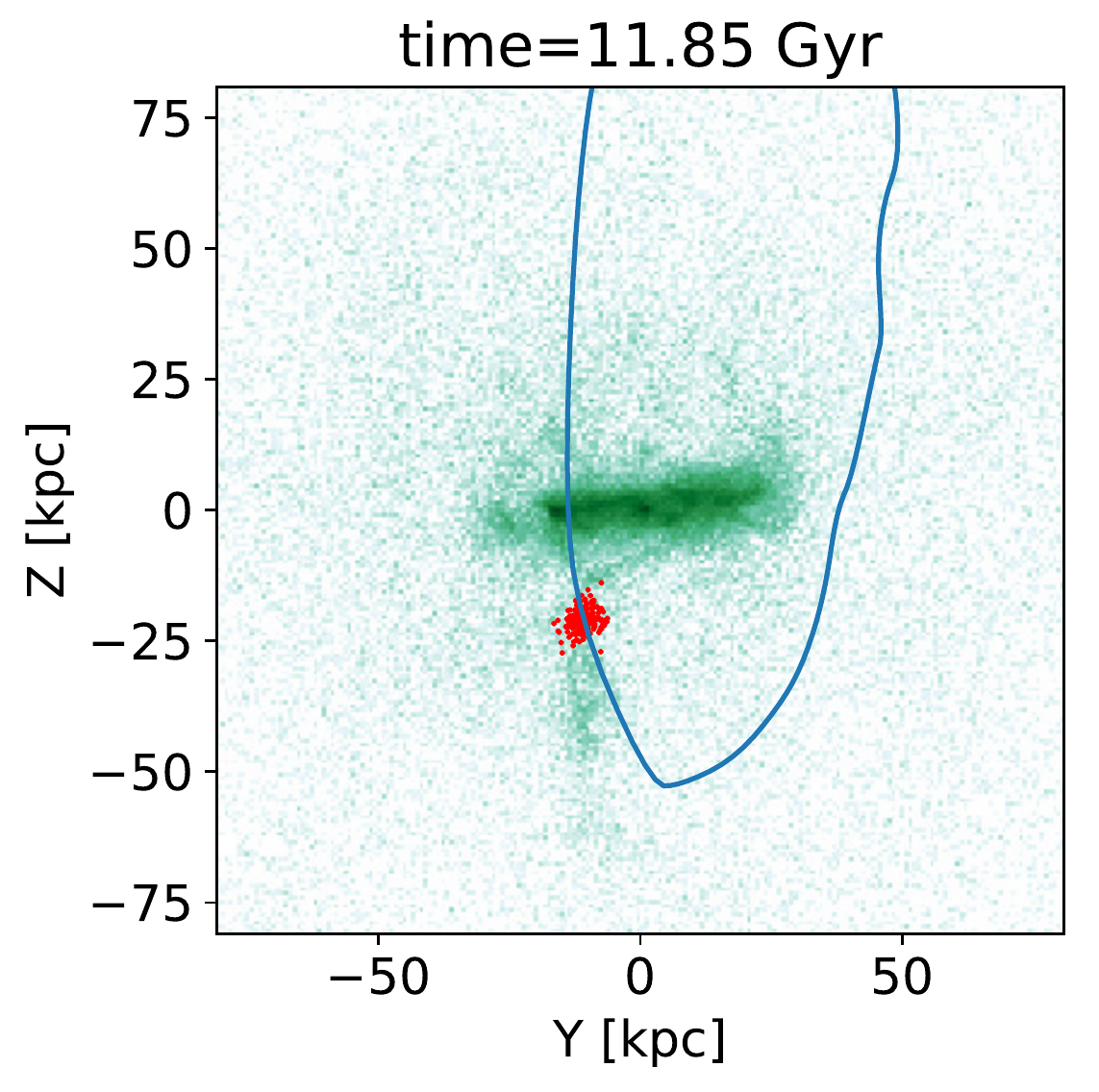}
\includegraphics[width=5.8cm]{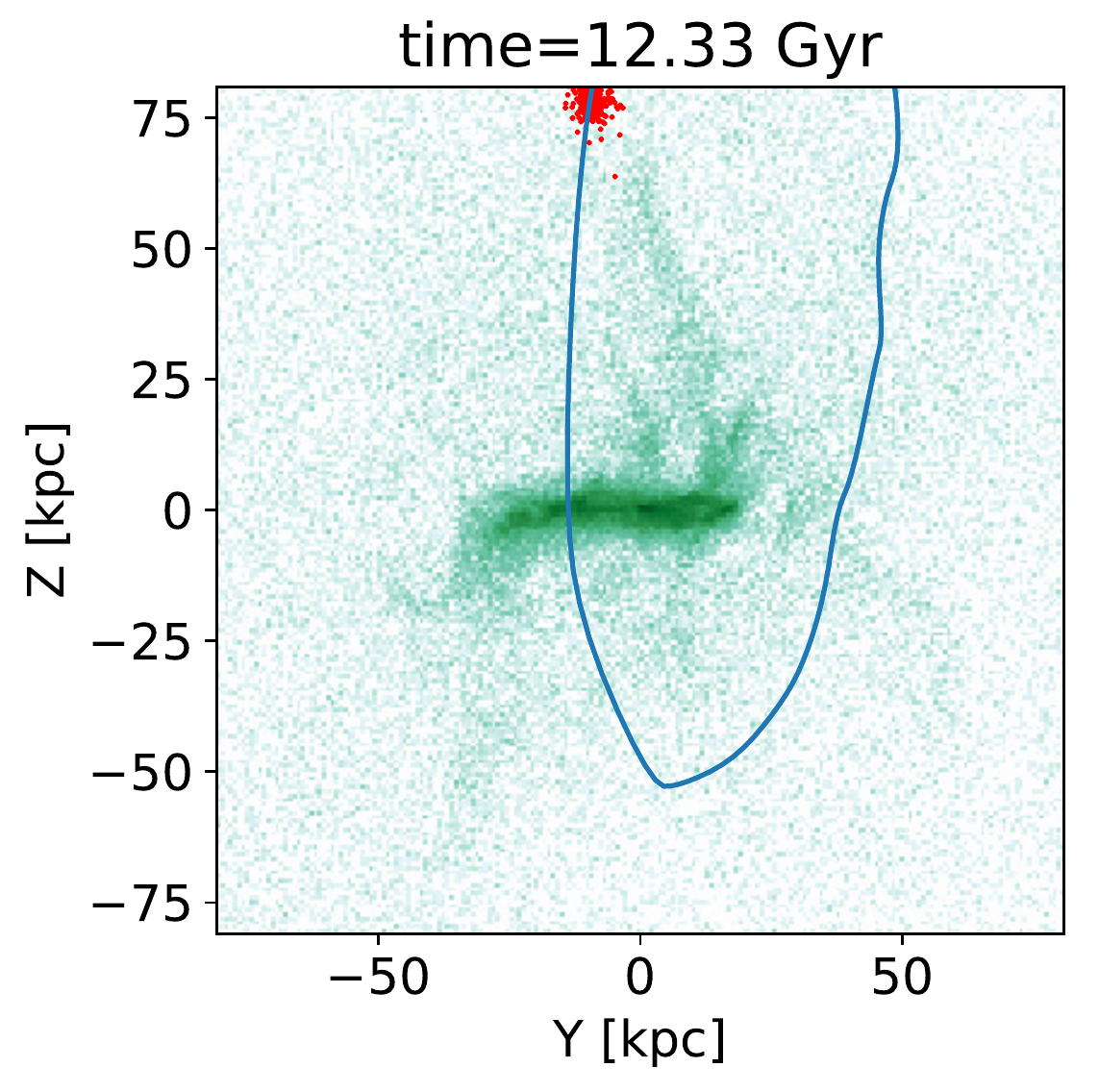}

\includegraphics[width=5.8cm]{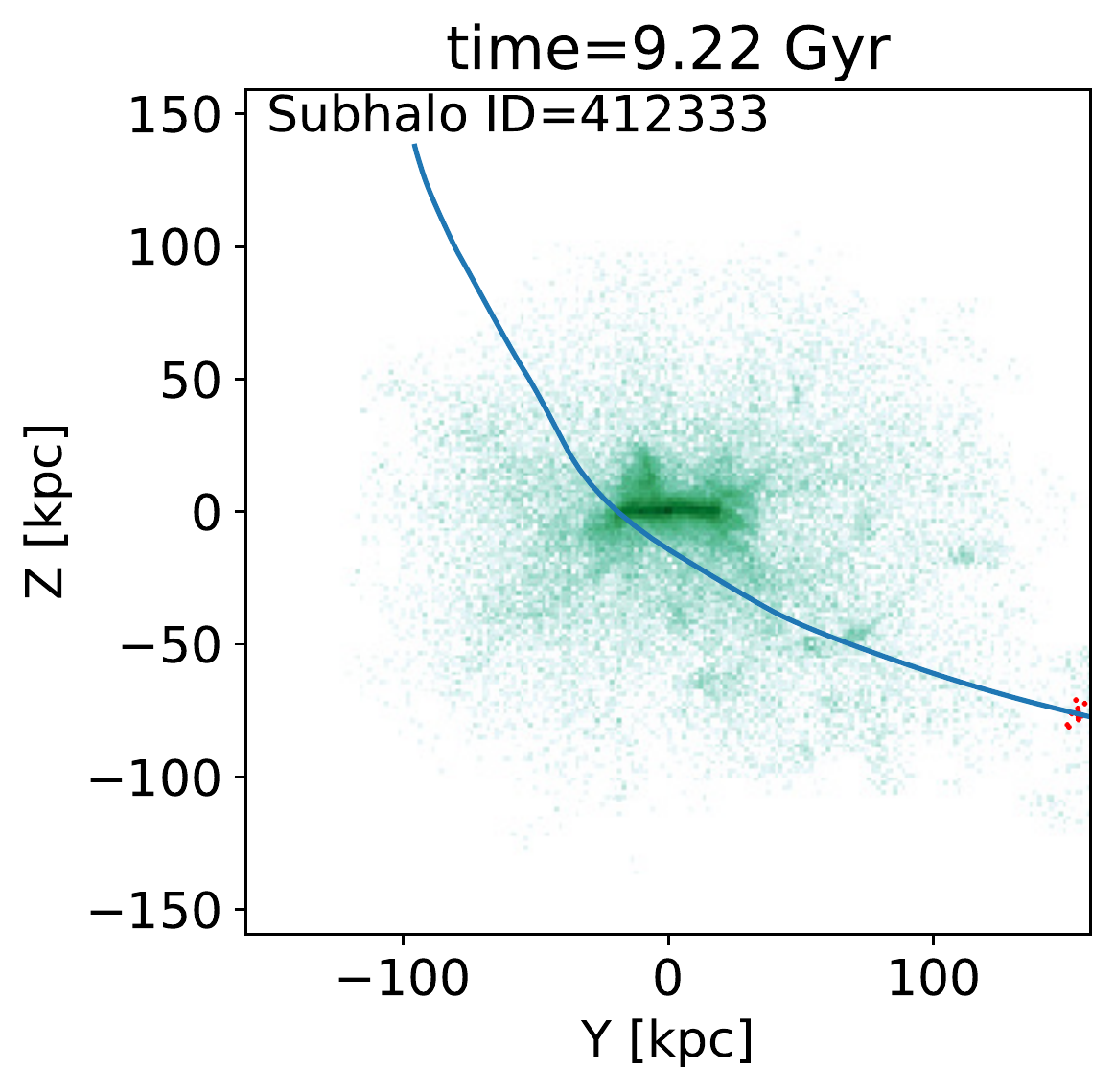}
\includegraphics[width=5.8cm]{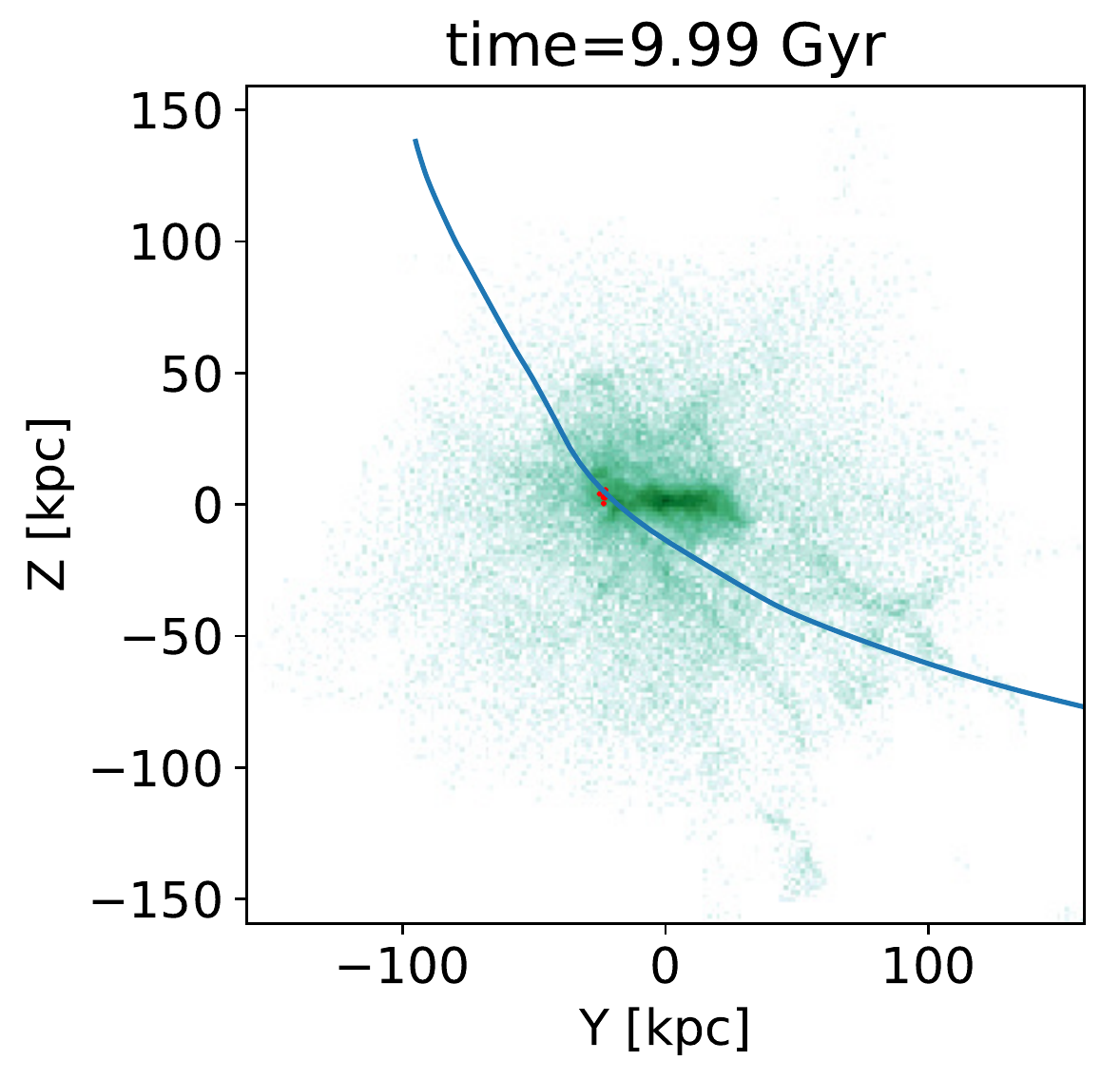}
\includegraphics[width=5.8cm]{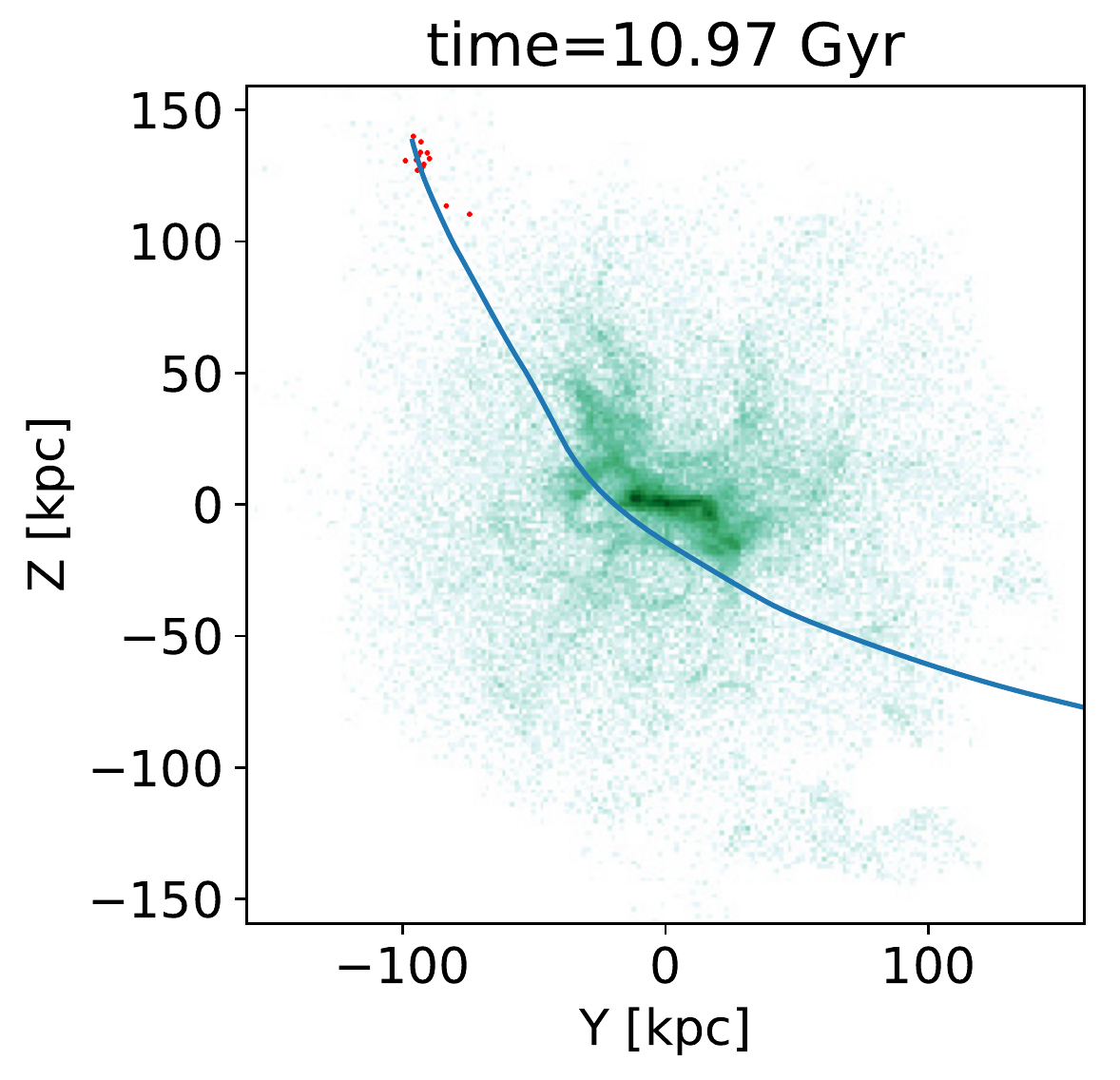}

\includegraphics[width=5.8cm]{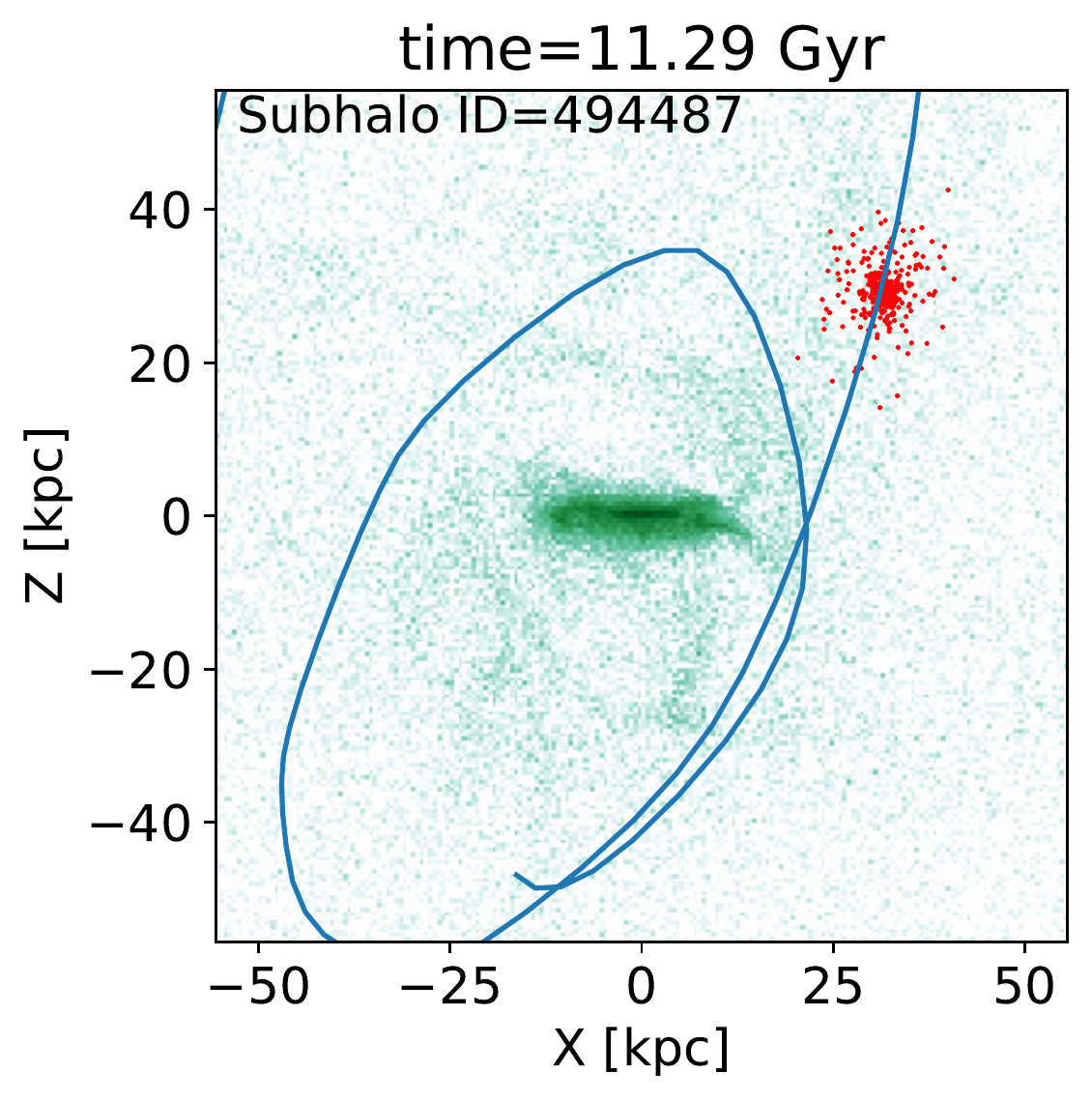}
\includegraphics[width=5.8cm]{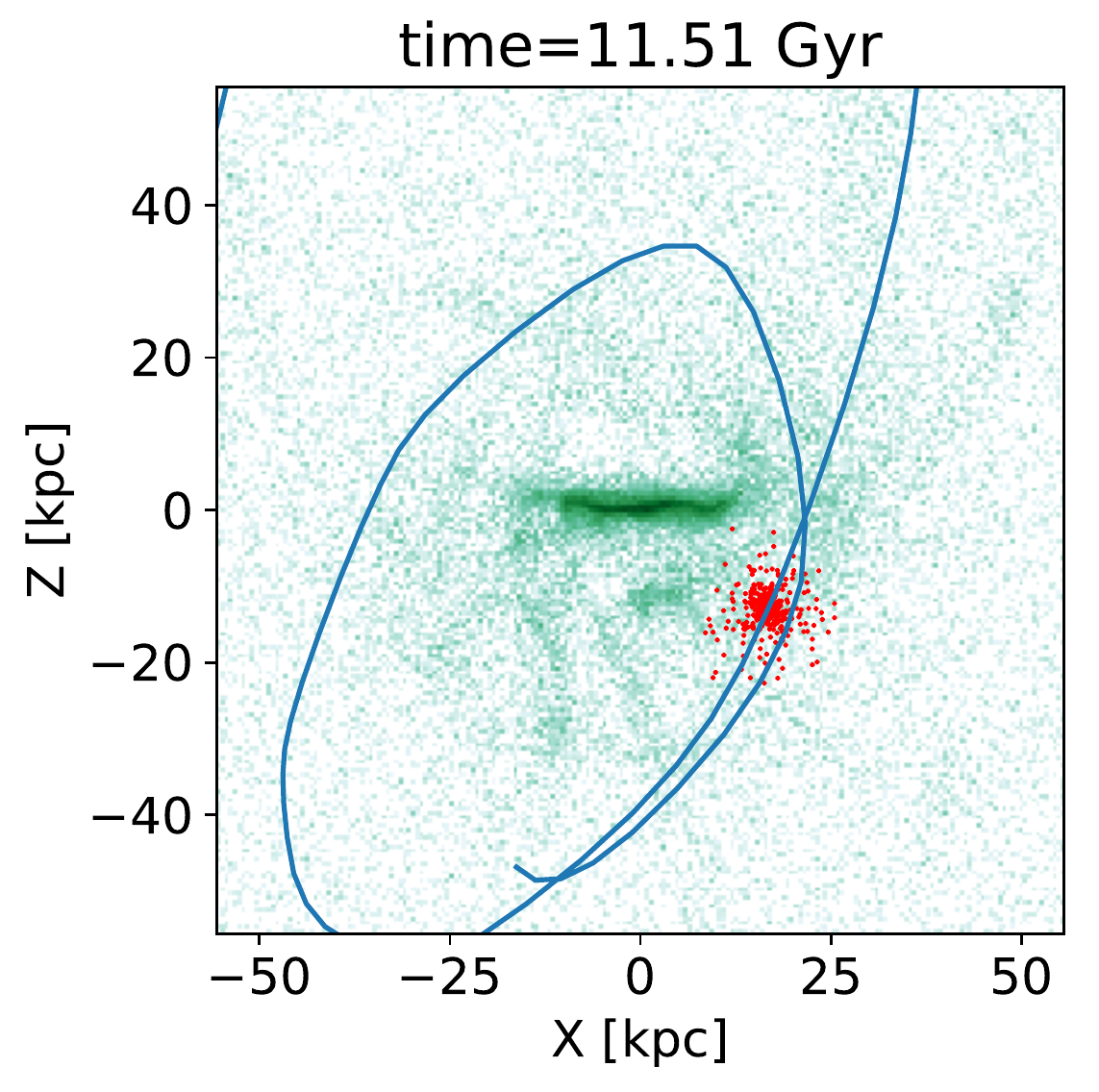}
\includegraphics[width=5.8cm]{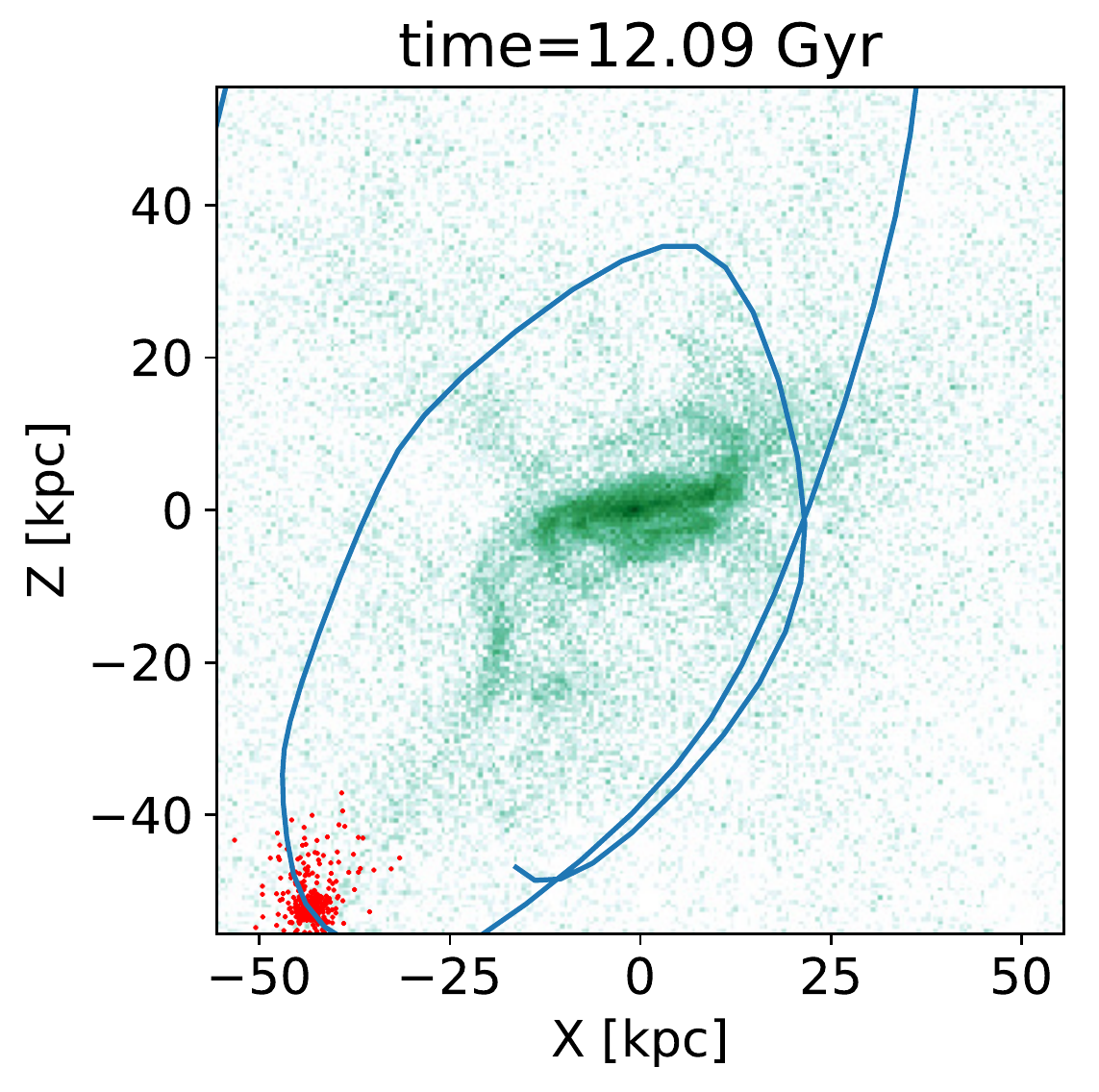}

\caption{Examples of tidal interactions between gaseous disks of galaxies and flying-by perturbers. Each row represents
a particular galaxy and each column a different snapshot. Middle column snapshots are snapshots closest to the pericenter
passage, while the left ones present images before the pericenter and the right ones after that. Red points are stellar
particles assigned to the perturber and blue lines are perturber orbits. 
For each case, the subhalo ID given is that of $z=0$.
Unlike Fig.~\ref{morph}, here the color
scaling of gas density is logarithmic to enhance the surrounding medium.}
\label{examples}
\end{figure*}

\begin{figure*}
\centering
\includegraphics[width=5.8cm]{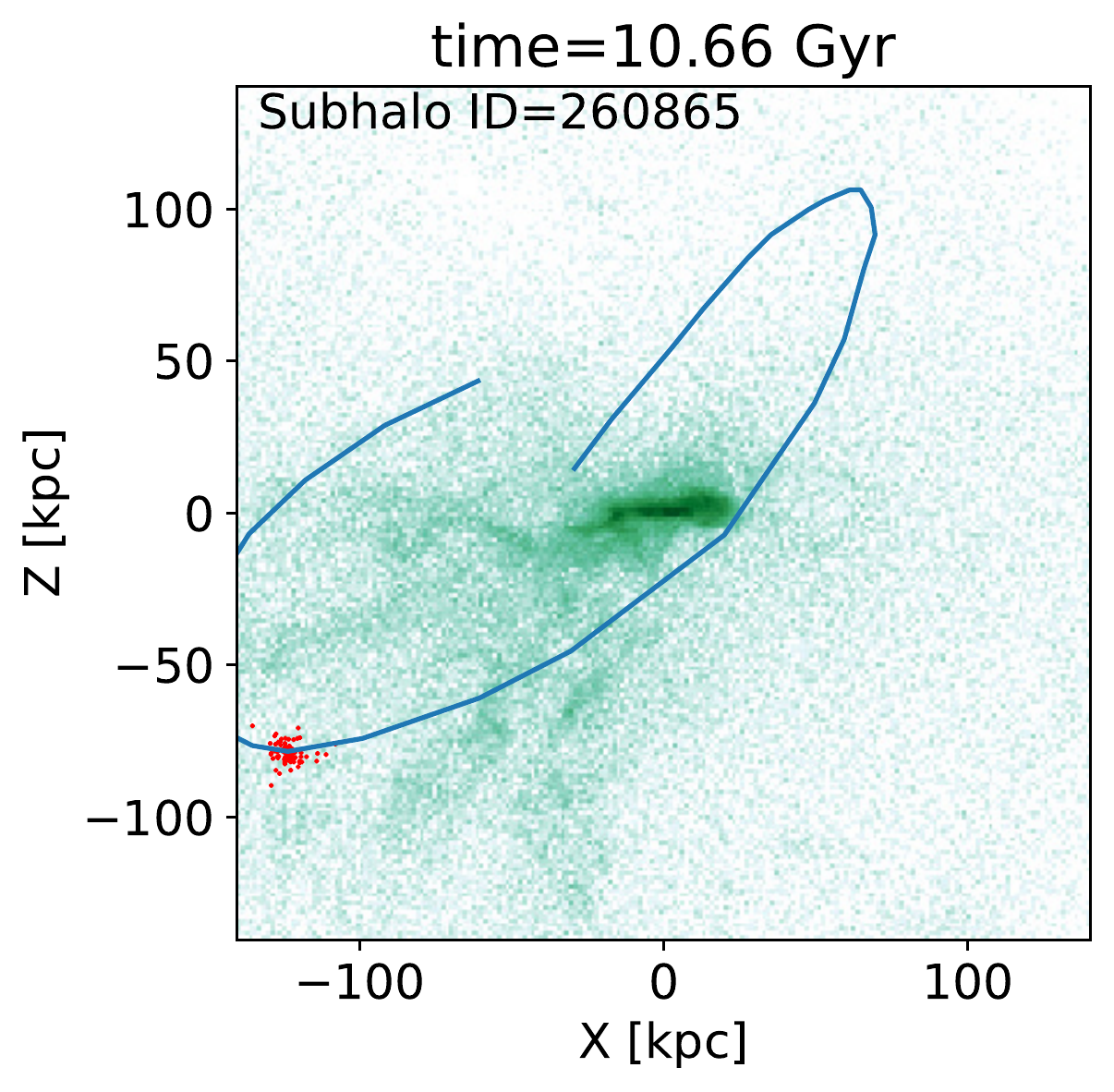}
\includegraphics[width=5.8cm]{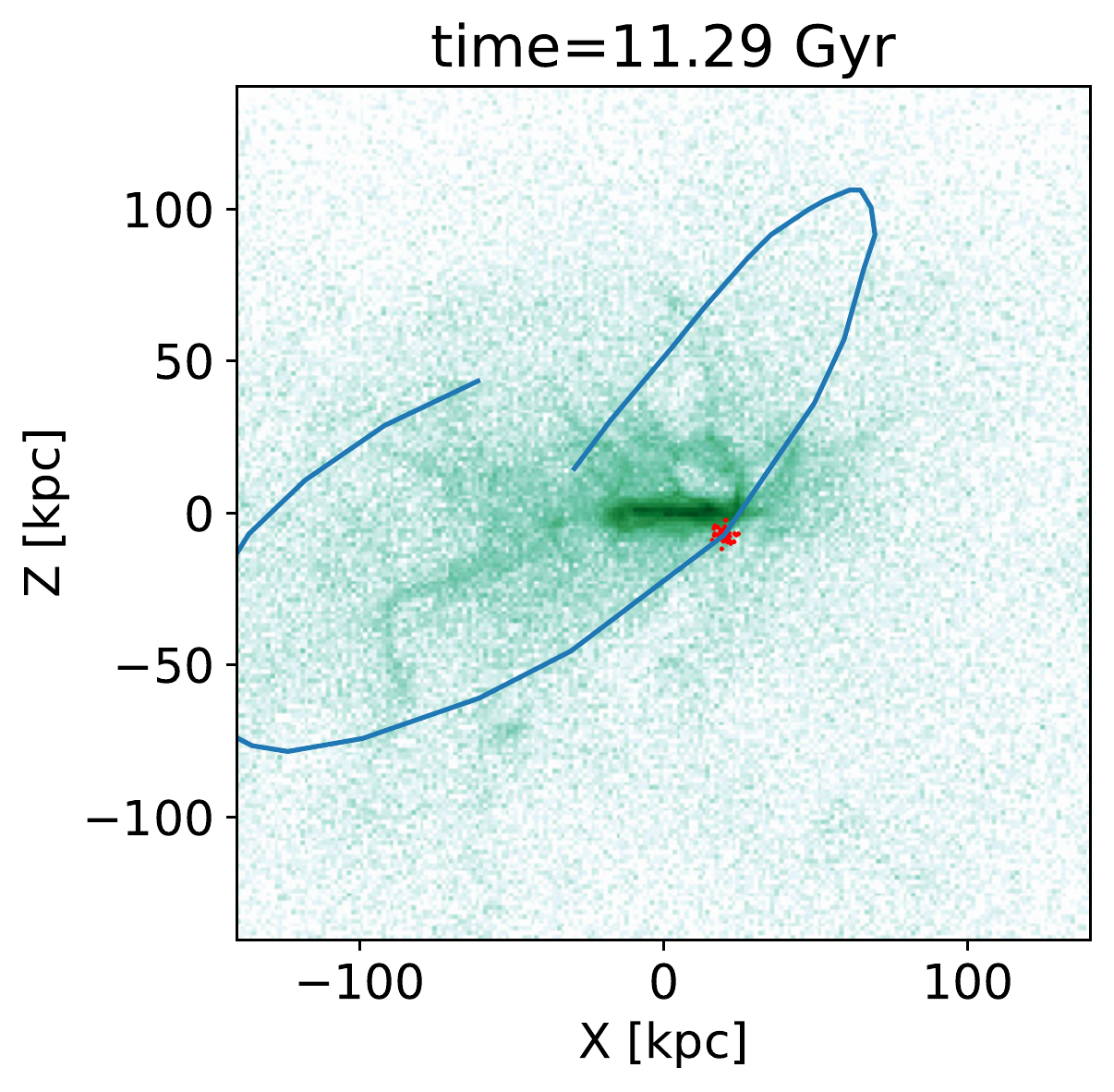}
\includegraphics[width=5.8cm]{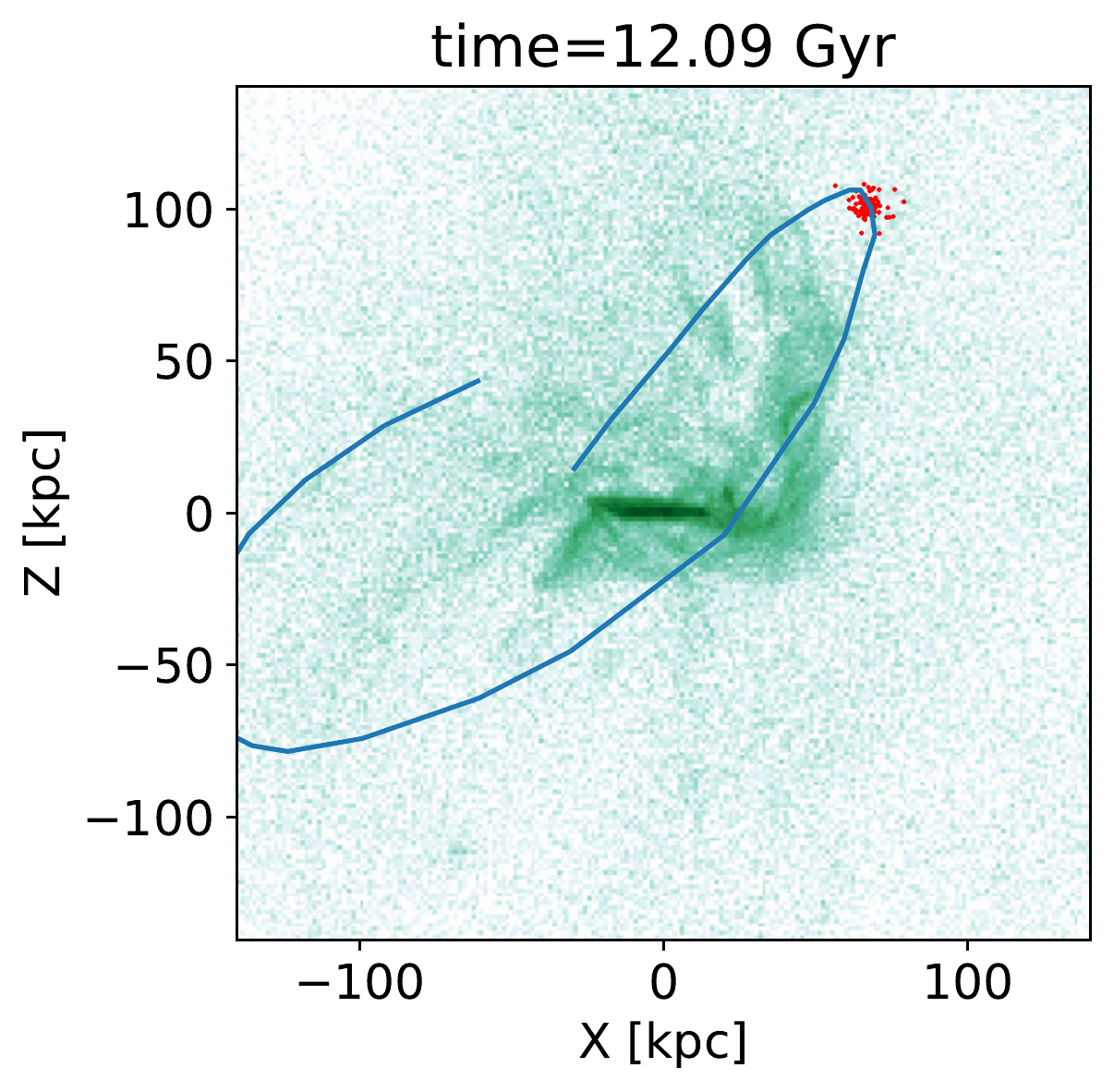}

\includegraphics[width=5.8cm]{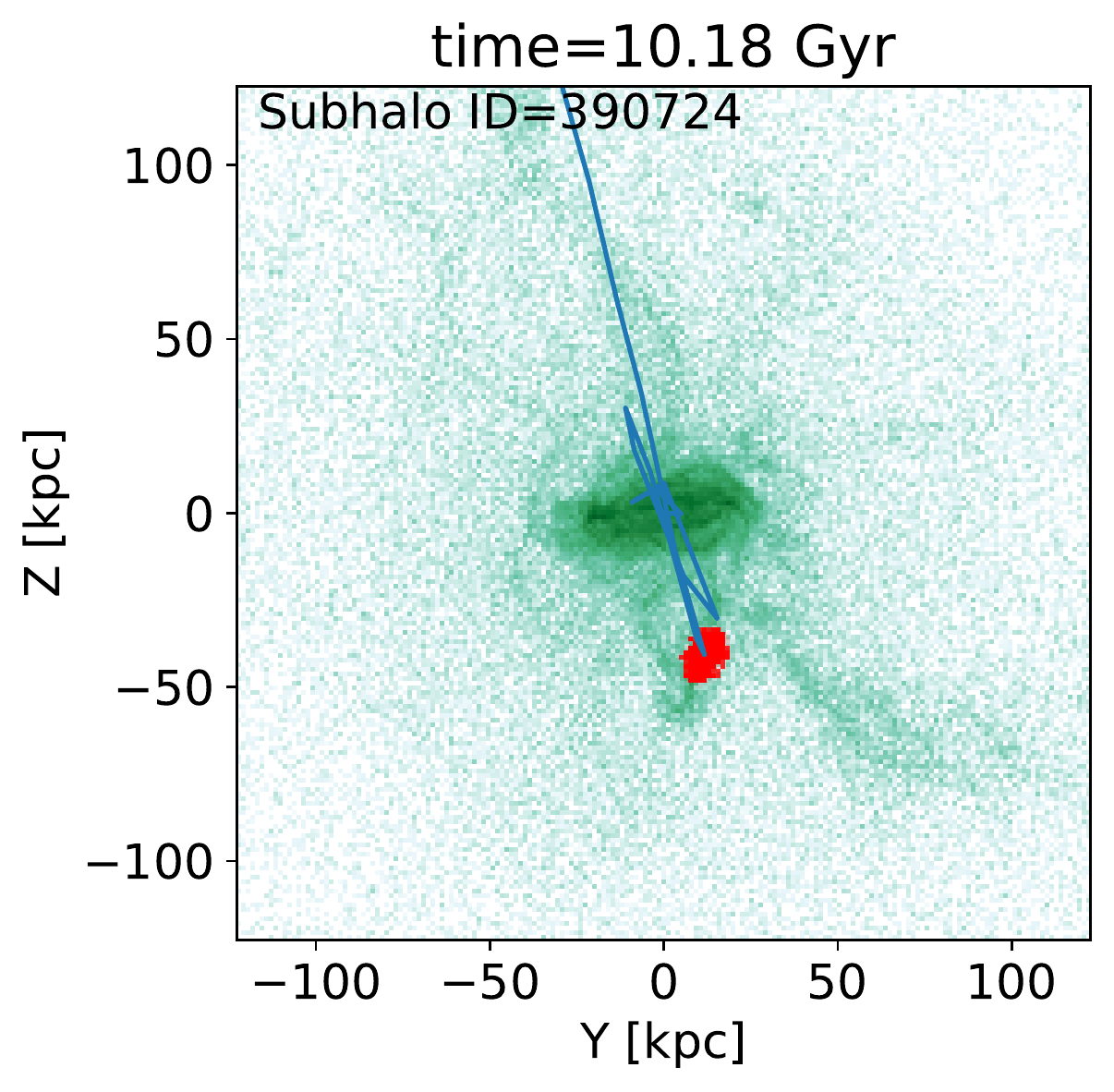}
\includegraphics[width=5.8cm]{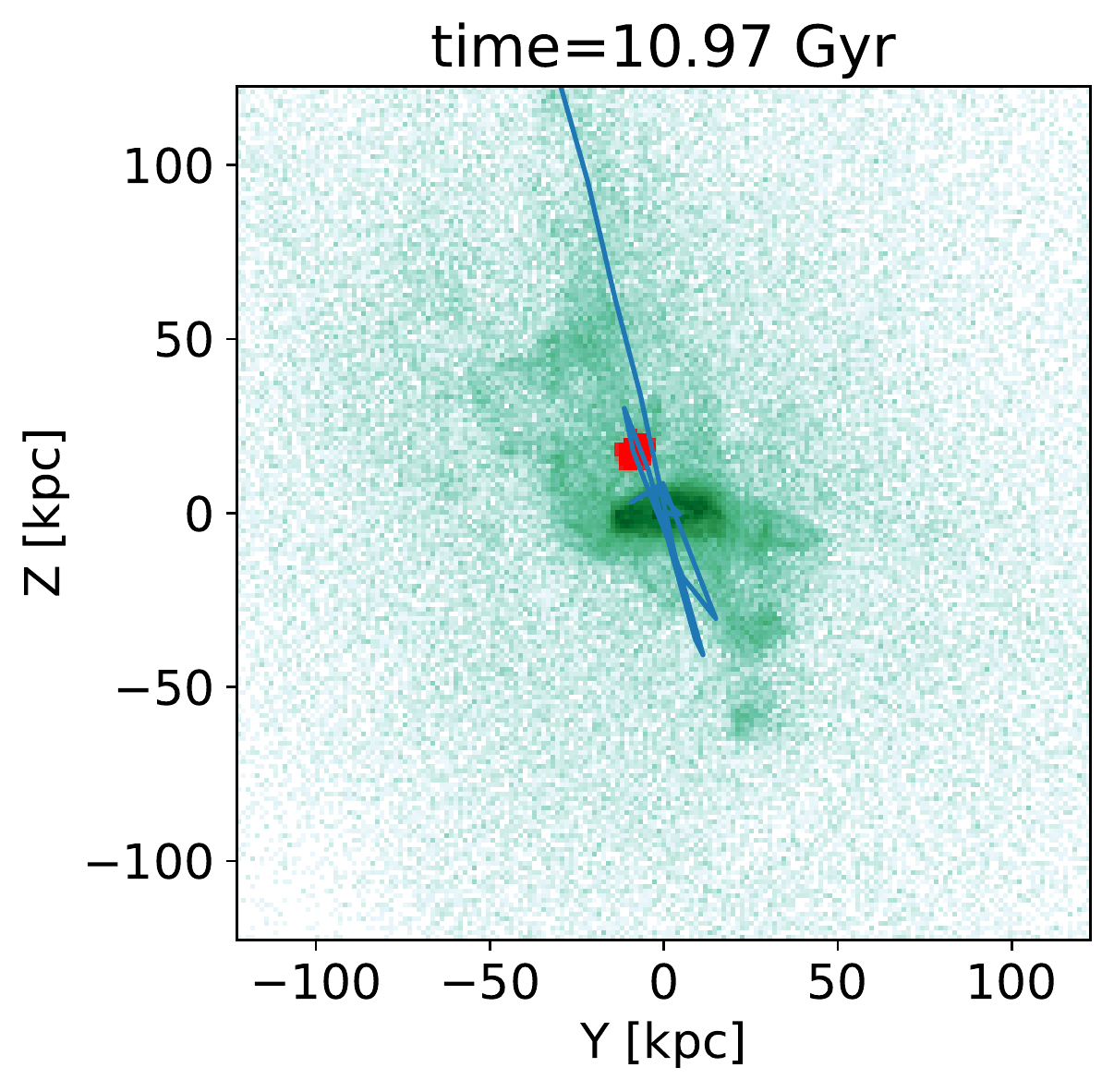}
\includegraphics[width=5.8cm]{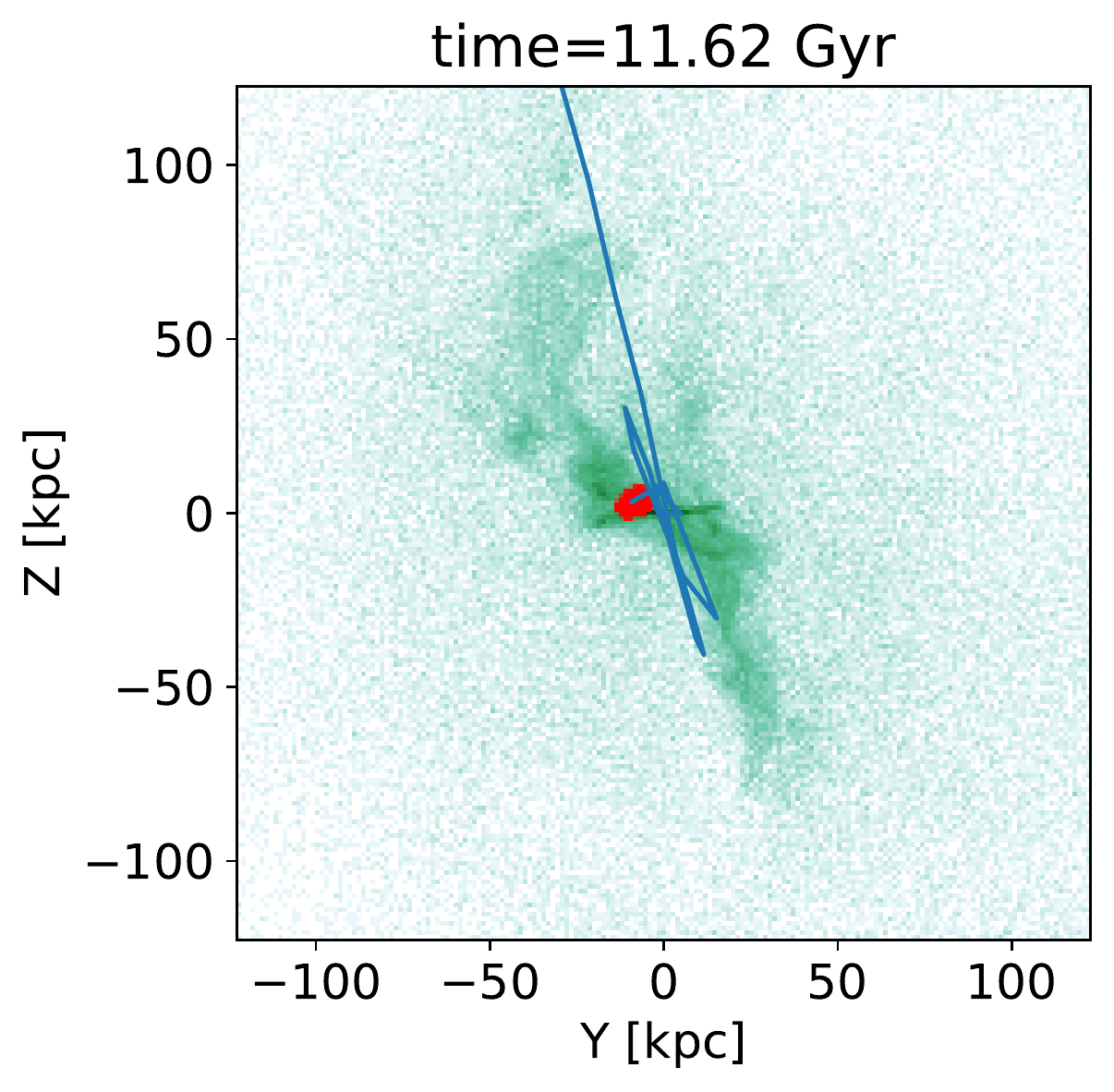}

\includegraphics[width=5.8cm]{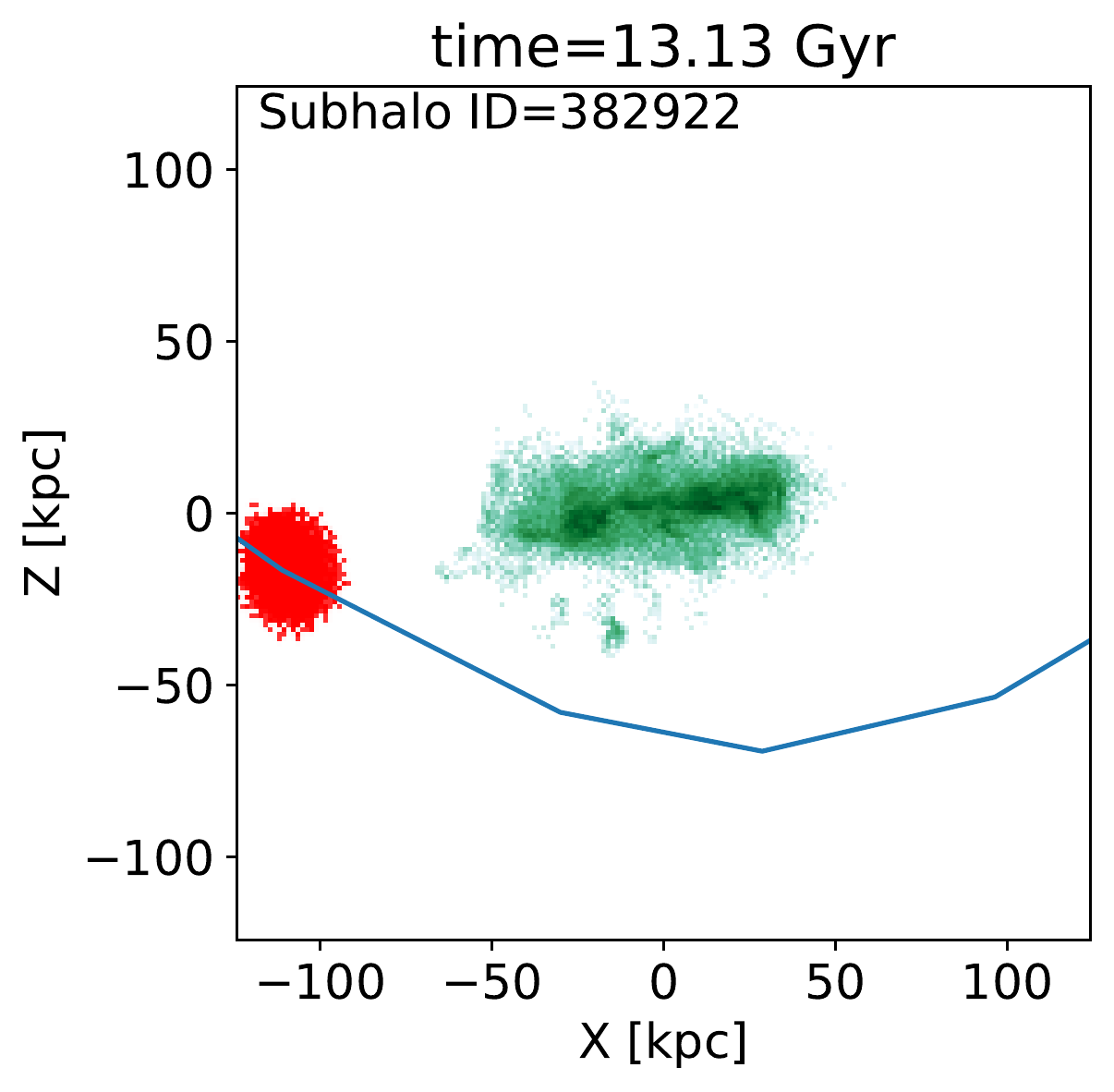}
\includegraphics[width=5.8cm]{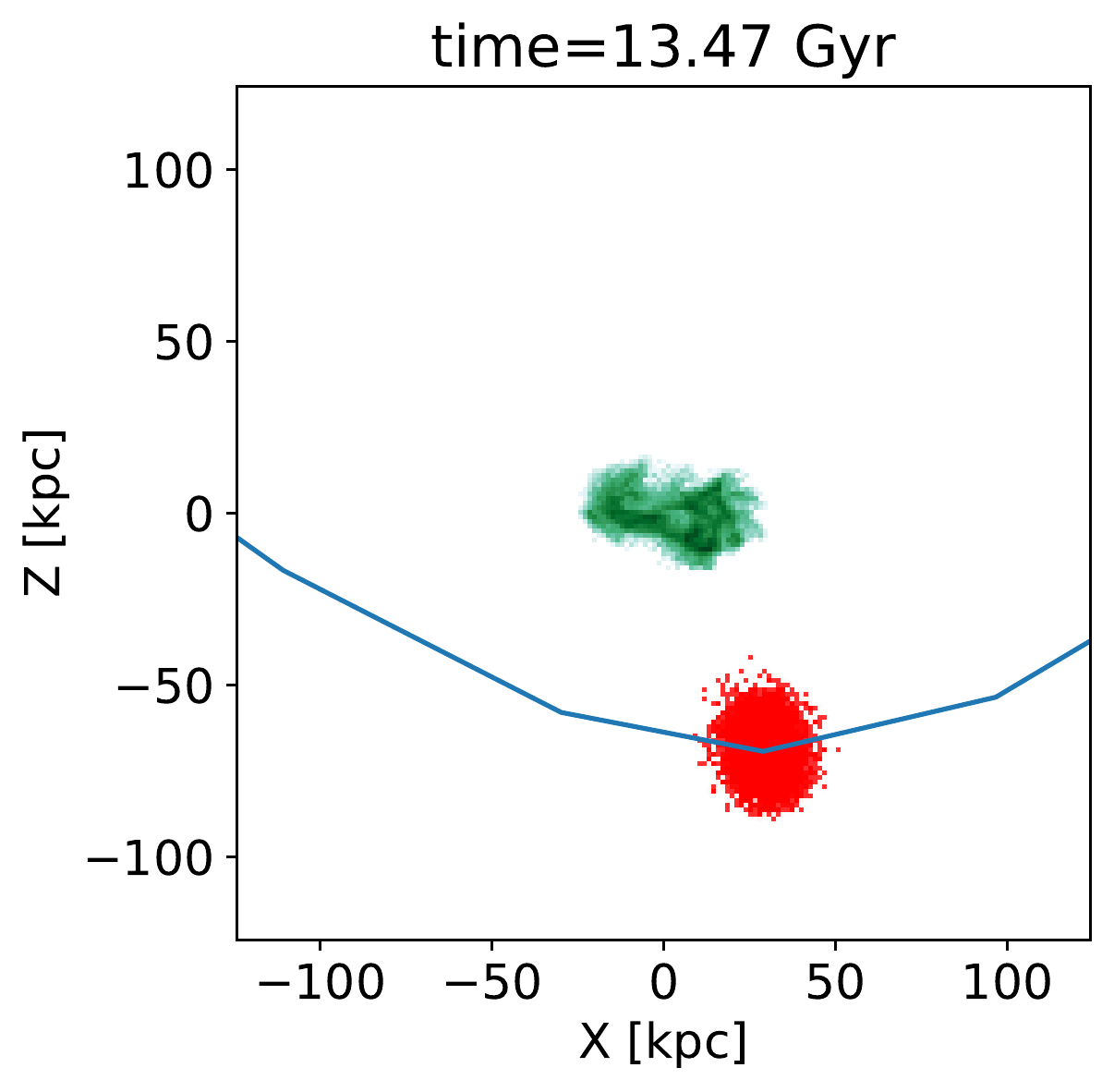}
\includegraphics[width=5.8cm]{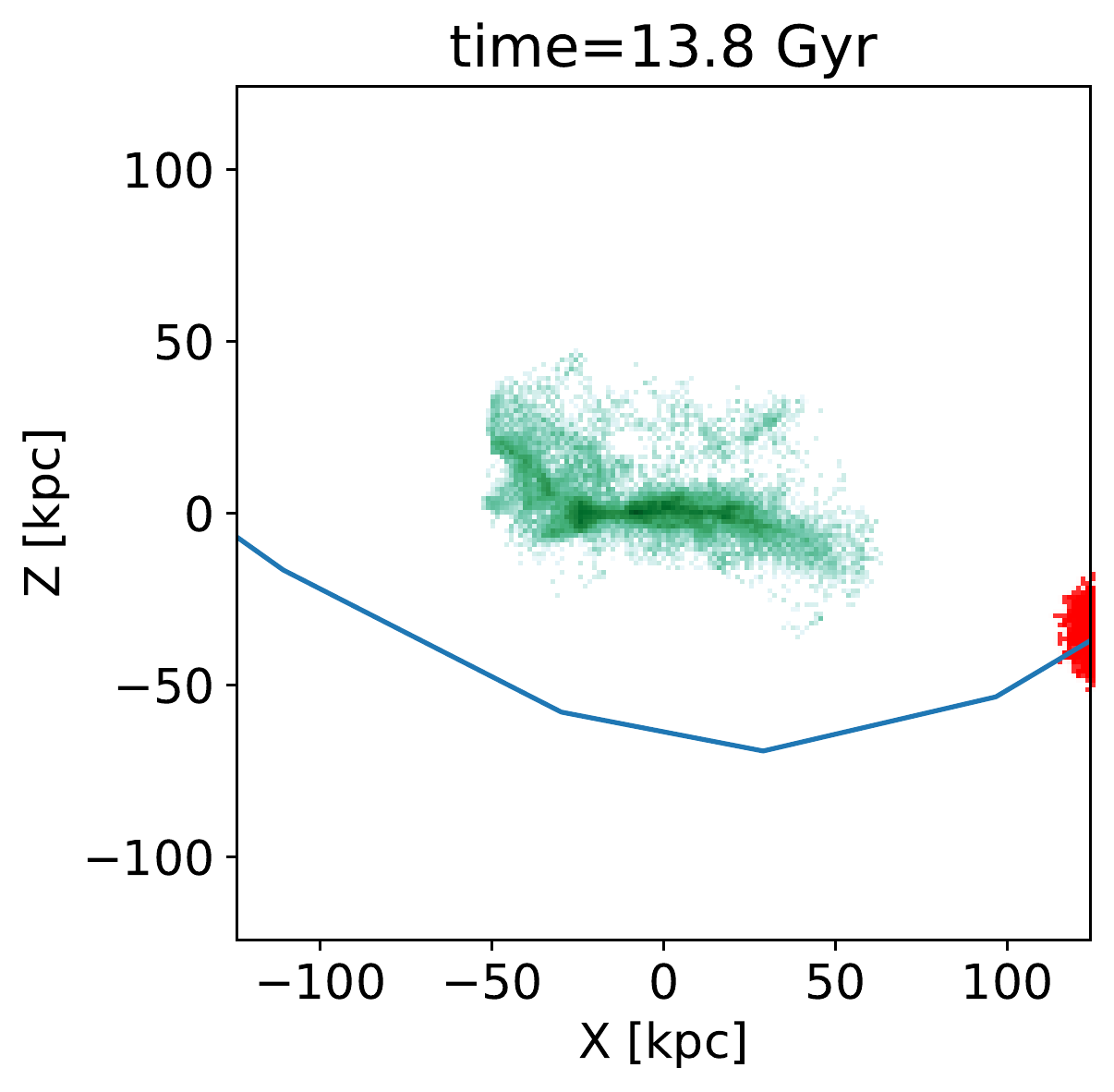}

\caption{Same as Fig.~\ref{examples}, but for three other galaxies.}
\label{examples2}
\end{figure*}

\begin{figure}
\centering
\includegraphics[width=8.5cm]{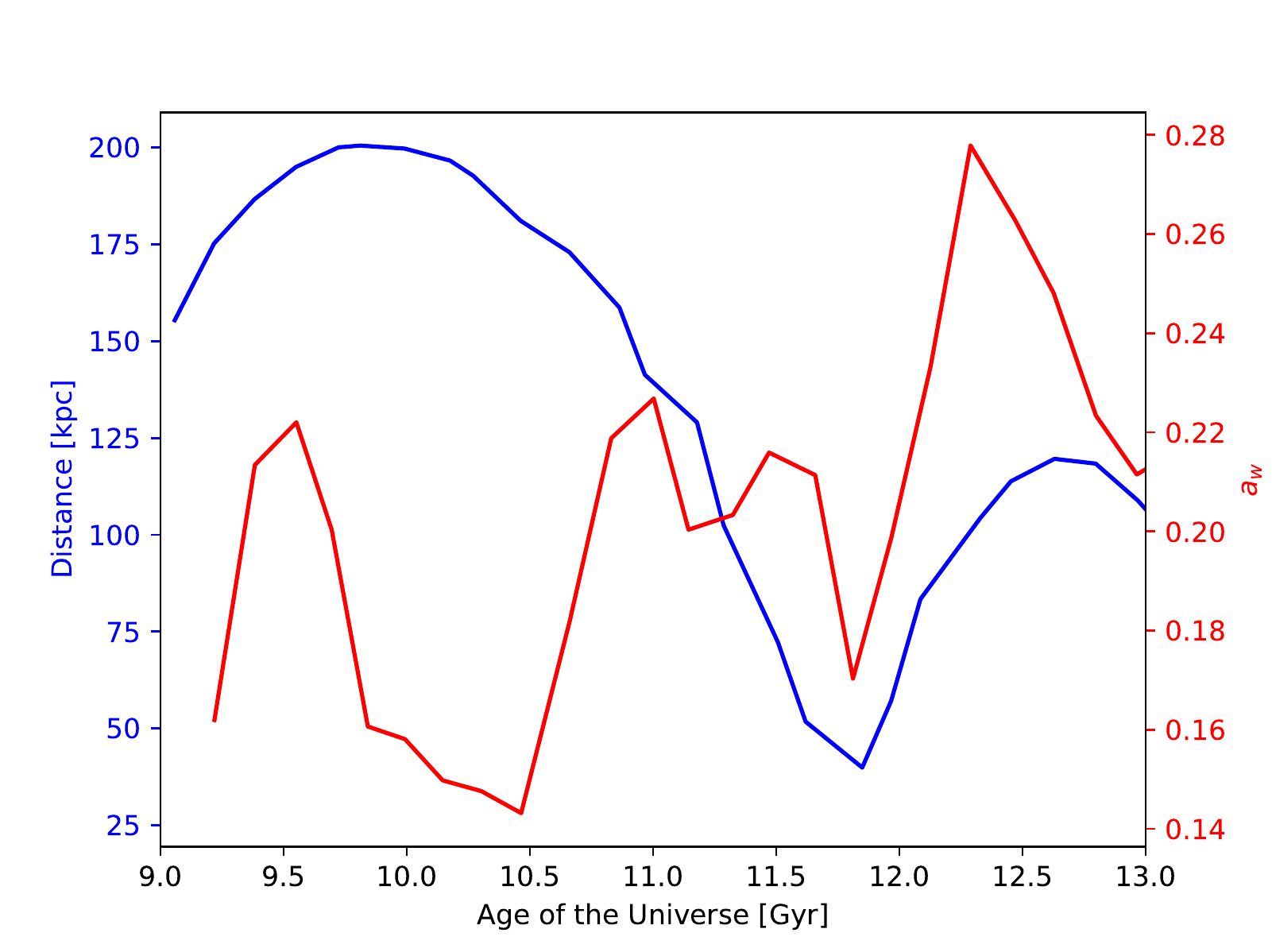}
\includegraphics[width=8.5cm]{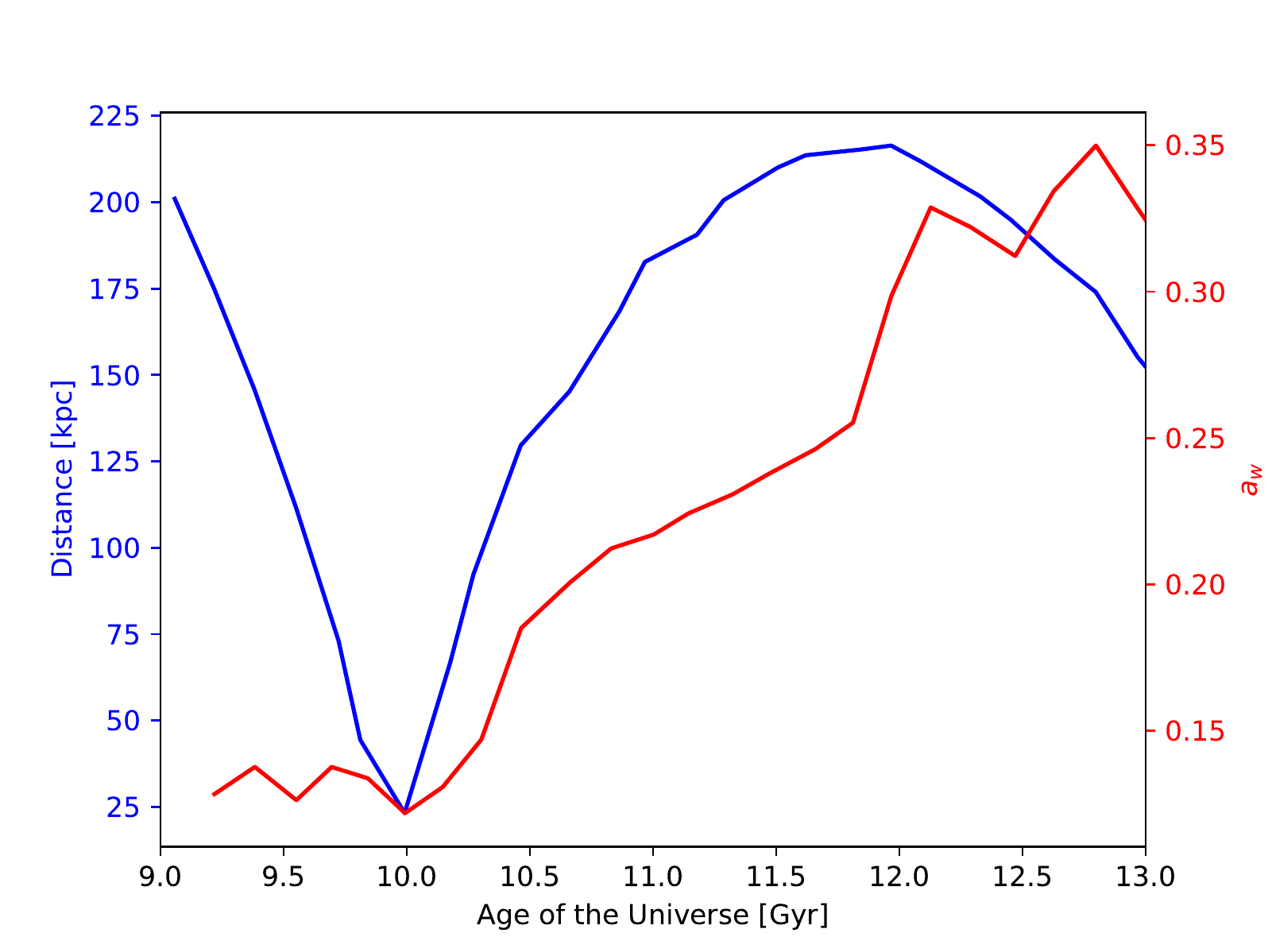}
\includegraphics[width=8.5cm]{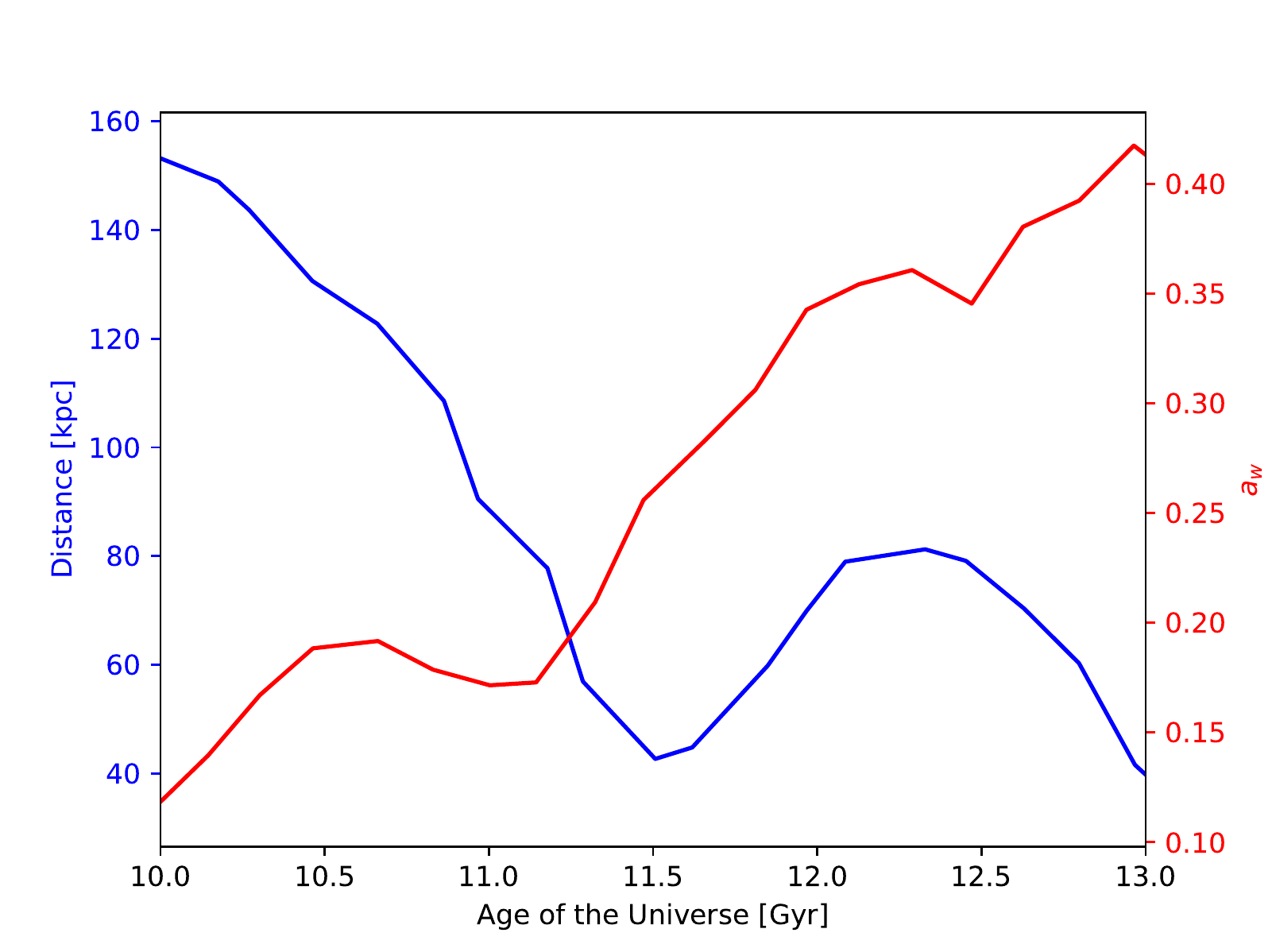}

\caption{Time dependence of distances between host galaxies and perturbers (blue lines) for the three examples presented in
Fig.~\ref{examples} compared with evolutions of the warp parameter $a_{\mathrm{w}}$ measured for the gas disks of the
hosts (red lines).}
\label{ad_ex}
\end{figure}

\begin{figure}
\centering
\includegraphics[width=8.5cm]{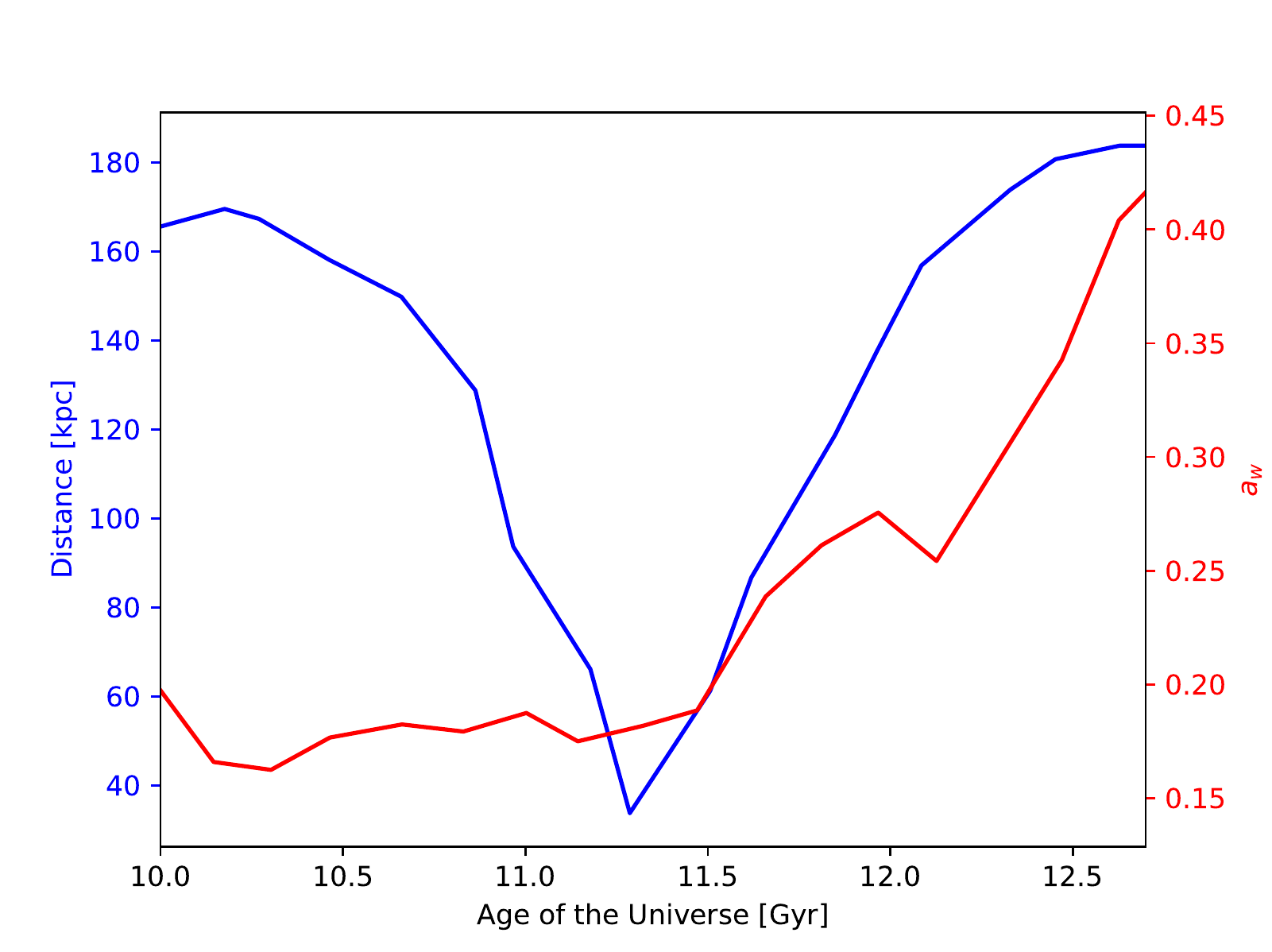}
\includegraphics[width=8.5cm]{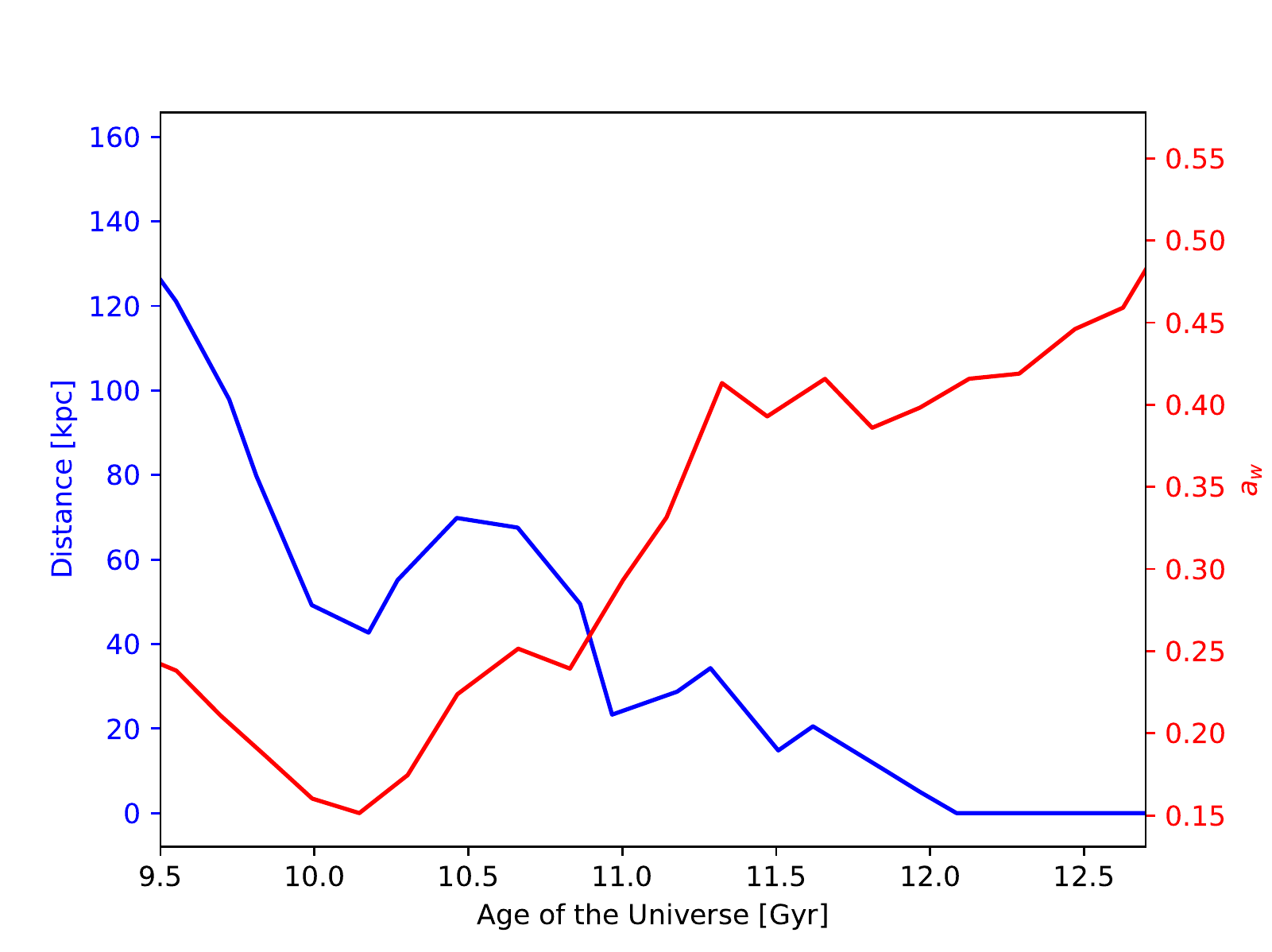}
\includegraphics[width=8.5cm]{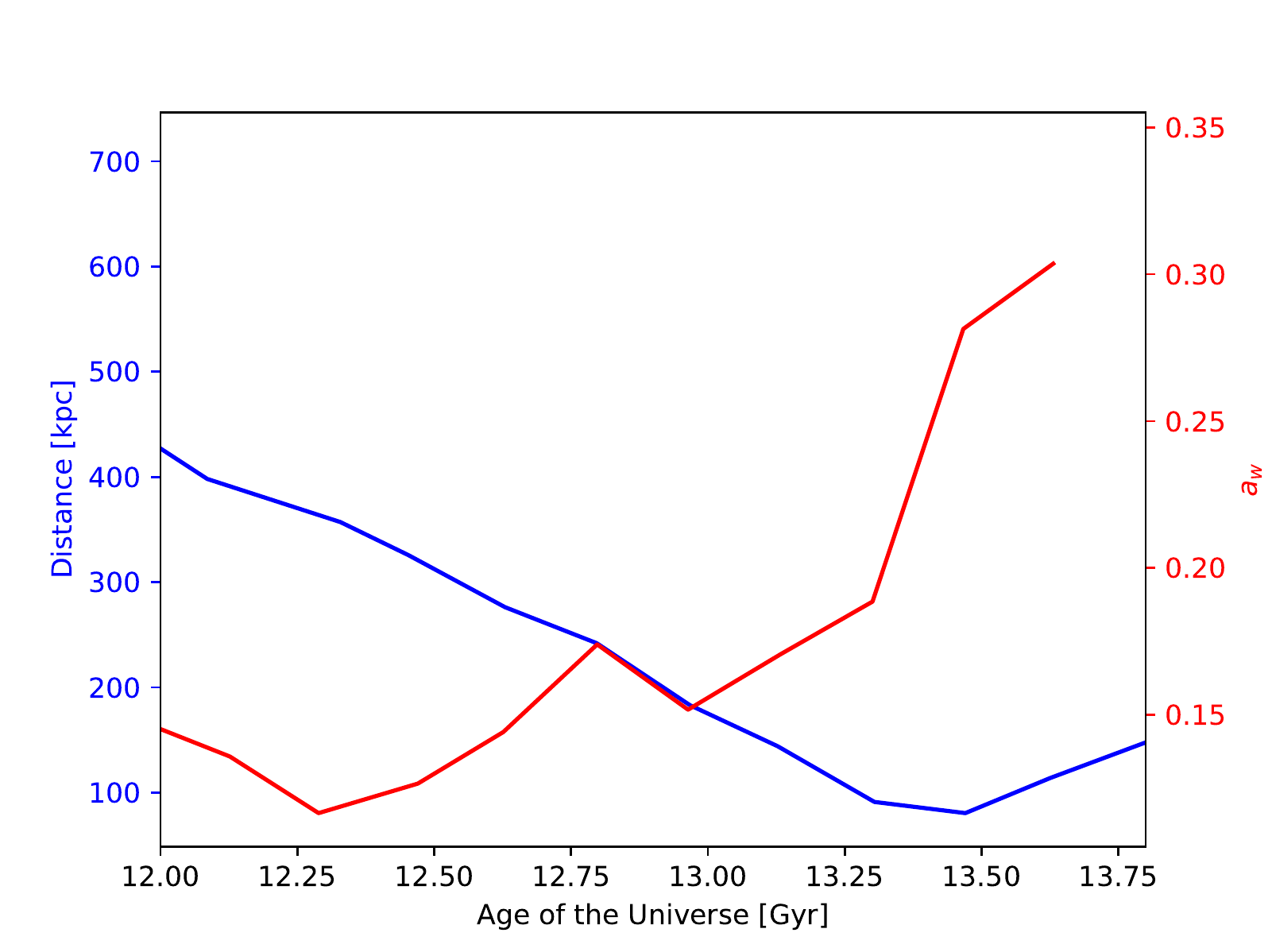}

\caption{Same as Fig.~\ref{ad_ex}, but three other galaxies whose encounter is shown in Fig.~\ref{examples2}.}
\label{ad_ex2}
\end{figure}

\begin{figure}
\centering
\includegraphics[width=9cm]{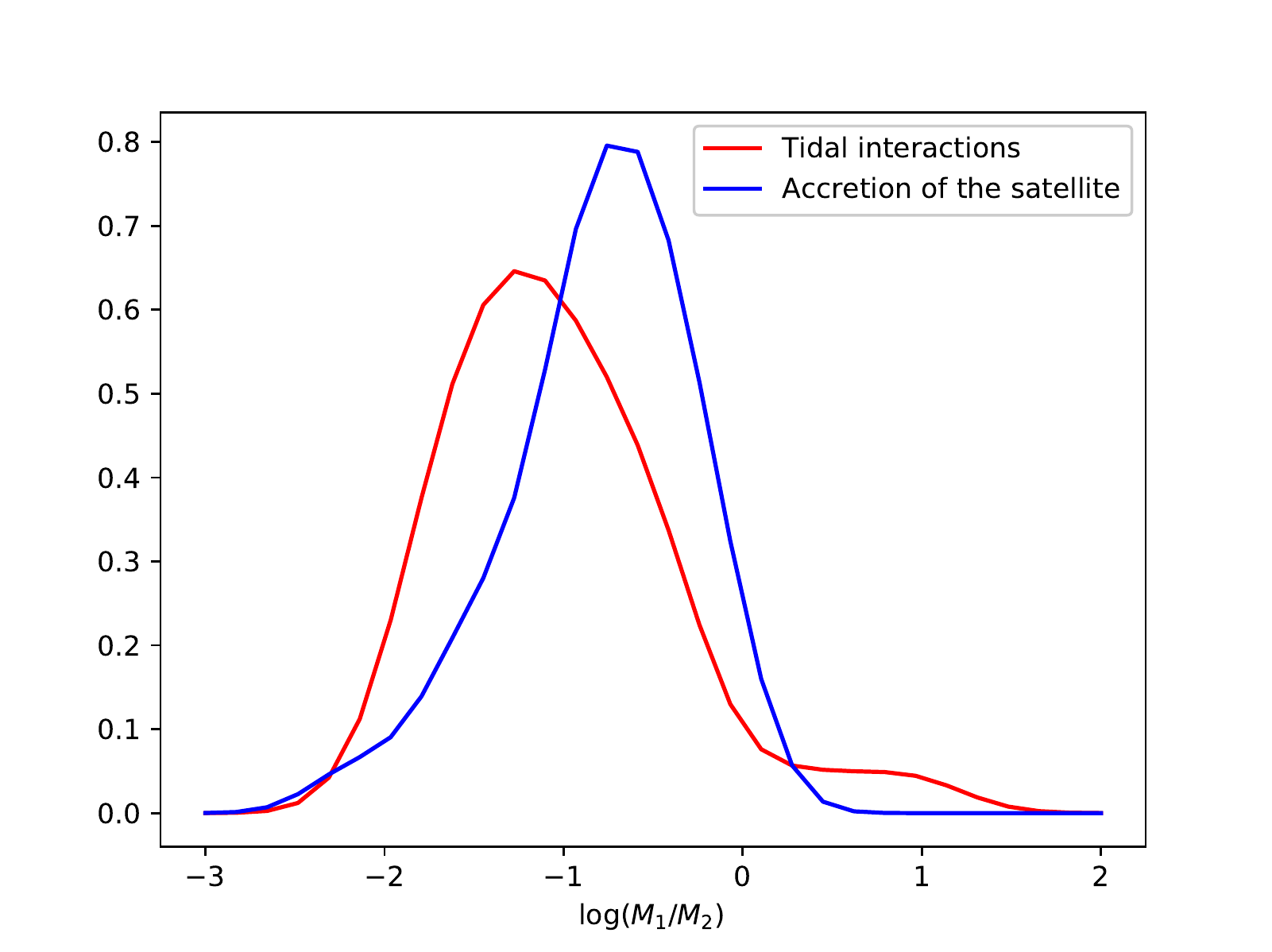}
\includegraphics[width=9cm]{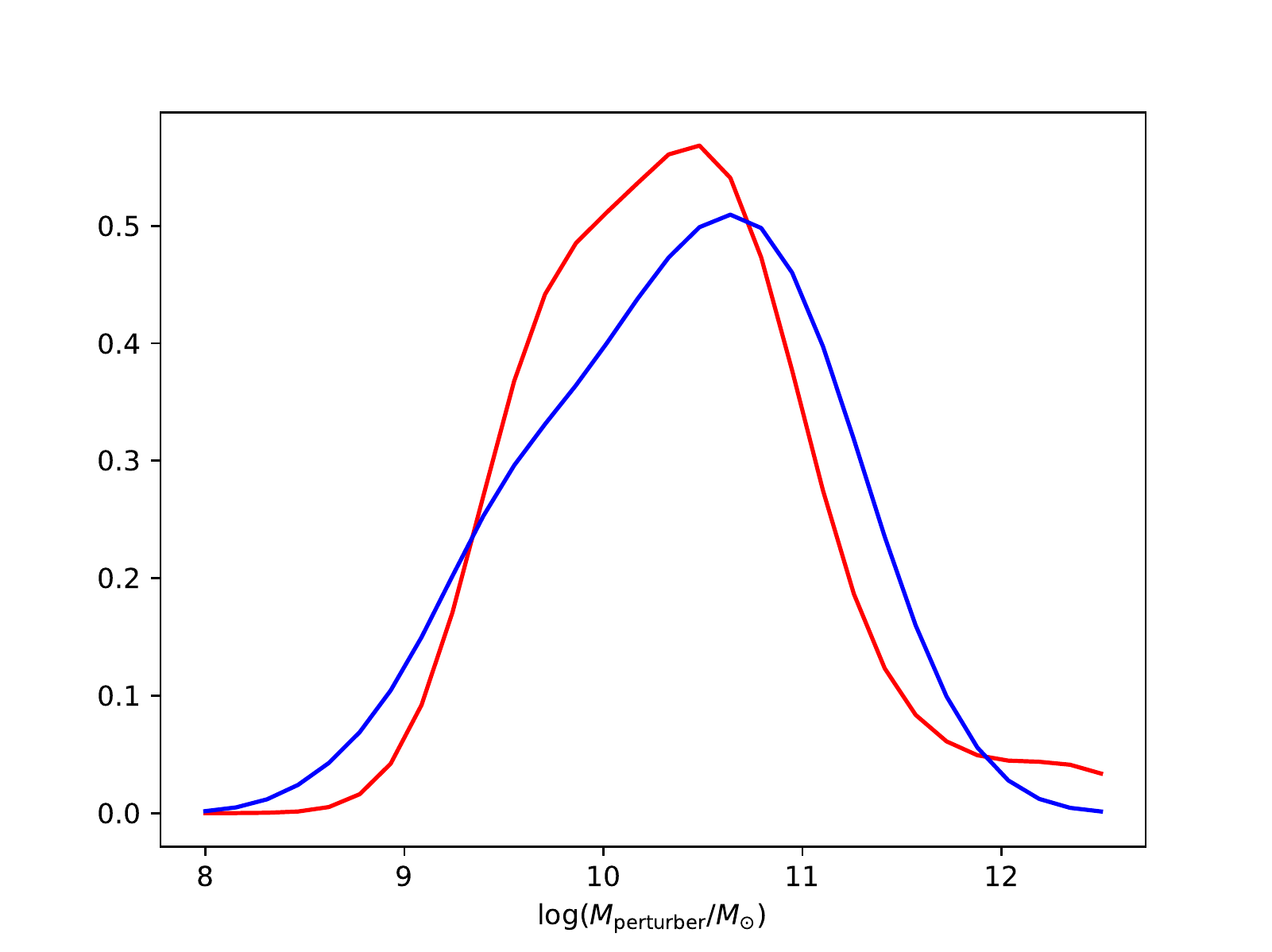}
\includegraphics[width=9cm]{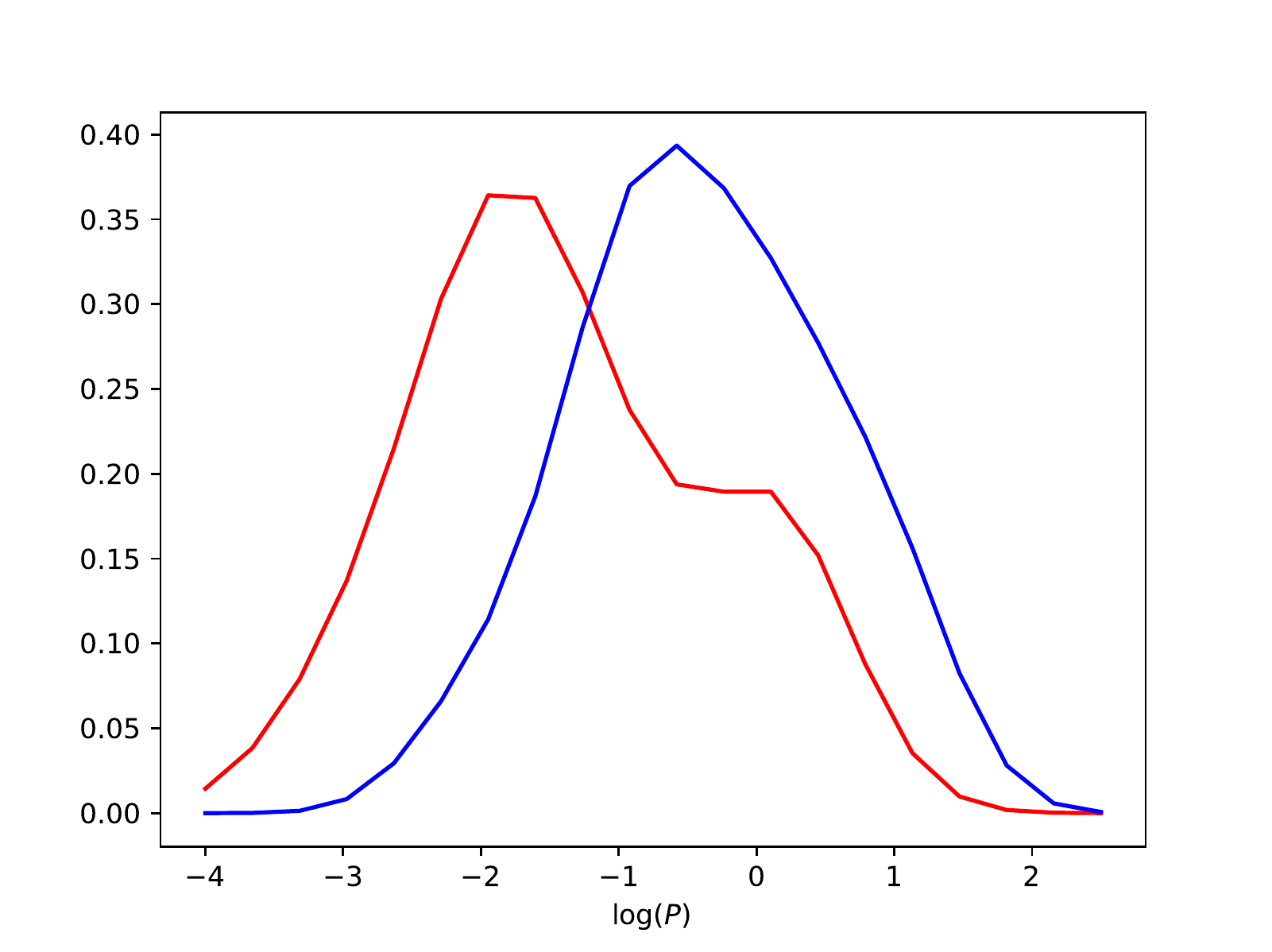}
\caption{Normalized histograms of (perturber to host) mass ratios (upper panel), masses of the perturbers (middle
panel) and tidal parameters (lower panel) for the warped galaxies that experienced tidal interaction with a flying-by
perturber (red lines) and those that accreted their satellite (blue lines).}
\label{stats}
\end{figure}

In order to check whether the warp of a given galaxy was tidally induced we looked for a relatively massive subhalo
appearing in the proximity of its host galaxy in the past 5 Gyr. Once the suspect for the perturber is identified, we
compare its orbit and pericenter passages with the evolution of $a_\mathrm{w}$. A coincidence between the time of the pericenter
passage of the perturber and the time of the rapid increase in $a_\mathrm{w}$ is an indicator of the tidal origin of the warp. For
cases where such a coincidence took place in order to confirm the tidal origin of the warp we then carefully inspected
edge-on gas maps including the flying-by perturber and measured the mass ratios and tidal parameters $P$ (as used
e.g. in \citealt{oh} and \citealt{m33})
\begin{equation}
  P=\Bigg( \frac{M_{\mathrm{pert}}}{M_{\mathrm{host}}}\Bigg) \Bigg( \frac{R_\mathrm{host}}{d_\mathrm{peri}} \Bigg)^3,
\end{equation}
where $M_{\mathrm{pert}}$ are the masses of perturbers measured at the last apocenter before the encounter,
$M_{\mathrm{host}}$ are the masses of host galaxies measured at the pericenter within pericentric distances, while
$d_\mathrm{peri}$ and $R_\mathrm{host}$ are linear sizes of hosts defined here to be $5 \; R_{1/2*}$. $P$ measures the 
tidal force exerted on the host and it was shown that the disk can respond to $P$ as low as 0.01 (\citealt{byrd}) in terms of spiral arm formation.   

Applying the procedures described above yielded that around 35\% (66 out of 187) of the warped galaxies from the
reduced S-shaped sample had their warps tidally induced (at some point in the last 5 Gyr). From this 35\%, half of the
encounters were caused by the flying-by galaxy or a satellite and the other half were caused by satellites that ended
up being accreted by the host (i.e. their identity in the Subfind catalogues was lost and passed to the host galaxy).
We note that out of all of these interaction cases it is difficult to discriminate whether the formation of the warps
was due entirely to the tidal force acting vertically on disks or it was an effect of the perturber collecting the
circumgalactic gas and dropping it at the host's disk causing non-planar perturbations resulting in S-shaped warps. We
suspect that for some cases, the origin of warps could be due to the combined effect of the two phenomena.

Six examples of tidally induced warps are presented in Fig.~\ref{examples}-~\ref{ad_ex2}. Figures~\ref{examples}
and~\ref{examples2} present three snapshots of flying-by perturbers on their orbits around hosts and the reactions of
the hosts' gaseous disks. The mass ratios between the perturbers and hosts are respectively 2.2\%, 3\%, 3.6\% for Figures~\ref{examples} and~\ref{ad_ex} from top to bottom, and 6\%, 23\%, 859\% for Figures~\ref{examples2} and~\ref{ad_ex2} from top to bottom. In the first three cases and the last one the scenario is very similar. Before and during the pericenter the gas disk is more
or less flat, with some irregular structures. Shortly after the pericenter, the edge-on profile is transformed into the characteristic S-shape. The tidal forces originating from a perturber that is passing by
on an orbital plane that is not parallel to the equatorial plane of the gaseous disk were able to induce these symmetrical
morphological features. The whole process is further illustrated in Fig.~\ref{ad_ex} and~\ref{ad_ex2}, where the values
of $a_\mathrm{w}$ tend to rapidly increase 
soon after pericenter. Gaseous disks
in these cases are not perfectly flat prior to the tidal perturbation, as constant accretion in cosmological
simulations rarely forms very regular disks. This is also reflected in Fig.~\ref{ad_ex}, where $a_\mathrm{w}$ has values
close to 0.15-0.2 prior to the pericenter (especially in the last example). However, the passage of the perturber increases
these values to make them sometimes twice as high as the average before the event.

For the fourth and fifth example the scenario seems to be a bit altered. 
For the fourth example we see that the gas
that was initially below the host's disk seems to be partially dragged upwards by the perturber. In the end, the gas
that remained below the disk, together with the part that was dragged in, both contribute to the asymmetric S-shape.
In the fifth case that ends in the accretion of the
perturber, non-planar gas seems to be both dragged by the perturber and accreted onto the host from above and below, which results in the S-shaped morphology.
These two cases illustrate well the difficulty of separating tidal effects from the accretion of the gas caused by the
satellite. While their tidal parameters seem to be very high ($P=0.17$ and 2.43), the analysis of the gas maps shows
that the satellites may play important roles in mixing and accreting the surrounding gas of the host. We test this
interpretation further in section 3.6.

Figure~\ref{stats} shows histograms of mass ratios between perturbers and hosts, masses of perturbers and tidal
parameters $P$ for the tidally induced warped galaxies divided into cases that did or did not end up in
accretion. The accretion sample seems to have higher mass ratios but this may be an artifact of measuring the
masses of hosts within the pericenter distances, which are naturally smaller for those cases. A similar distribution of
perturber masses seems to confirm that. Both higher mass ratios and small pericenters contribute to higher values of
$P$ for the accretion sample, which means that these encounters are more violent than the cases where the perturber
flies away and does not interact with the host so tightly.

\subsection{Lifetimes of tidally induced warps}

To find the rapid increase of $a_{\mathrm{W}}$ that was used to determine whether the warp formation was caused by a
flying-by perturber, we looked for a maximum of the derivative of the smoothed time evolution of $a_{\mathrm{W}}$ (or
for a moment when a threshold was exceeded). This helped us to establish when warps were created. We used a similar
approach to find out when warps start to dissolve by looking for a minimum in the derivative of the smoothed time
evolution of $a_{\mathrm{W}}$. We defined the time between the maximum of smoothed ${\rm d} a_{\mathrm{W}}/{\rm d} t$
associated with a pericenter passage of a perturber and the following minimum as a lifetime of the tidally induced warp
$\tau_{\mathrm{W}}$. Obtaining this kind of approximate value was possible for 51 out of 66 cases that we found to be
driven by interactions. For the remaining 15 cases it was impossible to achieve because the simulation ended before
the minimum of the derivative occurred, i.e. the warp was induced very recently. For our sample, we found an average
value of $\tau_{\mathrm{W}}=0.8\pm0.5$ Gyr. The numerical definition of $\tau_{\mathrm{W}}$ does not necessarily mean
that after that time the warp is dissolved since the minimum of the derivative means only that it starts to decrease,
but not necessarily disappear. Given that, it is fair to round up that number and say that the average lifetime is
$\sim1$ Gyr. This number is not an outlier in comparison e.g. with the values for stellar warps from \cite{kim}. They
derived lifetimes even up to 4-5 Gyr, but these were obtained in idealized simulations tailored to produce strong warps,
unlike the more realistic cosmological setup used in this paper.

\subsection{Incident angles of interactions that results in warped disks}

The angle between the angular momentum of a perturber on its orbit and the angular momentum of a rotating disk was
often discussed in the literature as an impactful factor on tidally induced morphologies (e.g. \citealt{donghia2010};
\citealt{lokas15}). The incident angle (also referred to as the inclination angle) $i$ was found to have the biggest
impact on galaxies were the perturber orbits were prograde $i=0^{\circ}$ and the smallest when they were retrograde
$i=180^{\circ}$. For these values, however, no warp would be induced as there is little or no tidal force
acting in the vertical direction on the host's disk.

We calculated incident angles for our sample of interaction-driven warps in the snapshots closest to the pericenter and plotted the normalized histograms of their
cosines for the whole sample and for the cases that end in the accretion of the perturber or not in the upper panel of
Fig.~\ref{prop}. From the analysis of the histograms one can see that most warps were induced with
$i\simeq50^{\circ}$ and in addition a small increase departing from the overall distribution can be found at
$i\simeq125^{\circ}$. This finding means that most warps were induced in encounters with incident angles close to these
two values (with much greater importance of the first one). This can be interpreted as the fact that for these
encounters warps were strong enough and lived long enough to be easily detectable, with the case lying close to the
prograde orientation having much bigger significance. This finding is consistent with the results of \cite{kim} who
found that the strongest and most long-lived warps are created for angles of $45^{\circ}$ and $135^{\circ}$. Their
interpretation of this finding was that these angles optimize the combined effect of two factors: the timescale
of the interaction, which is greatest for $i=0^{\circ},180^{\circ}$, and the strength of the vertical component of the
tidal force, which is greatest for $i=90^{\circ},270^{\circ}$. For the tidal warp formation, both of these factors
contribute. We agree with the interpretation of \cite{kim} and our results partly confirm their findings, with an
important modification that cases closer to prograde are more effective in warp formation in agreement with the
interpretation based on resonant behavior (\citealt{donghia2010}; \citealt{lokas15}).

\subsection{Asymmetry of tidally induced warps}

\begin{figure}
\centering
\includegraphics[width=9cm]{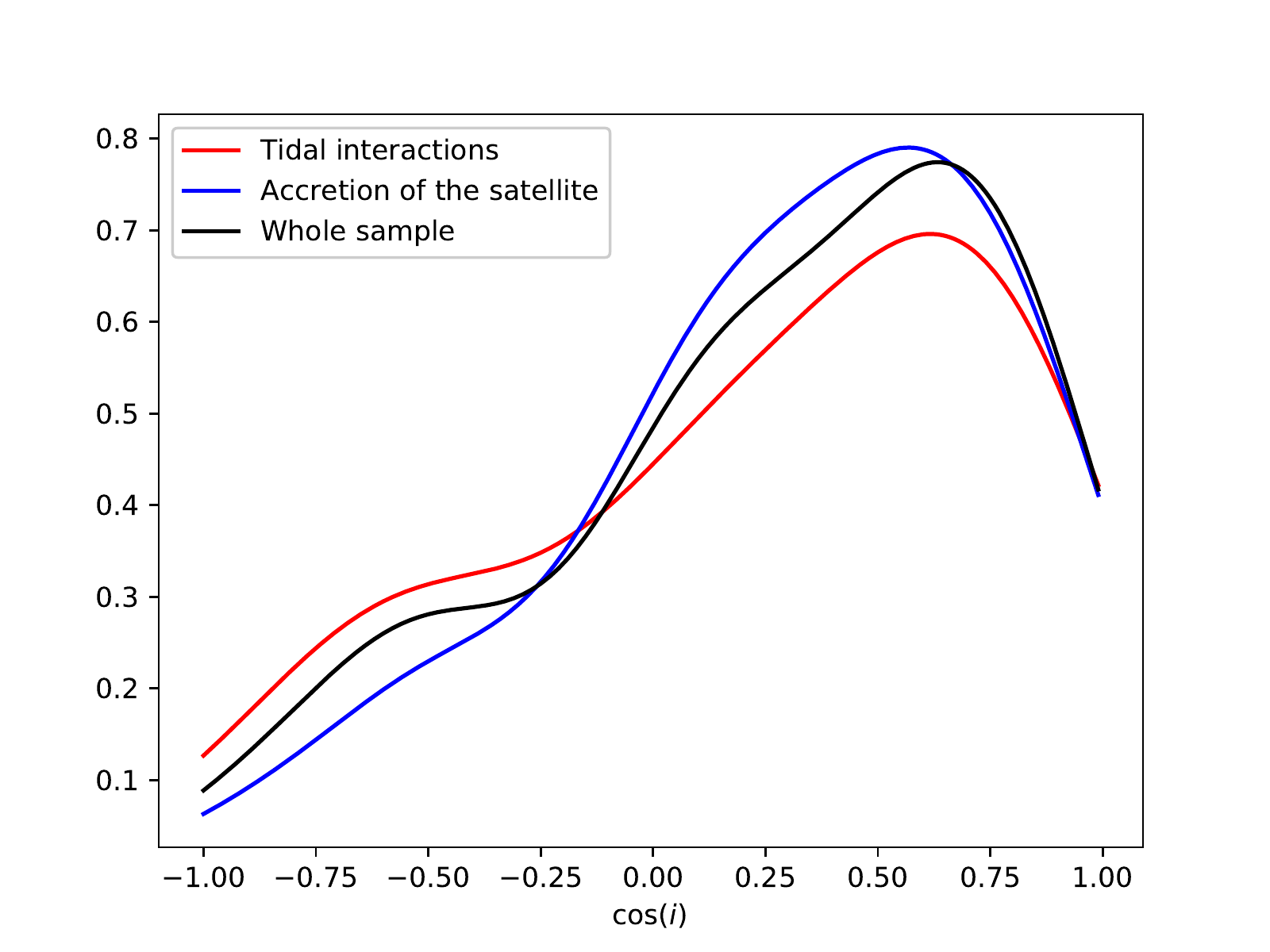}
\includegraphics[width=9cm]{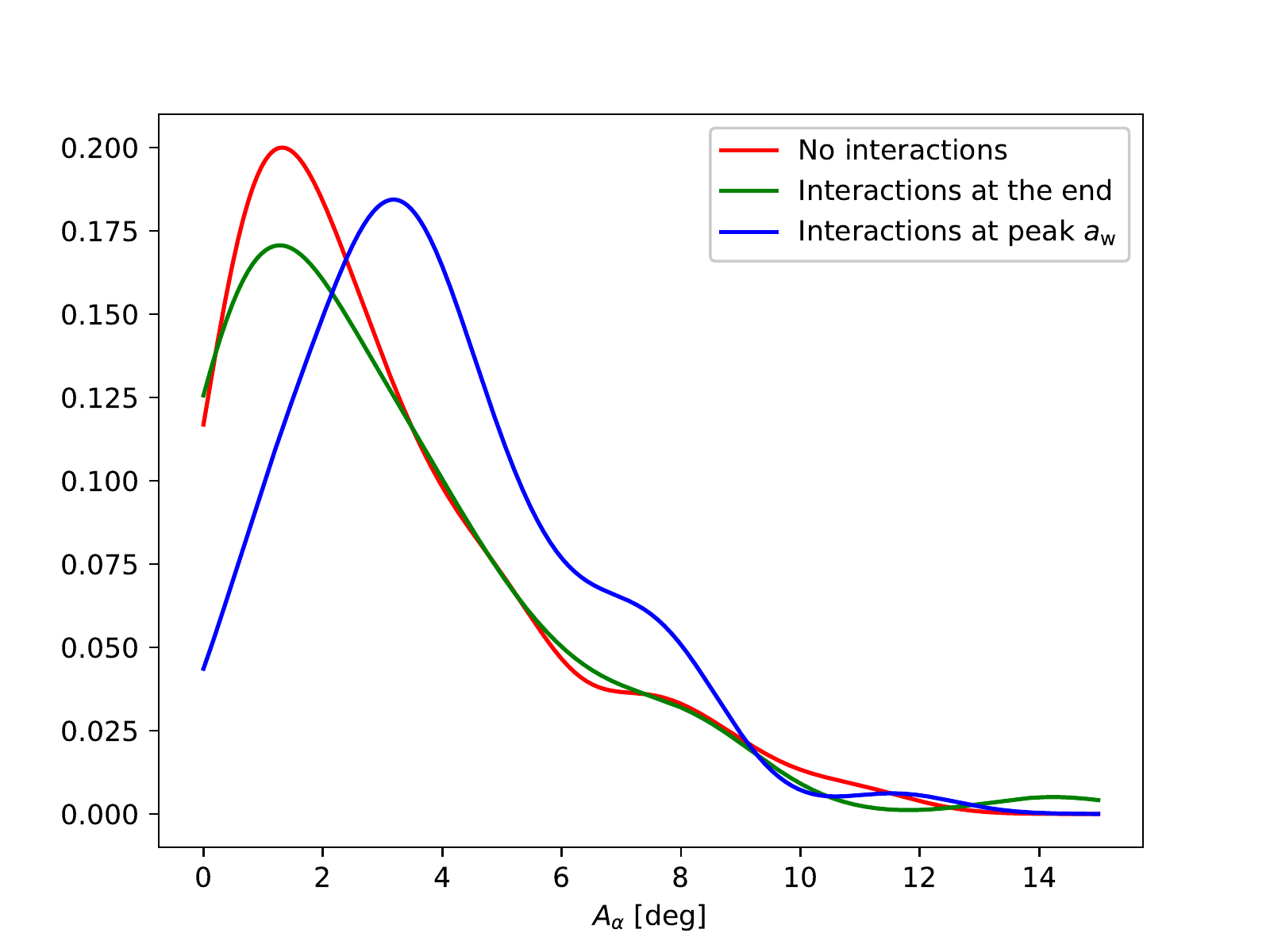}
\caption{Upper panel: Normalized histograms of the cosine of the incident angle $i$ for warped galaxies which experience
interaction with a flying-by perturber (red line), galaxies which are interacting with a perturber that is later accreted
(blue line) and for both of these samples combined. Lower panel: Normalized histograms of the asymmetry parameter
$A_{\alpha}$ measured for the sample of interacting galaxies at $z=0$ (green line) and at the time when
$a_{\mathrm{w}}$ peaks, i.e. shorty after the pericenter passage of the perturber (blue line) compared with the
histogram for all other S-shaped warps that were not interacting, measured at $z=0$ (red line).}
\label{prop}
\end{figure}

\cite{garciaruiz} and \cite{annpark} found that warps of galaxies that are interacting
are more asymmetrical, i.e. one side of them is longer than the other one. To verify if this trend is also present in
IllustrisTNG warps, we define the asymmetry parameter
\begin{equation}
	A_{\alpha}=|\arctan(|a_{\mathrm{up}}|)-\arctan(|a_{\mathrm{down}}|)|,
\end{equation}
where $a_{\mathrm{up}},\;a_{\mathrm{down}}$ are values of coefficients $a_{\mathrm{W}}$ for the two highest values
corresponding to the two tips of the warp (one pointing up and the other down). This parameter is inspired by the
asymmetry parameter from equation (4) of \cite{annpark}, however in our case $\arctan$ of
$a_{\mathrm{up}},\;a_{\mathrm{down}}$ are not exactly the warp angles $\alpha$ defined as the inclinations of
lines coming from the center of the galaxy to the edges of the warp (see Fig. 2 of \citealt{kim}). In section 3.1 we
showed that our parametrization of warps proved to be a good proxy for the deviation of warps from the equatorial
planes of disks and it is very useful for a large sample of investigated galaxies.

The lower panel of Fig.~\ref{prop} shows normalized histograms for $A_{\alpha}$ measured for our sample of 66 galaxies
that had their warps induced by interactions and for the remaining 121 warped galaxies for which we did not find any
connection between the warps and interactions. For those that had their warps caused by interactions we measured
$A_{\alpha}$ at two different epochs, at $z=0$ and at the time when $a_{\mathrm{W}}$ had a peak, which happens
shortly after the pericenter passage when the warp was just induced. Figure~\ref{prop} shows that at $z=0$ all the warps
are more or less symmetrical with $A_{\alpha}$ having a peak at $<2^{\circ}$. However, the warps from interactions show
a stronger signal ($A_{\alpha}\sim 4^{\circ}$) at the time shortly after the pericenter passages. 
This means that initially interaction-driven warps have one side more inclined than the other but with time this difference
vanishes as warps wind up and dissolve. 
The influence of the companion decreases and warps can be better characterized as self-consistent, rather than driven.
The source of the asymmetry could be similar to the case in the first example of
Fig.~\ref{examples2}, where one side of the warp is stronger due to the satellite gravitationally pulling out the
material from the host. Our findings presented in this section are therefore consistent with observational studies
claiming that warps coming from interactions are more asymmetric.

\subsection{Warps from the gas accreted by satellites}

\begin{figure}
\centering
\includegraphics[width=6cm]{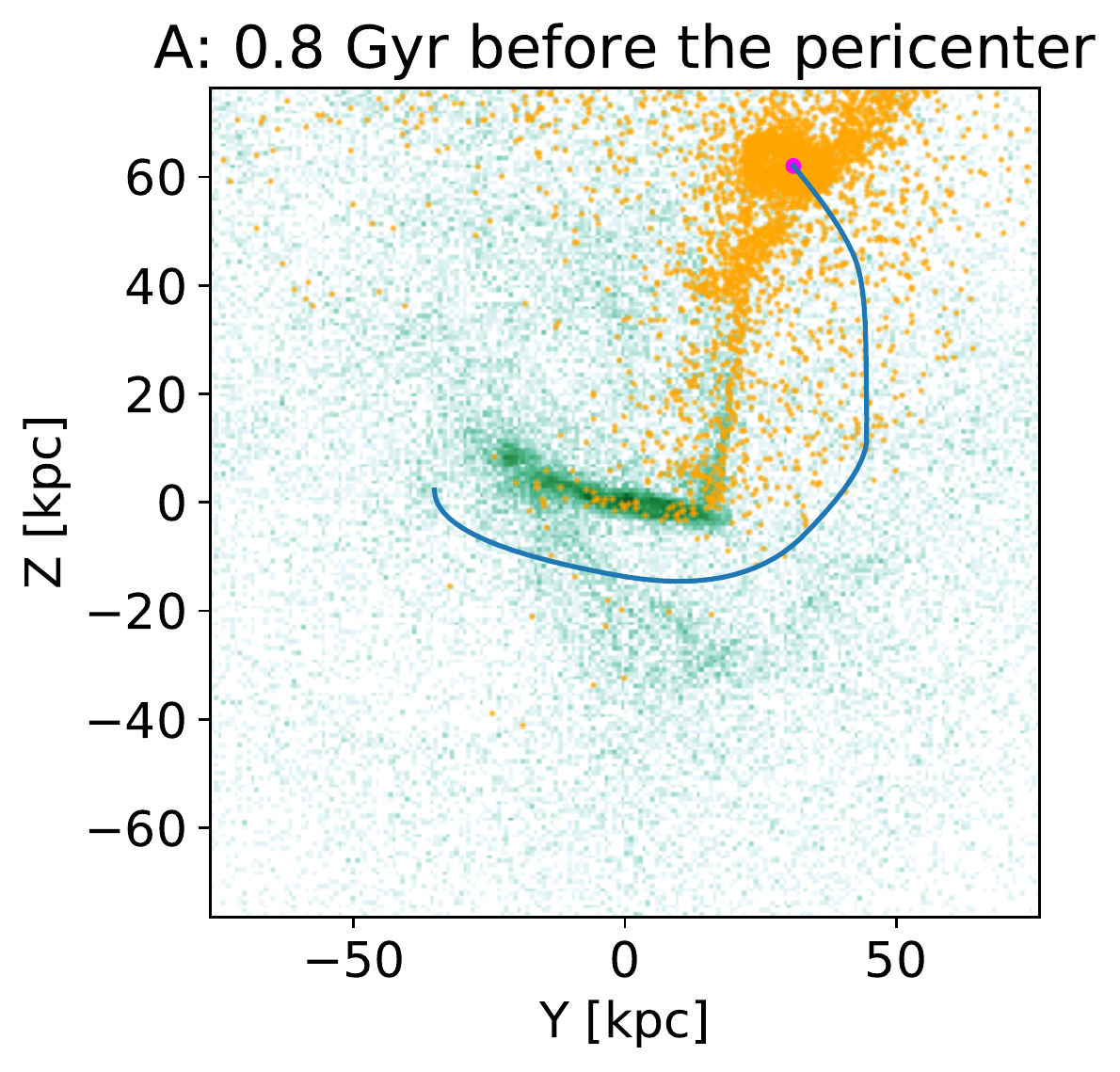}
\includegraphics[width=6cm]{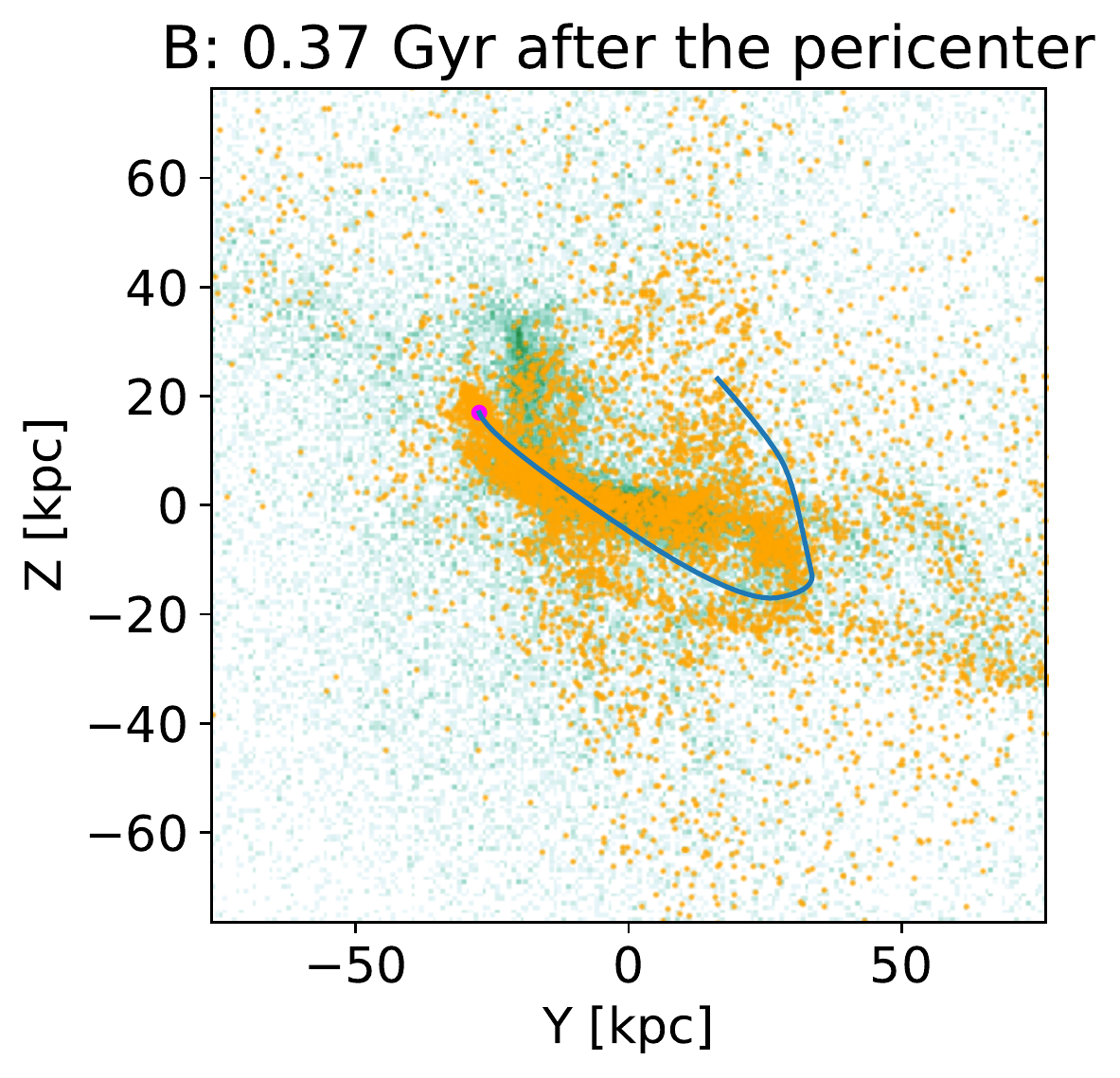}
\includegraphics[width=6cm]{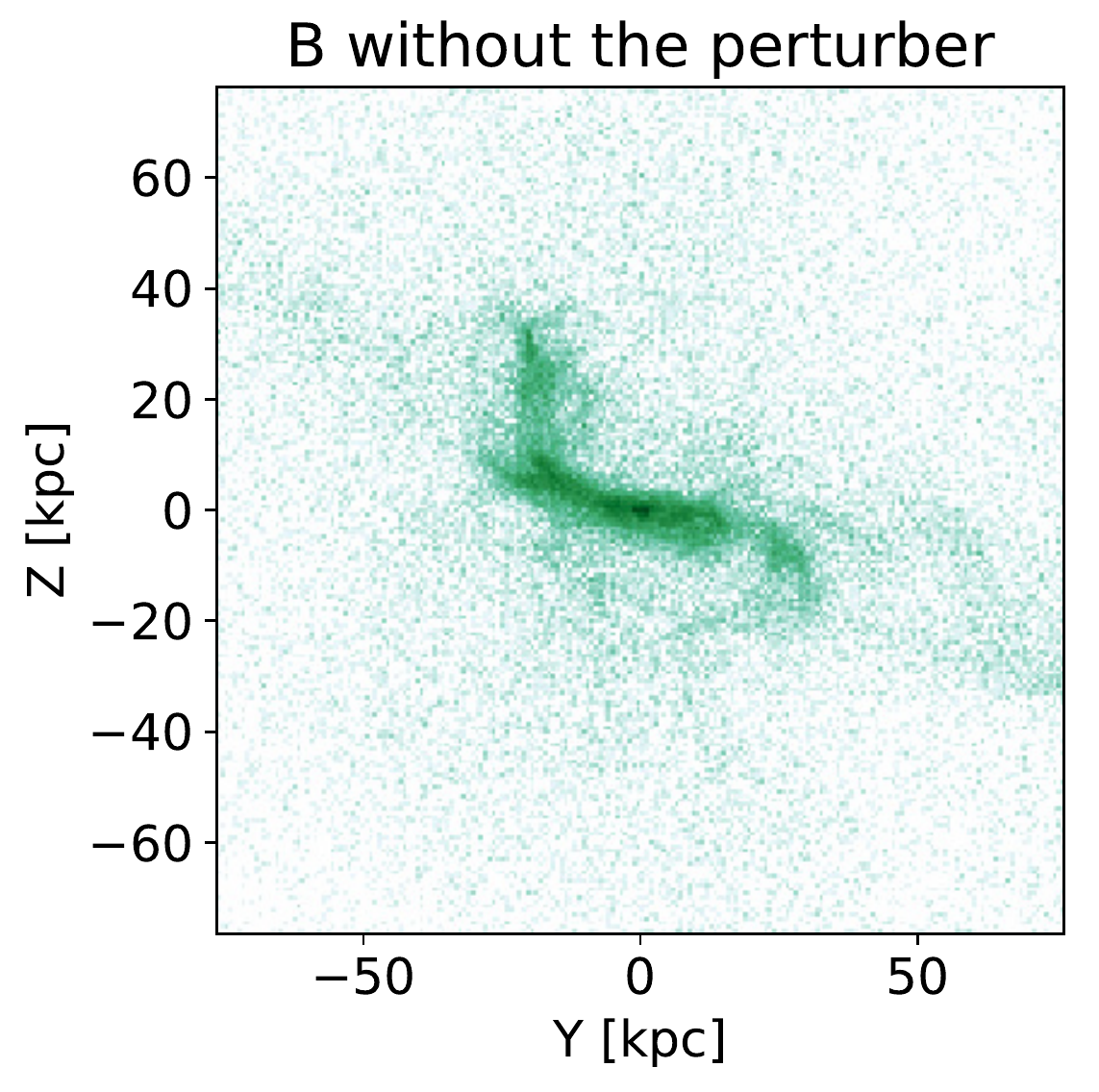}
\caption{Upper panel: Column density of gas of the host before the pericenter passage (color map) with gas cells
assigned to the perturber (orange points). Blue line is the orbit of the perturber and the magenta dot is its center.
Middle panel: Same as the first, but after the pericenter passage. Lower panel: Same as the second, but without the
tracer gas cells of the perturber. Subhalo ID of the host in this case at $z=0$ is 494091.}
\label{gas_acc}
\end{figure}

\begin{figure}
\centering
\includegraphics[width=8.5cm]{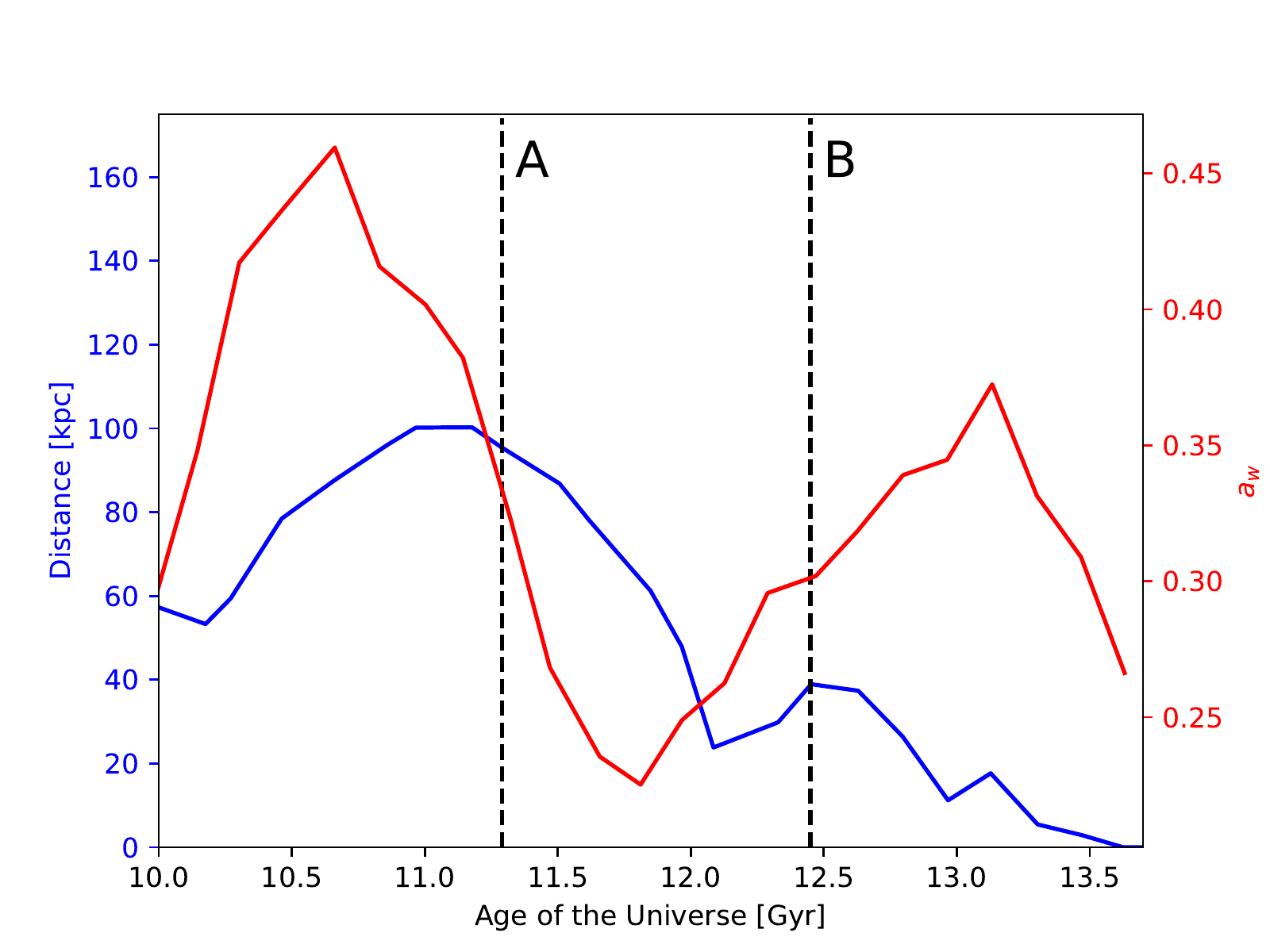}
\caption{Same as Fig.~\ref{ad_ex},~\ref{ad_ex2}, but for the case discussed in subsection 3.6. with marked times A
(before the pericenter) and B (after the pericenter), which are shown in Fig.~\ref{gas_acc}.}
\label{gas_acc2}
\end{figure}

In subsection 3.2 we interpreted the first two examples from Fig.~\ref{examples2} as having their S-shaped warps induced
not only by the tidal interaction but also by the satellite accreting gas onto the disk of the host. This
interpretation, however, was proposed based only on visual inspection of the maps like those in Fig.~\ref{examples2},
where some clouds of gas seem to be following the satellite and later the warp appears. This interpretation could be
wrong and only result from a misleading visual impression. To tell if the satellites are responsible for the gas
accretion that enhances warps, one needs to track positions of individual gas cells from snapshot to snapshot. This
task is particularly tricky and computationally expensive when using the IllustrisTNG data, where gas cells are
continually changing their identity and turning into stars. To track the gas it is
necessary to use tracer particles, which are not cataloged into any subgroups, so the loop over all tracers in the
snapshot has to be done to find the interesting subset. Tracking the gas is also difficult specifically in the TNG100
simulation, because tracers are saved only in so-called full snapshots, which are fewer and
saved less frequently than regular snapshots.

For the reasons described above, we chose to study only one example, that we suspected to have its warp significantly
enhanced by gas accretion. Using maps similar to those in Fig.~\ref{examples2} we chose this particular case after
seeing a lot of movement in the surrounding gas. Also the fact that the rise of the warp occurred between two full
snapshots was a factor that made this case interesting and feasible to study.

This example is presented in Fig.~\ref{gas_acc}, where the two top plots show the column density of the gas of the
host, with gas cells of the perturber overlapped in orange. These two plots were made using two full snapshots before (i.e. at time marked as A in Fig.~\ref{gas_acc})
and after (i.e. at time marked as B in Fig.~\ref{gas_acc}) the pericenter passage of the satellite and the time of the rise of the warp (see Fig.~\ref{gas_acc2}). It is
clear that the majority of gas cells of the perturber were correctly assigned to it before the pericenter and later
they end up being accreted on the host and contributing to the warp. We found that between the times A and B the mass
of the gas within the cylinder of the host limited by $|z|<2\; R_{1/2*}$ and $R<5 \; R_{1/2*}$ (see subsection 3.1)
increased by around 50\%. The third plot shows the host's gas distribution without the contribution from the satellite,
to discriminate which parts of the warp the perturber is contributing the most. After the time marked as B, the
satellite is quickly accreted and absorbed by the host (see Fig.~\ref{gas_acc2}), which guarantees that the gas
that contributes to the warp at B does not escape anywhere, which could be possible if for example the satellite was
leaving the host with a very high velocity. Figure~\ref{gas_acc2} also shows that previous passages of this satellite
induced the warp in the past and it is a recurrent event. However, we did not check if the previous passages also
accreted gas (surely this accretion could not be as strong as between time A and B because the previous pericenter was
$\sim 3$ times larger).

Due to technical limitations described at the beginning of this subsection we only checked for this one case how the
gas accreted by the satellite can influence the warp. We found here that it is an important factor and an alternative
driving force of the warping, together with tidal interactions. This scenario can be further investigated in a zoom-in
or idealized simulations to verify its importance (e.g. two simulations differing only in the gas content of the
satellite can show how much stronger the resulting warp is in the case with the gas accretion).

\section{Discussion and summary}

\subsection{Comparison with previous studies}

Several studies were already performed aiming to investigate tidally induced warps, however, it is difficult to compare our
results with these findings, as they focus mostly on stellar warps. On the other hand, studies that focus on disturbed
gaseous disks did not explore in great detail interactions as a possible origin of vertical morphologies, and therefore
we can only compare a few general results with them. For example, \cite{bahe} classified by eye 2200 HI gas disks from
the EAGLE simulations and found that around 2/3 of them are disturbed. This is in a rough agreement with our findings
that around half of gas disks are not flat and the discrepancy in value may arise from various effects (e.g. different
simulation methods or human bias in classification).

\cite{gomez1} investigated warps and corrugations in stellar disks of 16 MW-like analogs in a cosmological zoom-in
Auriga simulation. They found that the majority of their warped stellar disks had a significant past
tidal interaction, which agrees with our conclusions that interactions play an important role in warp formation. They
determined the minimal mass of a perturber to influence a MW-sized galaxy to be $\sim10^{10}\;\mathrm{M}_{\odot}$,
which, by assuming the mass of the MW around $\sim10^{12}\;\mathrm{M}_{\odot}$ gives a rough fraction of 1\% of the
host's mass. 
This is consistent with Fig.~\ref{stats} in this work, where almost no cases were found for mass ratios smaller than 1\%.
\cite{gomez1} also found that, for most of their cases, stellar warps were followed by the star-forming
gaseous disk. We checked if stellar counterparts of our interaction-driven warps show similar structures at $z=0$ and
found that 20 out of 66 stellar disks show similar warps but with smaller amplitudes than gaseous ones, which is
to some extent expected from observations. The remaining 46 stellar disks seemed flat at $z=0$ probably because the last
pericenter in those cases was on average $>0.8$ Gyr earlier than for the cases with stellar warps.

\cite{kim} performed idealized $N$-body simulations to study tidally induced stellar warps. Despite the difference in
the disk component that we were interested in, some of our conclusions seem to be consistent. In both \cite{kim} and
our study the preferred values of the incident angles were found to lie around $45^{\circ}$ and $135^{\circ}$ and
tidally induced warps were asymmetric with asymmetries not exceeding the values of $6^{\circ}$ in their case and
$8^{\circ}-9^{\circ}$ in ours. These agreements and the results of their only simulation with gas (presented in their
Fig. 11) suggest that warps induced by interactions behave similarly in gas and stars. The only difference seems to
be the amplitude, which we found to be higher in gas disks, and this also seems to agree with their Fig. 11. This
happens probably because gas disks are more extended and less bound in the outer parts and therefore more sensitive to
perturbations. Discrepancies between the particular numerical values of e.g. lifetimes of warps obtained in this study
and the work of \cite{kim} can be attributed to many differences among which the most important seem to be: the fact
of studying different components, different simulation setups and methods, and different ways of measuring
warp parameters.

\subsection{Summary}

Observational correlations and results of numerical simulations have already demonstrated the importance of tidal
interactions for warp formation in spiral galaxies. In this paper, we used the TNG100 simulation from the IllustrisTNG
set to investigate for the first time the importance of tidal interactions for the formation of S-shaped gas warps in
galaxies in a cosmological context. Using a sample of 1593 spiral galaxies that had sufficient resolution and gas
content to reproduce gas warps, we found that 15.6\% of them had specific S-shaped warps. The S-shaped and other
vertical distortions were characteristic of 46.3\% of all galaxies from this sample, which agrees with observational
findings on the frequency of warps (\citealt{bosma}; \citealt{garciaruiz}).

We found that around one third of the sample of S-shaped warped galaxies were induced by interactions with other
galaxies. Half of these interactions ended in the perturber being absorbed by its host, while in the other half
perturbers maintained their identity at the end of the simulation. Gas warps induced by interactions in our sample
tend to have a lifetime of $<1$ Gyr, however, recurrent passages of satellites can regenerate them. Similarly to
observations (\citealt{garciaruiz}), we found that warps originating from interactions are more asymmetric, however
this asymmetry tends to decrease with time after the pericenter passage. Consistently with simulations of \cite{kim},
we found that angles between the angular momentum of the host and the orbital angular momentum of the perturber that
preferentially lead to warp formation are around $40^{\circ}-50^{\circ}$ and to a lesser extent
$120^{\circ}-130^{\circ}$. These values are optimal to balance the time of the interaction and the vertical component
of the tidal force, which are both important for warp formation. Finally, we found that accretion of the gas
from perturbers also contributes to the formation of warps. However, the structure and time resolution of the
simulations did not allow us to investigate in greater detail how strong this effect is and in how many cases it
is comparable to the tidal evolution.

In future work, we plan to investigate the origin of the remaining two-thirds of the S-shaped warps found in this
study, to understand better the mechanisms of warp formation. 

\section*{Acknowledgements}

This work was supported in part by the Polish National
Science Centre under grant 2013/10/A/ST9/00023 and by the STFC grant \#~ST/S000453/1. We are grateful to B.-E. Semczuk for the contribution to the visual classification. 
We appreciate insightful discussions with I. Ebrova, N. Peschken, D. Nelson and A. Pillepich that contributed to this paper. 
E.D.O. acknowledges support from the Center for Computational Astrophysics at Flatiron Institute for the hospitality during the completion of this work.
V.P.D. is supported by STFC Consolidated grant \#~ST/R000786/1.
The simulation TNG100 that was used in this work, is the flagship run of the IllustrisTNG project and it was run on the HazelHen Cray XC40-system at the High
Performance Computing Center Stuttgart as part of the project
GCS-ILLU of the Gauss Centre for Supercomputing. We are thankful to the IllustrisTNG team for granting us early access to the simulation data.

\bibliographystyle{mnras}

 \bibliography{TNG_biblio}

\bsp	
\label{lastpage}
\end{document}